\documentclass[aps,prd,eqsecnum,notitlepage,superscriptaddress,nofootinbib,longbibliography]{revtex4-2}

\usepackage{amsmath,physics}
\usepackage{bm}
\usepackage{graphicx}
\usepackage{natbib}
\usepackage[caption=false]{subfig}
\usepackage{hyperref}
\usepackage{array}
\usepackage{float}
\usepackage{xcolor}
\usepackage[normalem]{ulem}
\usepackage{tikz}
\usetikzlibrary{automata,arrows,positioning,calc}
\usepackage{standalone}
\usepackage{sidecap}

\global\arraycolsep=2pt
\DeclareMathOperator{\sgn}{sgn}

\usepackage{tikz,xcolor,hyperref}
\definecolor{lime}{HTML}{A6CE39}
\DeclareRobustCommand{\orcidicon}{%
	\begin{tikzpicture}
	\draw[lime, fill=lime] (0,0) 
	circle [radius=0.16] 
	node[white] {{\fontfamily{qag}\selectfont \tiny ID}};	\draw[white, fill=white] (-0.0625,0.095) 
	circle [radius=0.007];	\end{tikzpicture}
	\hspace{-2mm}}
\foreach \x in {A, ..., Z}{%
	\expandafter\xdef\csname orcid\x\endcsname{\noexpand\href{https://orcid.org/\csname orcidauthor\x\endcsname}{\noexpand\orcidicon}}
	}

\newcommand{\be}{\begin{equation}} 
\newcommand{\ee}{\end{equation}}

\pdfoutput=1

\begin{document}

\title{The energetics of quantum vacuum friction: Field fluctuations}

\author{Xin Guo\orcidA{}}
\email{guoxinmike@ou.edu}
\affiliation{H. L. Dodge Department of Physics and Astronomy, University of Oklahoma, Norman, OK 73019, USA}

\author{Kimball A. Milton\orcidB{}}
\email{kmilton@ou.edu}
\affiliation{H. L. Dodge Department of Physics and Astronomy, University of Oklahoma, Norman, OK 73019, USA}

\author{Gerard Kennedy\orcidC{}}
\email{g.kennedy@soton.ac.uk}
\affiliation{School of Mathematical Sciences, University of Southampton, Southampton, SO17 1BJ, United Kingdom}

\author{William P. McNulty\orcidD{}}
\email{william.p.mcnulty-1@ou.edu}
\affiliation{H. L. Dodge Department of Physics and Astronomy, University of Oklahoma, Norman, OK 73019, USA}

\author{Nima Pourtolami\orcidF{}}
\email{nima.pourtolami@nbc.ca}
\affiliation{National Bank of Canada, Montreal, QC H3B 4S9, Canada}

\author{Yang Li\orcidE{}}
\email{leon@ncu.edu.cn}
\affiliation{Department of Physics, Nanchang University, Nanchang 330031, China}

\date{\today}

\begin{abstract}
Quantum fluctuations can induce a friction on a neutral but polarizable particle and cause it to radiate energy even if the particle is moving in free space filled with blackbody radiation, and is not in contact with or close to any surface or other object. We explore the energetics of such a particle moving uniformly in vacuum, continuing our previous investigations of quantum friction. The intrinsic polarizability of the particle is considered to be purely real before it is dressed by radiation. The particle is then guaranteed to be in the nonequilibrium steady state (NESS), where it absorbs and emits energy at the same rate. We first calculate the quantum frictional power and force on the particle in the rest frame of the blackbody radiation from first principles, namely the Maxwell-Heaviside equations and the Lorentz force law. Then we provide a simpler method of obtaining the same quantities in the rest frame of the particle by using the principle of virtual work. The equivalence of the two approaches is illustrated. The formulas we derive for quantum vacuum frictional power and force are fully relativistic and applicable to finite temperature. In NESS, the quantum vacuum frictional force on the particle is shown to be a true drag, independent of the model for polarizability and the polarization state of the particle. Finally, we give an estimate of the quantum vacuum friction on a gold atom and comment on the feasibility of detecting such quantum vacuum frictional effects.
\end{abstract}


\maketitle

\section{introduction}\label{introduction}
Friction has always been an intriguing subject to study. Recently, we have investigated the friction felt by a charged particle passing above a metallic surface \cite{Kim:charged} as well as by a moving neutral particle carrying either an electric or a magnetic dipole moment \cite{Kim:dipole}. What serves as a frictional force in these situations is just the classical electromagnetic force given by the Lorentz force law. However, friction can also be induced by quantum fluctuations. This idea has been around for decades, tracing back to Ref.~\cite{Pendry:1997} or even earlier Ref.~\cite{Teodorovich:1978, Levitov:1989, Hoye:1992, Hoye:1993}. For a brief review of the history, see Ref.~\cite{Kim:reality}. While most authors mainly consider quantum friction in some complicated background, the frictional effect in free space is often dismissed in the literature. This friction on matter due to its interaction with the surrounding blackbody radiation is what we term quantum vacuum friction in this paper. In Ref.~\cite{Kim:dipole}, we have already calculated the quantum vacuum friction on a neutral but polarizable particle with intrinsic dissipation moving through blackbody radiation. A Lorentz transformation to the rest frame of the particle (frame $ \mathcal{P} $) does not eliminate the quantum vacuum friction because it is the relative motion between the particle and the blackbody radiation that causes this frictional force. Mkrtchian et al.~\cite{Mkrtchian:universal} have argued that this ``universal drag'' is not without effect in various contexts, ranging from tungsten ovens to the cosmos. 

In this paper, we continue our efforts in Ref.~\cite{Kim:dipole} and calculate the quantum frictional power and force on a neutral but polarizable particle moving uniformly in vacuum. Unlike in Ref.~\cite{Kim:dipole}, the neutral particle considered in this paper is characterized by a real intrinsic polarizability $ \bm{\alpha}(\omega) $. But, as is demonstrated in Ref.~\cite{Kim:dipole}, some dissipative mechanism is always required for any frictional effect to occur. Here, it is through the particle's interaction with fluctuations of the electromagnetic field that the particle's effective polarizability $ \hat{\bm{\alpha}}(\omega) $ acquires an imaginary part, which is second order in $ \bm{\alpha}(\omega) $. Since we assume there is no  dissipation intrinsic to the particle, it must be in the nonequilibrium steady state (NESS), where its energy is conserved. In the rest frame of the blackbody radiation (frame $ \mathcal{R} $), it is precisely the radiation reaction on the particle that plays the role of the quantum friction. To keep the particle moving with constant velocity, the quantum friction has to be balanced by an external driving force. As a result, the energy dissipated to the vacuum through the quantum friction or blackbody radiation reaction is compensated by the positive work done by the external driving force. One may also choose to view NESS in frame $ \mathcal{P} $, where the energetics becomes even simpler. The quantum friction is still balanced by the external force to keep the particle fixed. But neither of the forces do any work on the particle. We explore the energetics of such a particle from both perspectives and derive formulas for the quantum frictional power and force. 

Various theoretical groups have also recently studied the quantum friction in similar contexts. Volokitin and Persson discussed the blackbody friction for a moving particle in Section 8.5 of their book \cite{Volokitin:book} and made a connection with the famous Einstein-Hopf effect. Similar to our treatment in Ref.~\cite{Kim:dipole}, they include dissipation in the particle's polarizability from the start and treat dipole fluctuations as independent from field fluctuations. As a result, the friction formula they derive involves both the temperature of the radiation and the temperature of the particle. But the latter can be eliminated by imposing the steady state condition. In Ref.~\cite{Dedkov:friction}, Dedkov and Kyasov obtained analytic expressions for nonrelativistic quantum friction at arbitrary temperature in the particle-plate and plate-plate configurations. In a series of papers, Intravaia et al.~\citep{Intravaia:2014PRA, Intravaia:2016PRL, Intravaia:2016PRA, Intravaia:NTQF} have extensively studied quantum friction on a particle moving above a surface. Similar to this paper, they consider the dipole fluctuations as entirely induced by the field fluctuations. However, their formulas are limited to the nonrelativistic and zero temperature regime. In contrast, our results for quantum frictional power and force are fully relativistic and applicable to arbitrary temperature. Even though we have focused on the vacuum situation in this paper, the formulation we give can be extended to a more general background.

As is well-known, the environment that an open quantum system interacts with can be modeled by a bath of harmonic oscillators \citep{Caldeira:Quantum}. Earlier works \cite{HuPaz,HuMatacz,Polonyi:Dissipation}
have been devoted to study quantum Brownian motion, where a Brownian particle couples linearly to such a bath of oscillators. Indeed, the dissipative mechanism behind the scenes in our work on quantum vacuum friction is similar. The neutral particle considered in this paper is an open quantum system interacting with the electromagnetic field, which can be thought of as an infinite set of harmonic oscillators. And we are here examining the manifestation of the induced dissipation as friction on the neutral particle moving through the thermal vacuum. 

The outline of the paper is as follows. In Sec.~\ref{power} and Sec.~\ref{force}, we quantize the classical formula for power and force on a moving classical dipole in frame $ \mathcal{R} $ by applying the fluctuation-dissipation theorem. The separate calculations for power and force confirm the power-force relation $ P=Fv $ explicitly. It is also shown that both the frictional power and force on the particle in NESS are negative definite, independent of the specific form of its intrinsic polarizability and its polarization state. In Sec.~\ref{rest}, we calculate the quantum frictional power and force in frame $ \mathcal{P} $ by differentiating the interaction free energy, using the principle of virtual work. The results obtained through this approach are shown to agree with those in Sec.~\ref{power} and Sec.~\ref{force}. The calculation in frame $ \mathcal{P} $ is not only simpler but more systematic in examining contributions from different polarization states of the particle. Also, by introducing the effective polarizability $ \hat{\bm{\alpha}}(\omega) $, we are able to extend our analysis to all orders in the intrinsic polarizability $ \bm{\alpha}(\omega) $. In Sec.~\ref{numerics}, the renormalized effective polarizability is used to calculate the quantum vacuum friction on a gold atom, the intrinsic polarizability of which is static and isotropic before being dressed by radiation. Only above $ T=6000 \rm{K} $ is the deviation from the low-frequency radiation reaction model seen for the effective dissipation of the gold atom. The frictional force on the gold atom is found to be too small to be observed experimentally around room temperature ($ T=300 \,\rm{K} $) but it becomes close to experimental reach when the temperature is raised by two orders of magnitude ($ T=30, 000 \,\rm{K} $). Concluding remarks can be found in Sec.~\ref{conclusions}.

In Appendix~\ref{apA}, the explicit form of both the general Green's dyadic and the vacuum Green's dyadic are given and the symmetries of the Green's functions are discussed. In Appendix~\ref{apB}, the Lorentz transformation properties for the dipole and the field are described both in the spacetime domain and in momentum space, since they are needed in the derivation of the quantum frictional power and force. Appendix~\ref{apC} defines the the momemtum distribution functions for different polarization states and collects their integrals often used in the formulas for the quantum vacuum frictional power and force in NESS. Appendix~\ref{apD} gives the formulas for quantum friction in a general background with translational symmetry in the $ x $ and $ y $ directions. Appendix~\ref{apE} provides a proof of the principle of virtual work to be applied to our calculation in frame $ \mathcal{P} $. Appendix~\ref{apF} proves that the imaginary parts of the diagonal elements of the effective polarizability are always nonnegative. Appendix~\ref{apG} illustrates how the emitted power coincides with the classical dipole radiation.

In this paper we use Heaviside-Lorentz (rationalized) electromagnetic units. We also set $ k_{B}=c=\hbar=1 $ in the derivation of formulas but the SI units are used in the numerical evaluations.

\section{quantum frictional power}\label{power}
\subsection{Quantization of the power in the rest frame of the blackbody radiation}\label{qpower}
In frame $ \mathcal{R} $, consider a neutral but polarizable particle moving uniformly in vacuum. The intrinsic polarizability of the neutral particle measured in its own rest frame is $ \bm{\alpha}(\omega) $.\footnote{It will be shown later that this polarizability must have been renormalized, absorbing the divergent part of the vacuum Green's dyadic.}  With no loss of generality, let us assume that it is moving in the $ x $ direction with velocity $ \vb{v}=v\hat{\vb{x}} $ and trajectory $ \vb{r}(t)=\vb{v}t $. The same physical situation can be transformed into frame $ \mathcal{P} $, where the particle sits in a fixed position, which we assume to be the origin $ \vb{r'}=\vb{0} $. Throughout the paper, primes are used to indicate quantities or spacetime coordinates in frame $ \mathcal{P} $, except that primes on the polarizabilities are all omitted as they are always evaluated in frame $ \mathcal{P} $ whenever they appear.

In general, the electromagnetic power into a system could be calculated by integrating the density of the rate at which the electromagnetic force does work on it,
\begin{equation}\label{eq2-1}
P(t)=\int d\vb{r} \, \vb{j}(t,\vb{r})\cdot\vb{E}(t,\vb{r}).
\end{equation}
To start with, let us first consider a time-dependent dipole moving with constant velocity $ \vb{v} $. The corresponding classical charge density and current density are
\begin{subequations}\label{eq2-2}
\begin{equation}\label{eq2-2a}
\rho(t,\vb{r})=-\div \vb{d}(t)\delta(\vb{r}-\vb{v}t),
\end{equation}
\begin{equation}\label{eq2-2b}
\vb{j}(t,\vb{r})=-\vb{v} \div{\vb{d}}(t)\delta(\vb{r}-\vb{v}t)+\dot{\vb{d}}(t)\delta(\vb{r}-\vb{v}t).
\end{equation}
\end{subequations}
When the current \eqref{eq2-2b} is inserted into Eq. \eqref{eq2-1}, we obtain
\begin{equation}\label{eq2-3}
P(t)=\vb{d}(t)\cdot\grad[\vb{v}\cdot\vb{E}(t,\vb{r}=\vb{v}t)]+\dot{\vb{d}}(t)\cdot\vb{E}(t,\vb{r}=\vb{v}t).
\end{equation}

Now consider the original problem we have in mind, where a neutral but polarizable particle is moving through vacuum. Classically, the particle does not possess any intrinsic dipole moment nor is any electromagnetic field present in the configuration. But quantum mechanically, electromagnetic field fluctuations are able to induce a dipole moment of the particle, which in turn induces an electromagnetic field through the following relations:
\begin{subequations}\label{induce}
\begin{equation}\label{eq2-6}
\vb{d}'(\omega)=\bm{\alpha}(\omega)\cdot \vb{E}'(\omega,\vb{0}),
\end{equation}
\begin{equation}\label{eq2-7}
\vb{E}(\omega;\vb{r})=\int d\tilde{\vb{r}} \left(-\frac{1}{i\omega}\right)\vb{\Gamma}(\omega;\vb{r},\tilde{\vb{r}}) \cdot \vb{j} (\omega,\tilde{\vb{r}}),
\end{equation}
\end{subequations}
where $ \vb{\Gamma} $ is the retarded Green's dyadic in vacuum, which has its explicit form detailed in Appendix~\ref{apA}. We note the first relation for the induced dipole is expressed in frame $ \mathcal{P} $ while the second relation for the induced field is more conveniently written down in frame $ \mathcal{R} $. As a result, the Lorentz transformation properties for the dipole and the fields are frequently used in the derivation of the frictional power and force. These are collected in Appendix~\ref{apB}. 

There are two leading contributions to the power, because $ P(t) $ in Eq.~\eqref{eq2-3} could be expanded to second order in $ \bm{\alpha} $ in two different ways:
\begin{equation}\label{eq2-7.5}
P=P_{\rm{I}}+P_{\rm{II}}.
\end{equation}
For the $ \rm{I} $ contribution, the $ \vb{d} $ operator in Eq.~\eqref{eq2-3} is expanded to second order in the intrinsic polarizability $ \bm{\alpha}(\omega) $ using Eq.~\eqref{induce} while the $ \vb{E} $ operator in Eq.~\eqref{eq2-3} is not expanded. For the $ \rm{II} $ contribution, the $ \vb{d} $ operator in Eq.~\eqref{eq2-3} is expanded to only first order in $ \bm{\alpha}(\omega) $ while the original $ \vb{E} $ operator in Eq.~\eqref{eq2-3} is expanded to first order in $ \bm{\alpha}(\omega) $ at the same time, with the resultant product of the two operators being second order in $ \bm{\alpha}(\omega) $. In frame $\mathcal{P}  $, the power $ P' $ is also broken into two contributions, which exactly correspond to the $ \vb{EE} $ and $ \vb{dd} $ contributions in Ref.~\cite{Kim:dipole}, as illustrated towards the end of Sec.~\ref{rest}.

After the expansion, each contribution contains a correlation of field operators. We then use the fluctuation-dissipation theorem (FDT) in frame $ \mathcal{R} $ to evaluate the correlation functions,
\begin{equation}\label{eq2-8}
\langle E_{i}(t_{1}, \vb{r}_{1})E_{j}(t_{2}, \vb{r}_{2}) \rangle=
\int_{-\infty}^{\infty} \frac{d\omega}{2\pi}e^{-i\omega (t_{1}-t_{2})}\Im\Gamma_{ij}(\omega; \vb{r}_{1},\vb{r}_{2})\coth{\left(\frac{\beta\omega}{2}\right)},
\end{equation}
where $ \beta $ is the inverse temperature of the blackbody radiation  and the field operators have been symmetrized so that the correlation function is real.\footnote{As discussed in Ref.~\cite{Kim:dipole},
 we use the correlation function for the symmetrized
product of operators, because this implies a Hermitian interaction.  Such
symmetrized correlation functions automatically occur when the closed-time-path
formalism is employed to derive the FDT
\cite{Hu:Nonequilibrium}.}

\subsection{X polarization}
Let us first calculate the contribution to the power due to a nonvanishing $ x $-$ x $ component of the polarizability $ \alpha_{xx} $. Then the induced dipole could only be polarized in the $ x $ direction. After a Fourier transform on the dipole operator, the power is written as
\begin{equation}\label{eq2-5a}
P^{X}(t)=\int \frac{d\omega}{2\pi}e^{-i\omega t} d_{x}(\omega)(v\partial_{x}-i\omega) E_{x}(t,\vb{r}=\vb{v}t).
\end{equation}
Here, we describe in detail how $ P_{\rm{I}}^{X} $ is derived.
First, expand the dipole in its rest frame using \eqref{eq2-6},
\begin{equation}\label{eq2-10}
d_{x}(\omega)=d'_{x}(\gamma \omega)=\alpha_{xx}(\gamma\omega)E'_{x}(\gamma \omega,\vb{0}),
\end{equation}
where $ \gamma=\frac{1}{\sqrt{1-v^{2}}} $ is the relativistic dilation factor.
Next, Lorentz transform the field to frame $ \mathcal{R} $,
\begin{equation}\label{eq2-11}
E'_{x}(\gamma \omega,\vb{0})=\int dt'_{1} e^{i\gamma\omega t'_{1}}E_{x}(t_{1}=\gamma t'_{1},x_{1}=\gamma v t'_{1}),
\end{equation}
where the $ y $ and $ z $ dependences of the field are suppressed as their Lorentz transformation is trivial.
Then keep expanding the field in \eqref{eq2-11} using \eqref{eq2-7}:\footnote{Here $ j_{x} $ is the only nonvanishing component of the induced current, we therefore only count the diagonal contribution. On the other hand, the off-diagonal contributions would vanish simply based on the structure of the Green's function even if other components of the current exists, which is confirmed by the explicit calculation in Sec.~\ref{rest}. }
\begin{equation}\label{eq2-12}
E_{x}(t_{1}=\gamma t'_{1},x_{1}=\gamma v t'_{1})=\int\frac{d\omega_{1}}{2\pi}e^{-i\omega_{1}\gamma t'_{1}}\int d\tilde{\vb{r}} \left(-\frac{1}{i\omega_{1}}\right)\Gamma_{xx}(\omega_{1};\vb{r}_{1},\tilde{\vb{r}})j_{x}(\omega_{1};\tilde{\vb{r}}),
\end{equation}
where the Fourier transformed current is
\begin{equation}\label{eq2-12.2}
j_{x}(\omega_{1};\tilde{\vb{r}})=-i\frac{\omega_{1}}{v} d_{x}\left(\frac{\tilde{x}}{v}\right) e^{i\omega_{1} \tilde{x}/v }\delta\left(\tilde{y}\right)\delta\left(\tilde{z}\right).
\end{equation}
The Green's function could be further Fourier transformed in the $ x $ direction,
\begin{equation}\label{eq2-12.4}
\Gamma_{xx}(\omega_{1};\vb{r}_{1},\tilde{\vb{r}})=\int\frac{dk_{x}}{2\pi} e^{ik_{x}(x_{1}-\tilde{x})} G_{xx}(\omega_{1},k_{x}),
\end{equation}
where the $ y $ and $ z $ dependence of $  G_{xx}(\omega_{1},k_{x}) $ is also suppressed.
Once again, expand the dipole operator inside the current in Eq.~\eqref{eq2-12.2},
\begin{equation}\label{eq2-13}
d_{x}\left(\frac{\tilde{x}}{v}\right)=\frac{1}{\gamma}d'_{x}\left(\frac{\tilde{x}}{\gamma v}\right)=\frac{1}{\gamma}\int\frac{d\omega_{2}}{2\pi} e^{-i\omega_{2}\tilde{x}/\gamma v} \alpha_{xx}(\omega_{2})E'_{x}(\omega_{2},\vb{0}).
\end{equation}
Finally, Lorentz transform the field in \eqref{eq2-13} back to frame $ \mathcal{R} $,
\begin{equation}\label{eq2-14}
E'_{x}(\omega_{2},\vb{0})=\int dt'_{2} e^{i\omega_{2}t'_{2}} E_{x}(t_{2}=\gamma t'_{2}, x_{2}=\gamma v t'_{2}).
\end{equation}
Apply the FDT to evaluate the correlation between the field operator appearing in \eqref{eq2-14} and the field operator in \eqref{eq2-5a} using the Fourier transformed Green's function  in the $ x $ direction,
\begin{equation}\label{eq2-15}
\langle E_{x}(t_{2}, x_{2})E_{x}(t, x) \rangle=
\int_{-\infty}^{\infty} \frac{d\omega_{3}}{2\pi}e^{-i\omega_{3} (t_{2}-t)} \int\frac{d\bar{k}_{x}}{2\pi} e^{i\bar{k}_{x}(x_{2}-x)}\Im G_{xx}(\omega_{3},\bar{k}_{x})\coth{\left(\frac{\beta\omega_{3}}{2}\right)}.
\end{equation}
When \eqref{eq2-10}--\eqref{eq2-15} are assembled into \eqref{eq2-5a} and the integrals on $ t_{1}'$, $t_{2}'$ , $ \tilde{x} $ are carried out, we find $ P_{\rm{I}}^{X} $ is actually time independent, as a consequence of the time translational symmetry of the FDT shown in \eqref{eq2-8},
\begin{equation}\label{eq2-16}
P_{\rm{I}}^{X}=-\frac{i}{\gamma^{2}}\int\frac{d\omega}{2\pi}\int\frac{dk_{x}}{2\pi}\int\frac{d\bar{k}_{x}}{2\pi} \,\alpha_{xx}^{2}(\gamma\omega)(\omega+\bar{k}_{x}v)
 G_{xx}(\omega+k_{x}v, k_{x}) \Im G_{xx}(\omega+\bar{k}_{x}v, \bar{k}_{x})\coth\left(\frac{\beta(\omega+\bar{k}_{x}v )}{2}\right).
\end{equation}

Taking into account the symmetry of the real polarizability in frequency $ \bm{\alpha}(-\omega)=\bm{\alpha}(\omega) $ together with the symmetry properties of the Green's functions discussed in Appendix~\ref{apA}, only the imaginary part of the first Green's function in Eq.~\eqref{eq2-16} will be picked out. Rescaling the frequency $ \gamma\omega=\tilde{\omega} $, the power $P_{\rm{I}}^{X}$ may be written as
\begin{align}\label{eq2-17}
P_{\rm{I}}^{X}=\frac{1}{4\pi^{3}\gamma^{3}}\int_{0}^{\infty} d\tilde{\omega} \int dk_{x} d\bar{k}_{x}
\,\alpha_{xx}^{2}(\tilde{\omega})\left(\frac{\tilde{\omega}}{\gamma}+\bar{k}_{x}v\right)
&\Im G_{xx}\left(\frac{\tilde{\omega}}{\gamma}+k_{x}v, k_{x}\right)\nonumber\\ 
\cross &\Im G_{xx}\left(\frac{\tilde{\omega}}{\gamma}+\bar{k}_{x}v, \bar{k}_{x}\right)\coth\left[\frac{\beta}{2}\left(\frac{\tilde{\omega}}{\gamma}+\bar{k}_{x}v\right)\right],
\end{align}
where we have used the fact that the integrand is even in $ \omega $ after picking out the imaginary part of the first Green's function. 

The explicit form for the vacuum Green's functions is shown in Eq.~\eqref{eqA-7},
\begin{equation}\label{eq2-17.2}
G_{xx}(\nu, k_{x})=\int\frac{dk_{y}}{2\pi} \frac{\nu^{2}-k_{x}^{2}}{2\kappa},
\end{equation}
where $ \kappa=\sqrt{k^{2}-\nu^{2}} $ becomes imaginary when $ \nu^{2}>k^{2} $,
\begin{equation}\label{eq2-17.4}
\kappa=-i\sgn(\nu)\sqrt{\nu^{2}-k^{2}}.
\end{equation}
Since $ \kappa $ is the only place where an imaginary part could arise in the Green's function, $ \Im G_{xx}(\nu,k_{x}) $ emerges in the region $ \nu^{2}-k_{x}^{2}>k_{y}^{2}>0 $ but vanishes otherwise. Defining $ k_{y}^{0}=\sqrt{\nu^{2}-k_{x}^{2}} $, $ \Im G_{xx}(\nu,k_{x}) $ can then be written as
\begin{equation}\label{eq2-17.6}
 \addtolength{\arraycolsep}{-3pt}
 \Im G_{xx}(\nu,k_{x})=\left\{%
 \begin{array}{lcrcl}
 \sgn(\nu)\int_{-k_{y}^{0}}^{k_{y}^{0}}\frac{dk_{y}}{2\pi} \frac{(k_{y}^{0})^{2}}{2\sqrt{(k_{y}^{0})^{2}-k_{y}^{2}}}=\frac{1}{4}(k_{y}^{0})^{2}\sgn(\nu),& \qquad k_{x}^{2}<\nu^{2},\\
0, & \qquad k_{x}^{2}>\nu^{2}.
\end{array}\right.
 \end{equation}
When we insert \eqref{eq2-17.6}  into \eqref{eq2-17}, the power reads
\begin{align}\label{eq2-18}
P_{\rm{I}}^{X}=&\frac{1}{64\pi^{3}\gamma^{3}}\int_{0}^{\infty} d\tilde{\omega}\,\alpha_{xx}^{2}(\tilde{\omega}) \int dk_{x}  \, \sgn\left(\frac{\tilde{\omega}}{\gamma}+k_{x}v\right) \left[\left(\frac{\tilde{\omega}}{\gamma}+k_{x}v\right)^{2}-k_{x}^{2}\right]\nonumber \\
&\cross \int d\bar{k}_{x} \sgn\left(\frac{\tilde{\omega}}{\gamma}+\bar{k}_{x}v\right)\left[\left(\frac{\tilde{\omega}}{\gamma}+\bar{k}_{x}v\right)^{2}-\bar{k}_{x}^{2}\right]
\left(\frac{\tilde{\omega}}{\gamma}+\bar{k}_{x}v\right)
\coth\left[\frac{\beta}{2}\left(\frac{\tilde{\omega}}{\gamma}+\bar{k}_{x}v\right)\right].
\end{align}
The integral limits on $ k_{x} $ and $ \bar{k}_{x} $ are now over a finite range determined by $ (\frac{\tilde{\omega}}{\gamma}+k_{x}v)^{2}-k_{x}^{2}>0 $ and $ (\frac{\tilde{\omega}}{\gamma}+\bar{k}_{x}v)^{2}-\bar{k}_{x}^{2}>0 $ respectively.
It is convenient to make changes of variables $ (\frac{\tilde{\omega}}{\gamma}+k_{x}v)=\tilde{\omega} y $ and $ (\frac{\tilde{\omega}}{\gamma}+\bar{k}_{x}v)=\tilde{\omega}\bar{y} $ and use the definition of the momentum distribution function for $ x $ polarization in Appendix \ref{apC}
\begin{equation}\label{eq2-18.5}
f^{X}(y)=\frac{3}{4\gamma v} \left[y^{2}-\left(y-\frac{1}{\gamma}\right)^{2} \frac{1}{v^{2}}\right]=\frac{3}{4 \gamma v} \left[1-\frac{1}{\gamma^{2}v^{2}}(y-\gamma)^{2}\right].
\end{equation}
Then Eq.~\eqref{eq2-18} can be expressed as
\begin{equation}\label{eq2-19}
P_{\rm{I}}^{X}=\frac{1}{36\pi^{3}\gamma}\int_{0}^{\infty} d\tilde{\omega}\,\alpha_{xx}^{2}(\tilde{\omega})\,\tilde{\omega}^{7} \int_{y_{-}}^{y_{+}} dy\, f^{X}(y) \int_{y_{-}}^{y_{+}} d\bar{y}\:\bar{y}f^{X}(\bar{y}) \coth(\frac{\beta\tilde{\omega}}{2}\bar{y}),
\end{equation}
where $ y_{+}=\sqrt{\frac{1+v}{1-v}} $ and $ y_{-}=\sqrt{\frac{1-v}{1+v}} $.
With $ f^{X}(y) $ being normalized as detailed in Appendix \ref{apC}, the integral on $ y $ is readily carried out, leading to
\begin{align}\label{eq2-20}
P_{\rm{I}}^{X}=\frac{1}{36\pi^{3}\gamma}\int_{0}^{\infty} d\tilde{\omega}\,\alpha_{xx}^{2}(\tilde{\omega})\,\tilde{\omega}^{7} 
\int_{y_{-}}^{y_{+}} d\bar{y}\,\bar{y}f^{X}(\bar{y})\coth(\frac{\beta\tilde{\omega}}{2}\bar{y}).
\end{align}
Although the integral on $ \bar{y} $ can be carried out explicitly and expressed in terms of polylogarithms, it is not very illuminating, so we leave the result in the current form.

Following the second expansion scheme described after Eq.~\eqref{eq2-7.5}, we likewise find
\begin{align}\label{eq2-20.5}
P_{\rm{II}}^{X}=-\frac{1}{4\pi^{3}\gamma^{3}}\int_{0}^{\infty} d\tilde{\omega} \int dk_{x} d\bar{k}_{x}\, \alpha_{xx}^{2}(\tilde{\omega})\left(\frac{\tilde{\omega}}{\gamma}+k_{x}v\right)
&\Im G_{xx}\left(\frac{\tilde{\omega}}{\gamma}+k_{x}v, k_{x}\right)\nonumber\\
 \cross &\Im G_{xx}\left(\frac{\tilde{\omega}}{\gamma}+\bar{k}_{x}v, \bar{k}_{x}\right)\coth\left[\frac{\beta}{2}\left(\frac{\tilde{\omega}}{\gamma}+\bar{k}_{x}v\right)\right].
\end{align}
Comparing \eqref{eq2-17} and \eqref{eq2-20.5}, we see that $  P_{\rm{II}}^{X}$ can be immediately obtained from $  P_{\rm{I}}^{X}$ by changing the sign of  $  P_{\rm{I}}^{X} $ and trading the factor $ (\frac{\tilde{\omega}}{\gamma}+\bar{k}_{x}v) $ for $ (\frac{\tilde{\omega}}{\gamma}+k_{x}v )$. This turns out to be a general symmetry that applies between $ \rm{I} $ and $ \rm{II} $ contributions.
Making similar variable changes as is done on $ P_{\rm{I}}^{X} $ and inserting the vacuum Green's functions, we can write $ P_{\rm{II}}^{X} $ as
\begin{equation}\label{eq2-21}
P_{\rm{II}}^{X}=-\frac{1}{36\pi^{3}\gamma}\int_{0}^{\infty} d\tilde{\omega}\,\alpha_{xx}^{2}(\tilde{\omega})\,\tilde{\omega}^{7} \int_{y_{-}}^{y_{+}} dy\, yf^{X}(y) \int_{y_{-}}^{y_{+}} d\bar{y}\,f^{X}(\bar{y}) \coth(\frac{\beta\tilde{\omega}}{2}\bar{y}).
\end{equation}
When the integral on $ y $ is carried out using the integrals provided in Appendix \ref{apC}, we find
\begin{equation}\label{eq2-22}
P^{X}_{\rm{II}}=-\frac{1}{36\pi^{3}}\int_{0}^{\infty} d\tilde{\omega}\,\alpha_{xx}^{2}(\tilde{\omega})\,\tilde{\omega}^{7} 
\int_{y_{-}}^{y_{+}} d\bar{y}\, f^{X}(\bar{y}) \coth(\frac{\beta\tilde{\omega}}{2}\bar{y}).
\end{equation} 

We find the quantum frictional power due to the $ x $ polarization by adding Eq.~\eqref{eq2-17} and Eq.~\eqref{eq2-20.5},
\begin{align}\label{eq2-22.5}
P^{X}=\frac{v}{4\pi^{3}\gamma^{3}}\int_{0}^{\infty} d\tilde{\omega} \int dk_{x} d\bar{k}_{x}\, \alpha_{xx}^{2}(\tilde{\omega})\,(\bar{k}_{x}-k_{x})
&\Im G_{xx}\left(\frac{\tilde{\omega}}{\gamma}+k_{x}v, k_{x}\right)\nonumber\\
 \cross &\Im G_{xx}\left(\frac{\tilde{\omega}}{\gamma}+\bar{k}_{x}v, \bar{k}_{x}\right)\coth\left[\frac{\beta}{2}\left(\frac{\tilde{\omega}}{\gamma}+\bar{k}_{x}v\right)\right].
\end{align}
This formula can be applied to general backgrounds with translational symmetry in the $ x $-$ y $ plane, since we have not specified the Green's functions.

In the vacuum situation, we may obtain $ P^{X} $ by simply adding Eq.~\eqref{eq2-20} and Eq.~\eqref{eq2-22}. Even though the $ \omega $ integrals involved in $  P_{\rm{I}}^{X} $ and $ P_{\rm{II}}^{X} $ have ultraviolet divergences unless $ \alpha(\omega) $ falls off faster than $ 1/\omega^{4} $, the quantum vacuum frictional power $ P^{X} $ is free from such divergences because the potentially divergent pieces cancel when doing the integration on $ k_{x} $ and $ \bar{k}_{x} $.
The quantum vacuum frictional power due to the $ x $ polarization is then
\begin{equation}\label{eq2-23}
P^{X}=\frac{1}{18\pi^{3}\gamma}\int_{0}^{\infty} d\tilde{\omega}\,\alpha_{xx}^{2}(\tilde{\omega})\,\tilde{\omega}^{7} 
\int_{y_{-}}^{y_{+}} d\bar{y}\,(\bar{y}-\gamma)f^{X}(\bar{y}) \frac{1}{e^{\beta\tilde{\omega}\bar{y}}-1},
\end{equation}
noting the integrand on $ \bar{y} $ without the exponential factor is odd with respect to $ \bar{y}=\gamma $.

\subsection{Other polarizations}
Let us now turn to the contribution to the frictional power due to other polarizations. If $ \alpha_{yy} $ is the only nonvanishing component of the polarizability, the power becomes
\begin{equation}\label{eq2-5b}
P^{Y}(t)=\int \frac{d\omega}{2\pi}e^{-i\omega t} d_{y}(\omega)[v\partial_{y}E_{x}(t,\vb{r}=\vb{v}t)-i\omega E_{y}(t,\vb{r}=\vb{v}t)].
\end{equation}
The only complication is that the magnetic field will appear when the electric field is transformed from $ \mathcal{P} $ to $ \mathcal{R} $. This requires the use of Faraday's law in Fourier space, $ \curl \vb{E}(\omega;\vb{r})=i\omega \vb{B}(\omega;\vb{r}) $.
Then following the same quantization procedure as outlined for $ P^{X} $, we find the two contributions to the vacuum frictional power $ P^{Y} $ are respectively,
\begin{subequations}\label{2-24}
\begin{equation}\label{eq2-24a}
P_{\rm{I}}^{Y}=\frac{1}{36\pi^{3}\gamma}\int_{0}^{\infty} d\tilde{\omega}\,\alpha_{yy}^{2}(\tilde{\omega})\,\tilde{\omega}^{7} \int_{y_{-}}^{y_{+}} d\bar{y} \, \bar{y} f^{Y}(\bar{y})\coth(\frac{\beta\tilde{\omega}}{2}\bar{y}),
\end{equation}
\begin{equation}\label{eq2-24b}
P_{\rm{II}}^{Y}=-\frac{1}{36\pi^{3}}\int_{0}^{\infty} d\tilde{\omega}\,\alpha_{yy}^{2}(\tilde{\omega})\,\tilde{\omega}^{7} \int_{y_{-}}^{y_{+}} d\bar{y}\, f^{Y}(\bar{y})\coth(\frac{\beta\tilde{\omega}}{2}\bar{y}),
\end{equation}
\end{subequations}
where we have used the momentum distribution function for the $ Y $ polarization defined in Appendix \ref{apC},
\begin{equation}\label{eq2-24.5}
f^{Y}(y)=\frac{3}{4\gamma v}\left\lbrace 1-\frac{1}{2}\left[y^{2}-\left(y-\frac{1}{\gamma}\right)^{2} \frac{1}{v^{2}}\right]\right\rbrace=\frac{3}{4 \gamma v}\left\lbrace 1-\frac{1}{2}\left[1-\frac{1}{\gamma^{2}v^{2}}\left(y-\gamma\right)^{2}\right]\right\rbrace.
\end{equation}
These formulas are just like Eq.~\eqref{eq2-20} and \eqref{eq2-22}, only replacing $ f^{X} $ with $ f^{Y} $. As a result, the total contribution to the power due to the $ y $-$ y $ component of polarizability $ P^{Y} $ has the same form as $ P^{X} $ in \eqref{eq2-23},
\begin{equation}\label{eq2-25}
P^{Y}=\frac{1}{18\pi^{3}\gamma}\int_{0}^{\infty} d\tilde{\omega}\,\alpha_{yy}^{2}(\tilde{\omega})\,\tilde{\omega}^{7} 
\int_{y_{-}}^{y_{+}} d\bar{y}\,(\bar{y}-\gamma)f^{Y}(\bar{y}) \frac{1}{e^{\beta\tilde{\omega}\bar{y}}-1}.
\end{equation}

From the symmetry of the problem considered, the contribution to the power due to the $ z $ polarization would be like the $ y $ polarization, with $ \alpha_{zz} $ replacing the $ \alpha_{yy} $,
\begin{equation}\label{eq2-26}
P^{Z}=\frac{1}{18\pi^{3}\gamma}\int _{0}^{\infty} d\tilde{\omega}\,\alpha_{zz}^{2}(\tilde{\omega})\,\tilde{\omega}^{7} 
\int_{y_{-}}^{y_{+}} d\bar{y}\,(\bar{y}-\gamma) f^{Y}(\bar{y}) \frac{1}{e^{\beta\tilde{\omega}\bar{y}}-1}.
\end{equation}

If the neutral particle is isotropic $ \alpha_{ij}=\alpha \delta_{ij} $, the contributions from all three polarizations are easily summed. It is convenient to use the momentum distribution function for isotropic polarization defined in Appendix \ref{apC} 
\begin{equation}\label{fiso}
f^{\rm{ISO}}(\bar{y})=f^{X}(\bar{y})+2f^{Y}(\bar{y})=\frac{3}{2\gamma v}.
\end{equation}
The frictional power on the isotropic particle can then be expressed as
\begin{equation}\label{eq2-27}
P^{\rm{ISO}}=\frac{1}{18\pi^{3}\gamma}\int_{0}^{\infty} d\tilde{\omega}\,\alpha^{2}(\tilde{\omega})\,\tilde{\omega}^{7} 
\int_{y_{-}}^{y_{+}} d\bar{y}\,(\bar{y}-\gamma)f^{\rm{ISO}}(\bar{y}) \frac{1}{e^{\beta\tilde{\omega}\bar{y}}-1}.
\end{equation}
As a result, the formulas for frictional power from the different polarization states have exactly the same structure. In Eqs.~\eqref{eq2-23},\,\eqref{eq2-25},\,\eqref{eq2-26} and \eqref{eq2-27}, the formulas for frictional power all involve $ \bar{y}-\gamma $, the momentum distribution function for the specific polarization state $ f^{\rm{P}}(\bar{y}) $, which is even with respect to $ \bar{y}=\gamma $, and an exponential factor, which is monotonically decreasing with $ \bar{y} $. Since the integrals on $ \bar{y} $ are all taken over $ [y_{-}, y_{+}] $, an interval even with respect to $ \bar{y}=\gamma$, the quantum vacuum frictional power is negative definite in all cases.  Therefore, the neutral particle always radiates net energy out through interaction with field fluctuations regardless of the model for its polarizability or its polarization state.

For the more general situation when the atom is anisotropic, there will be contributions from off-diagonal components of the polarizability. We will discuss those contributions in Sec.~\ref{rest}, where the system is quantized in frame $ \mathcal{P} $.

\section{quantum frictional force}\label{force}
\subsection{Quantization of the force in the rest frame of the blackbody radiation}
The Lorentz force density is
\begin{equation}\label{eq3-1}
\vb{f}(t,\vb{r})=\rho(t,\vb{r})\vb{E}(t,\vb{r})+\vb{j}(t,\vb{r})\cross\vb{B}(t,\vb{r}).
\end{equation}
The frictional force on a system moving in the $ x $ direction could be obtained by integrating the $ x $ component of the Lorentz force density,
\begin{equation}\label{eq3-2}
F(t)=\int d\vb{r} \, [\rho(t,\vb{r})E_{x}(t,\vb{r})+j_{y}(t,\vb{r})B_{z}(t,\vb{r})-j_{z}(t,\vb{r})B_{y}(t,\vb{r})].
\end{equation}
Inserting the charge and current densities associated with a moving classical dipole in Eq.~\eqref{eq2-2}, we find
\begin{equation}\label{eq3-3}
F(t)=\vb{d}(t)\cdot\grad E_{x}(t,\vb{v}t)+\dot{d}_{y}(t)B_{z}(t,\vb{v}t)-\dot{d}_{z}(t)B_{y}(t,\vb{v}t).
\end{equation}
The strategy of quantizing the force formula Eq.~\eqref{eq3-3} is the same as detailed for the power in Sec.~\ref{qpower}. The quantum frictional force from each diagonal polarization state of the neutral particle is broken into two contributions, 
\begin{equation}\label{eq3-3.5}
F^{\rm{P}}=F_{\rm{I}}^{\rm{P}}+F_{\rm{II}}^{\rm{P}},
\end{equation}
where $ \rm{P} $ denotes the different polarization states.
\subsection{X polarization}
The contribution to the frictional force due to a nonvanishing $ x $-$ x $ component of the polarizability $ \alpha_{xx} $ is
\begin{equation}\label{eq3-4}
F^{X}(t)=d_{x}(t)\partial_{x}E_{x}(t,\vb{v}t).
\end{equation}
As an example, we work out $ F_{\rm{I}}^{X} $ explicitly in momentum space, where the physics is more transparent than in the spacetime domain.
First, let us Fourier transform the dipole,
\begin{equation}\label{eq3-5}
d_{x}(t)=\int\frac{d\omega}{2\pi} e^{-i\omega t} d_{x}(\omega).
\end{equation}
Then expand it in frame $ \mathcal{P} $,
\begin{equation}\label{eq3-5.5}
d_{x}(\omega)=d'_{x}(\gamma\omega)=\alpha_{xx}(\gamma\omega)E_{x}'(\gamma\omega;\vb{0}),
\end{equation}
followed by Lorentz transforming the field into $ \mathcal{R} $ in momentum space,
\begin{equation}\label{eq3-6}
E_{x}'(\gamma\omega;\vb{0})=\frac{1}{\gamma}\int\frac{d^{2}\vb{k}_{\perp}}{(2\pi)^{2}} \, E_{x}(\omega+k_{x}v,\vb{k}_{\perp};z=0).
\end{equation}
Next, expand the field in terms of the current in momentum space,
\begin{equation}\label{eq3-7}
E_{x}(\omega+k_{x}v,\vb{k}_{\perp};z=0)=\int d\tilde{z}\,\frac{1}{-i(\omega+k_{x}v)}g_{xx}(\omega+k_{x}v,\vb{k}_{\perp};z=0,\tilde{z})j_{x}(\omega+k_{x}v,\vb{k}_{\perp};\tilde{z}),
\end{equation}
where the Fourier transformed current is
\begin{equation}\label{eq3-7.5}
j_{x}(\omega+k_{x}v,\vb{k}_{\perp};\tilde{z})=-i(\omega+k_{x}v)d_{x}(\omega)\delta(\tilde{z}).
\end{equation}
After expanding the dipole inside the current as we did in \eqref{eq3-5.5} and \eqref{eq3-6}, $ d_{x}(t) $ in \eqref{eq3-4} is eventually written as
\begin{equation}\label{eq3-8}
d_{x}(t)=\frac{1}{\gamma^{2}}\int\frac{d\omega}{2\pi}\frac{d^{2}\vb{k}_{\perp}}{(2\pi)^{2}}\frac{d^{2}\bar{\vb{k}}_{\perp}}{(2\pi)^{2}}\, e^{-i\omega t} \alpha_{xx}^{2}(\gamma\omega)g_{xx}(\omega+k_{x}v,\vb{k}_{\perp})E_{x}(\omega+\bar{k}_{x}v,\bar{\vb{k}}_{\perp}),
\end{equation}
where we have suppressed the $ z $ and $ \tilde{z} $ dependence in $ g_{xx} $ and $ E_{x} $. They are all evaluated at $ z=\tilde{z}=0 $.
For the original field operator in Eq.~\eqref{eq3-4}, we just need to Fourier transform it and take its $ x $ derivative,
\begin{equation}\label{eq3-10}
\partial_{x} E_{x}(t,\vb{r}=\vb{v}t)=\int\frac{d\nu}{2\pi}\frac{d^{2}\tilde{\vb{k}}_{\perp}}{(2\pi)^{2}}\,e^{-i\nu t}e^{i\tilde{k}_{x}vt} \, i\tilde{k}_{x} E_{x}(\nu,\tilde{\vb{k}}_{\perp}).
\end{equation}
Now correlate the field operator in Eq.~\eqref{eq3-8} with the one in Eq.~\eqref{eq3-10} and evaluate the correlation function using FDT in momentum space,
\begin{equation}\label{eq3-11}
\langle E_{x}(\omega+\bar{k}_{x}v,\bar{\vb{k}}_{\perp})E_{x}(\nu,\tilde{\vb{k}}_{\perp}) \rangle=
(2\pi)^{3}\delta(\omega+\bar{k}_{x}v+\nu)\delta^{(2)}(\bar{\vb{k}}_{\perp}+\tilde{\vb{k}}_{\perp})\Im g_{xx}(\omega+\bar{k}_{x}v,\bar{\vb{k}}_{\perp})\coth{\left(\frac{\beta}{2}(\omega+\bar{k}_{x}v)\right)}.
\end{equation}
Combining \eqref{eq3-8} and \eqref{eq3-10} with the use of \eqref{eq3-11}, we find the quantum vacuum frictional force on the polarizable atom is also time independent,
\begin{align}\label{eq3-12}
F_{\rm{I}}^{X}=\frac{1}{16\pi^{5}\gamma^{3}}\int_{0}^{\infty} d\tilde{\omega}\int d^{2}\vb{k}_{\perp} \,d^{2}\bar{\vb{k}}_{\perp}\, \alpha_{xx}^{2}(\tilde{\omega})\,\bar{k}_{x} &\Im g_{xx}\left(\frac{\tilde{\omega}}{\gamma}+\bar{k}_{x}v,\vb{k}_{\perp}\right) \nonumber\\
\cross &\Im g_{xx}\left(\frac{\tilde{\omega}}{\gamma}+\bar{k}_{x}v,\bar{\vb{k}}_{\perp}\right) 
\coth\left[\frac{\beta}{2}\left(\frac{\tilde{\omega}}{\gamma}+\bar{k}_{x}v\right)\right].
\end{align}
Here we have again utilized the symmetry properties of the integrand and rescaled the frequency dependence, $ \tilde{\omega}=\gamma\omega $.

Following the second expansion scheme described after Eq.~\eqref{eq2-7.5}, we find the II contribution to the force to be
\begin{align}\label{eq3-14}
F_{\rm{II}}^{X}=-\frac{1}{16\pi^{5}\gamma^{3}}\int_{0}^{\infty} d\tilde{\omega}\int d^{2}\vb{k}_{\perp} \,d^{2}\bar{\vb{k}}_{\perp}\, \alpha_{xx}^{2}(\tilde{\omega})\, k_{x} &\Im g_{xx}\left(\frac{\tilde{\omega}}{\gamma}+k_{x}v,\vb{k}_{\perp}\right) \nonumber\\
\cross &\Im g_{xx}\left(\frac{\tilde{\omega}}{\gamma}+\bar{k}_{x}v,\bar{\vb{k}}_{\perp}\right) 
\coth\left[\frac{\beta}{2}\left(\frac{\tilde{\omega}}{\gamma}+\bar{k}_{x}v\right)\right].
\end{align}
The reflection rule that works for the two contributions to the frictional power also applies to the two contributions to the frictional force: changing the sign of $ F_{\rm{I}}^{X} $ and exchanging the momentum factor $ k_{x}' $ for $ k_{x} $ give $ F_{\rm{II}}^{X} $ directly.

The sum of the two contributions from the $ x $ polarization is manifestly convergent,
\begin{align}\label{eq3-15}
F^{X}=\frac{1}{8\pi^{5}\gamma^{3}}\int_{0}^{\infty} d\tilde{\omega}\int d^{2}\vb{k}_{\perp}d^{2}\bar{\vb{k}}_{\perp}\, \alpha_{xx}^{2}(\tilde{\omega})(\bar{k}_{x}-k_{x}) \Im g_{xx}\left(\frac{\tilde{\omega}}{\gamma}+k_{x}v,\vb{k}_{\perp}\right) \Im g_{xx}\left(\frac{\tilde{\omega}}{\gamma}+\bar{k}_{x}v,\bar{\vb{k}}_{\perp}\right) \frac{1}{e^{\beta(\frac{\tilde{\omega}}{\gamma}+\bar{k}_{x}v)}-1}.
\end{align}
In Eq.~\eqref{eq3-15}, we have not specified the background. So, Eq.~\eqref{eq3-15} applies to the more complicated situations. For example, it could be used to investigate the quantum friction experienced by a neutral particle moving above a planar surface, which we hope to revisit in the near future. 

Let us now specify the background to be merely blackbody radiation. Inserting the vacuum Green's functions in Appendix \ref{apA} and making the same change of variables, Eq.~\eqref{eq3-15} becomes
\begin{equation}\label{eq3-16}
F^{X}=\frac{1}{18\pi^{3}\gamma v}\int_{0}^{\infty} d\tilde{\omega}\, \alpha_{xx}^{2}(\tilde{\omega})\,\tilde{\omega}^{7}  \int_{y_{-}}^{y_{+}} d\bar{y}\, (\bar{y}-\gamma)f^{X}(\bar{y})\frac{1}{e^{\beta\tilde{\omega}\bar{y}}-1},
\end{equation}
where $ f^{X} $ is defined in Eq.~\eqref{eq2-18.5}. Comparing Eq.~\eqref{eq3-16} with Eq.~\eqref{eq2-23}, we see that the power-force relation $ P^{X}=F^{X}v $ is verified explicitly.

\subsection{Other polarizations}
We now turn to the contributions to the quantum frictional force from other polarizations. The frictional force for a neutral particle only polarizable in the $ y $ direction is
\begin{equation}\label{eq3-16.2}
F^{Y}(t)=d_{y}(t)\partial_{y}E_{x}(\vb{v}t,t)+\dot{d}_{y}(t)B_{z}(\vb{v}t,t).
\end{equation}
To work in the momentum space, we need Faraday's law written in that space,
\begin{equation}\label{eq3-16.4}
i\nu B_{z}(\nu,\vb{k}_{\perp})=(ik_{x}E_{y}-ik_{y}E_{x})(\nu,\vb{k}_{\perp}).
\end{equation}
Following the same quantization procedure outlined for $ F^{X} $, we find the quantum vacuum frictional force due to the nonvanishing $ y $-$ y $ component of polarizability to be
\begin{equation}\label{eq3-17}
F^{Y}=\frac{1}{18\pi^{3}\gamma v}\int_{0}^{\infty} d\tilde{\omega}\, \alpha_{yy}^{2}(\tilde{\omega})\,\tilde{\omega}^{7}  \int_{y_{-}}^{y_{+}} d\bar{y} \,(\bar{y}-\gamma)f^{Y}(\bar{y})\frac{1}{e^{\beta\tilde{\omega}\bar{y}}-1},
\end{equation}
where $ f^{Y} $ is defined in Eq.~\eqref{eq2-24.5}.
Apparently,  the quantum vacuum frictional force $ F^{Y} $ has exactly the same structure as $ F^{X} $, only with $ f^{Y} $ replacing $ f^{X} $. 

Using the symmetry between the $ y $ and $ z $ direction for the vacuum problem, $ F^{Z} $ can be readily inferred to be
\begin{equation}\label{eq3-17.5}
F^{Z}=\frac{1}{18\pi^{3}\gamma v}\int_{0}^{\infty} d\tilde{\omega}\, \alpha_{zz}^{2}(\tilde{\omega})\,\tilde{\omega}^{7}  \int_{y_{-}}^{y_{+}} d\bar{y} \,(\bar{y}-\gamma)f^{Y}(\bar{y})\frac{1}{e^{\beta\tilde{\omega}\bar{y}}-1}.
\end{equation}

Finally, if the neutral particle is isotropic, the total force acting on it in the moving direction is
\begin{equation}\label{eq3-18}
F^{\rm{ISO}}=\frac{1}{18\pi^{3}\gamma v}\int_{0}^{\infty} d\tilde{\omega}\,\alpha^{2}(\tilde{\omega})\,\tilde{\omega}^{7} 
\int_{y_{-}}^{y_{+}} d\bar{y} \,f^{\rm{ISO}}(\bar{y})(\bar{y}-\gamma)\frac{1}{e^{\beta\tilde{\omega}\bar{y}}-1}.
\end{equation}
where we used again the definition introduced in Sec.~\ref{power}, $ f^{\rm{ISO}}(\bar{y})=f^{X}(\bar{y})+2 f^{Y}(\bar{y})=\frac{3}{2 \gamma v}$. Comparing the formulas for force in this section with the formulas for power in Sec.~\ref{power},  we verify explicitly the power-force relation 
\begin{equation}
P^{\rm{P}}=F^{\rm{P}}v
\end{equation}
holds true for each polarization $ \rm{P} $. Since the frictional powers are all negative, the frictional forces are true drags on the particle, opposing its motion. In frame $ \mathcal{R} $, the moving particle loses energy to the electromagnetic vacuum because of the negative work done by the quantum friction $ F $. In the meantime, to keep the particle moving in constant velocity, an external driving force $ F_{\rm{ext}}=-F $ is needed to balance the quantum friction. The particle therefore gains exactly the same amount of energy through the external force doing positive work as it loses through the quantum friction doing negative work on it. Overall, the energy of the neutral particle is conserved in the nonequilibrium steady state. 


So far, we have obtained formulas of quantum vacuum frictional power and force for each polarization. These formulas clearly exhibit symmetries between the $ \rm{I} $ and $ \rm{II} $ contributions and between different polarizations. In the process of working out the quantum vacuum friction, we have in fact derived the formulas for the quantum frictional force in a more general background for all diagonal polarizations. These formulas are recorded in Appendix \ref{apD} where the symmetries between the $ \rm{I} $ and $ \rm{II} $ contributions and between different polarizations are still obvious. With these formulas, we could in principle calculate the quantum friction on a neutral particle moving uniformly above a planar surface lying in the $ x$-$y $ plane, in which case we would insert the general Green's function displayed in Eq.~\eqref{eqA-4} instead of the vacuum Green's function into the general formulas in Appendix \ref{apD}. 

We note that our formulas for quantum friction in a general background do not reduce to those obtained by Intravaia et al., Eq.~(1) together with Eq.~(S3) of Ref.~\citep{Intravaia:2016PRL} in particular, in the nonrelativistic and zero temperature limits. There are several points of distinction between their approach and ours. First, in spite of the claim in Ref.~\cite{Intravaia:2016PRA} that the ordering of the operators does not matter, we believe symmetric ordering is required in order to have a Hermitian interaction; see footnote 2. Technically, the employment of symmetric ordering would change the Heaviside functions in the correlation functions to $\sgn$ functions, which corresponds to the zero temperature limit of the $ \coth $ in our formulas. And only with the sgn functions is it possible to restrict the frequency integration in the friction formula to positive frequencies utilizing the symmetry of the integrand. Second, even if symmetric ordering is employed, their modified formula for quantum friction only yields what we call the $ \rm{II} $ contribution but leaves out the $ \rm{I} $ contribution. The source of this omission might be that they have only considered the induced dipole fluctuations, as they argue in Ref.~\cite{Intravaia:2016PRA}. Finally, their claim in Ref.~\citep{Intravaia:2016PRL} that the vacuum part of the Green's dyadic does not contribute to the frictional force is indeed true at zero temperature because of the exponentially decreasing factor in the formulas for quantum friction. At finite temperature, however, the vacuum part of the Green's dyadic does give rise to the blackbody friction (quantum vacuum friction), which is precisely the subject of this paper. 

The quantum vacuum frictional force plays a crucial role in the energetics of a neutral particle maintained in NESS and moving through blackbody radiation. It is particularly in the vacuum situation that the importance of keeping both the I and the II contributions becomes obvious; without keeping both contributions, the force would have been divergent. As a consequence of not symmetrically ordering the operators and not including the I contribution, it is not possible to obtain the well-known nonrelativistic Einstein-Hopf drag \cite{Einstein:Hopf} from the approach advocated in Ref.~\cite{Intravaia:2016PRL}.  In contrast, our relativistic, finite-temperature formulas for quantum vacuum friction correctly generalize the Einstein-Hopf formula to the relativistic regime. 

For instance, in the nonrelativistic limit, Eq.~\eqref{eq3-18} reduces to 
\begin{equation}\label{EH}
F^{\rm{ISO}}\sim -\frac{v}{72\pi^{3}}\int_{0}^{\infty} \, d\omega\, \alpha^{2}(\omega)\, \omega^{7} \frac{\beta\omega}{\sinh^{2}(\beta\omega/2)}.
\end{equation}
This is just the Einstein-Hopf drag~\cite{Einstein:Hopf} felt by the neutral particle. It exactly coincides with the formula for Einstein-Hopf friction in \cite{Kim:dipole} if we use the radiation reaction model for the effective dissipation
\begin{equation}\label{mapping}
\Im\hat{\alpha}(\omega)=\frac{\omega^{3}}{6\pi}\alpha^{2}(\omega).
\end{equation}
In fact, Eq.~\eqref{mapping} is only a lowest order (second order in $ \alpha $) approximation for the imaginary part of the full effective polarizability, $ \hat{\alpha}(\omega) $, when the particle does not possess any intrinsic dissipation, as will be seen more clearly in the next section. 



\section{quantization in the rest frame of the particle}\label{rest}
%
%
%
%
\subsection{Relationship between power and force in the rest frame of radiation and in the rest frame of particle}
The derivations in the preceding two sections of expressions for the power, $P$, and the frictional force, $F$, in frame $ \mathcal{R} $, are complicated by the need to Lorentz transform the electromagnetic field and the dipole back and forth several times between that frame and frame $ \mathcal{P} $. In this section, we demonstrate a rather more efficient approach to establishing and generalizing these expressions, by quantizing the electromagnetic field directly in frame $ \mathcal{P} $. 

The power, $P'$, and the frictional force, $F'$, in frame $ \mathcal{P} $, are related to $P$ and $F$ in frame $ \mathcal{R} $ by
\begin{subequations}\label{4.1}
\begin{equation}\label{4.1a}
P=\frac{\partial}{\partial t} \mathcal F=\gamma\left(\frac{\partial}{\partial t'} - v\frac{\partial}{\partial x'}\right) \frac{1}{\gamma} \mathcal F'=\left(\frac{\partial}{\partial t'} - v \frac{\partial}{\partial x'}\right) \mathcal F'=P'+vF'
\end{equation}
and
\begin{equation}\label{4.1b}
F=-\frac{\partial}{\partial x} \mathcal F=-\gamma\left(\frac{\partial}{\partial x'} - v\frac{\partial}{\partial t'}\right) \frac{1}{\gamma} \mathcal F'=-\left(\frac{\partial}{\partial x'} - v \frac{\partial}{\partial t'}\right) \mathcal F'=F'+vP',
\end{equation}
\end{subequations}
where $\mathcal F$ and $\mathcal F'$ denote the particle-field interaction free energy in $ \mathcal{R} $ and $ \mathcal{P} $, respectively, which are themselves related\footnote{From the Lorentz invariance of the effective action, $W=-\mathcal F \mathcal T=-\mathcal F' \mathcal T'$, and the Lorentz transformation of the duration of the steady state configuration, $\mathcal T=\gamma \mathcal T'$, it follows that $\mathcal F'=\gamma \mathcal F$} by $\mathcal F'=\gamma \mathcal F$. Generation of the power and the frictional force in this way, by differentiation of the particle-field interaction free energy, is a statement of the principle of virtual work, a proof of which, applicable to the current context, is provided in Appendix E. We note, in passing, that, as an immediate consequence of the relationships in Eqs.~\eqref{4.1}, the NESS condition may be expressed in any one of the following equivalent forms:
\begin{subequations}\label{4.2}
\begin{equation}\label{4.2a}
P'=0,
\end{equation}
\begin{equation}\label{4.2b}
F=F',
\end{equation}
and
\begin{equation}\label{4.2c}
P=Fv,
\end{equation}
\end{subequations}
which, respectively, pertain to only frame $ \mathcal{P} $, to both $ \mathcal{P} $ and $ \mathcal{R} $, and to only $ \mathcal{R} $.

\subsection{Expressions to second order in the intrinsic polarizability}
$\mathcal F'$ may be formally written as the symmetrized expectation of the (induced) dipole-field interaction Hamiltonian,
\begin{equation}\label{4.3}
\mathcal F' =-\frac12  \left\langle \mathbf{E}'^T \hat{\mathbf{d}}'\right\rangle =-\frac12  \left\langle \mathbf{E}'^T \boldsymbol{\alpha}\, \mathbf{E}'\right\rangle =-\frac12 \left\langle\mathbf{E}'^{fT} \frac{1}{\mathbf{1}-\boldsymbol{\alpha} \,\boldsymbol{\Gamma}'^T}\,\boldsymbol{\alpha}\,\frac{1}{\mathbf{1}-\boldsymbol{\Gamma}'\boldsymbol{\alpha}} \, \mathbf{E}'^f \right\rangle,
\end{equation}
where the relationship
\begin{equation}\label{4.4}
\mathbf{E}'=\mathbf{E}'^f + \boldsymbol{\Gamma}' \boldsymbol{\alpha} \,\mathbf{E}'
\end{equation}
has been used to express the effective, or interacting, electric field, $\mathbf{E}'$, in terms of the free, or fluctuating (rather than induced),  electric field, $\mathbf{E}'^f$, 
the (real, symmetric, bare) intrinsic polarizability, $\boldsymbol{\alpha}$, and the vacuum Green's dyadic for the electric field in frame $ \mathcal{P} $, $\boldsymbol{\Gamma}'$. In relation to the expectation, $\expval{ \mathbf{E}'^T \hat{\mathbf{d}}'}$, in \eqref{4.3}, we will refer to the location of the dipole, $\hat{\mathbf{d}}'$, as the dipole point, and the location of the field, $\mathbf{E}'^T$, as the field point.

In order to be able to generate $P'$ and $F'$ by differentiation of $\mathcal F'$, it is  first necessary to identify and separate the dipole point, $(t_0', \mathbf{r}_0')=(t_0', x'_0, 0, 0)$, and the field point, $(t_1', \mathbf{r}_1')=(t_1',x_1',0,0)$. Doing so in Eq.~\eqref{4.3}, and keeping terms only through second order in $\boldsymbol{\alpha}$, we obtain
\begin{equation}\label{4.5}
\mathcal F'=\mathcal F_0' + \mathcal F_{\text{I}}'+\mathcal F_{\text{II}}',
\end{equation}
where
\begin{subequations}\label{4.6}
\begin{equation}\label{4.6a}
\mathcal F_0'\equiv-\frac12\int\frac{d\omega}{2\pi}\int\frac{d\omega'}{2\pi}e^{-i(\omega t_0'+\omega' t_1')} \tr\left.\left[\boldsymbol{\alpha}(\omega) \,\mathbf{C}'(\omega, \omega'; \mathbf{r}_0', \mathbf{r}_1')\right]\right|_{t_1'\to t_0', \mathbf{r}_1'\to \mathbf{r}_0'},
\end{equation}
\begin{equation}\label{4.6b}
\mathcal F_{\text{I}}'\equiv-\frac12\int\frac{d\omega}{2\pi}\int\frac{d\omega'}{2\pi}e^{-i(\omega t_0'+\omega' t_1')} \tr\left.\left[\boldsymbol{\alpha}(\omega) \,\boldsymbol{\Gamma}'(\omega;\mathbf{r}_0',\mathbf{r}_0') \,\boldsymbol{\alpha}(\omega)\,\mathbf{C}'(\omega, \omega'; \mathbf{r}_0', \mathbf{r}_1')\right]\right|_{t_1'\to t_0', \mathbf{r}_1'\to \mathbf{r}_0'},
\end{equation}
and
\begin{equation}\label{4.6c}
\mathcal F_{\text{II}}'\equiv-\frac12\int\frac{d\omega}{2\pi}\int\frac{d\omega'}{2\pi}e^{-i(\omega t_0'+\omega' t_1')} \tr\left.\left[\boldsymbol{\alpha}(\omega') \,\boldsymbol{\Gamma}'(\omega';\mathbf{r}_0',\mathbf{r}_1') \,\boldsymbol{\alpha}(\omega)\,\mathbf{C}'(\omega, \omega'; \mathbf{r}_0', \mathbf{r}_0')\right]\right|_{t_1'\to t_0', \mathbf{r}_1'\to \mathbf{r}_0'}.
\end{equation}
\end{subequations}
Here,
\begin{equation}\label{4.7}
\mathbf{C}'(\omega, \omega'; \mathbf{r}_0', \mathbf{r}_1')\equiv \left\langle \mathbf{E}'^f(\omega; \mathbf{r}_0') \otimes \mathbf{E}'^f(\omega';\mathbf{r}_1')\right\rangle
\end{equation}
denotes the symmetrized correlation function of the corresponding free electric field operators, which, using the FDT in frame $ \mathcal{P} $, may be written as 
\begin{equation}\label{4.8}
\mathbf{C}'(\omega, \omega'; \mathbf{r}_0', \mathbf{r}_1')=2\pi \delta(\omega+\omega') \int \frac{dk_x}{2\pi} e^{ik_x(x_0'-x_1')} \operatorname{Im}\mathbf{G}'(\omega, k_x) \coth\left(\frac{\beta\gamma}{2}(\omega+vk_x)\right),
\end{equation}
with
\begin{equation}\label{4.9}
\mathbf{G}'(\omega,k_x)\equiv \int\frac{dk_y}{2\pi}\, \mathbf{g}'(\omega, k_x, k_y).
\end{equation}
Substituting Eq.~\eqref{4.8} into Eq.~\eqref{4.6}, there result
\begin{subequations}\label{4.10}
\begin{equation}\label{4.10a}
\mathcal F_0'=-\frac12 \int\frac{d\omega}{2\pi}\int\frac{dk_x}{2\pi}e^{-i\omega(t_0'-t_1')}e^{ik_x(x_0'-x_1')}\tr\left.\left[\boldsymbol{\alpha}(\omega)\operatorname{Im} \mathbf{G}'(\omega, k_x)\right]\right|_{t_1'\to t_0', \mathbf{r}_1'\to \mathbf{r}_0'}\coth\left(\frac{\beta\gamma}{2}(\omega+vk_x)\right),
\end{equation}
\begin{equation}\label{4.10b}
\begin{split}
\mathcal F_{\text{I}}'=-\frac12 \int\frac{d\omega}{2\pi}\int\frac{dk_x}{2\pi}e^{-i\omega(t_0'-t_1')}e^{ik_x(x_0'-x_1')}\tr&\left.\left[\boldsymbol{\alpha}(\omega)\,\boldsymbol{\Gamma}'(\omega; \mathbf{r}_0', \mathbf{r}_0')\,\boldsymbol{\alpha}(\omega)\operatorname{Im} \mathbf{G}'(\omega, k_x)\right]\right|_{t_1'\to t_0', \mathbf{r}_1'\to \mathbf{r}_0'}\\
&\times \coth\left(\frac{\beta\gamma}{2}(\omega+vk_x)\right), 
\end{split}
\end{equation}
and 
\begin{equation}\label{4.10c}
\begin{split}
\mathcal F_{\text{II}}'=-\frac12 \int\frac{d\omega}{2\pi}\int\frac{dk_x}{2\pi}e^{-i\omega(t_0'-t_1')}e^{ik_x(x_0'-x_0')}\tr&\left.\left[\boldsymbol{\alpha}(\omega)\,\boldsymbol{\Gamma}'{}^*(\omega; \mathbf{r}_0', \mathbf{r}_1')\,\boldsymbol{\alpha}(\omega)\operatorname{Im} \mathbf{G}'(\omega, k_x)\right]\right|_{t_1'\to t_0', \mathbf{r}_1'\to \mathbf{r}_0'}\\
&\times \coth\left(\frac{\beta\gamma}{2}(\omega+vk_x)\right), 
\end{split}
\end{equation}
\end{subequations}
where, in the last of these, we have used the reflection properties $\boldsymbol{\alpha}(-\omega)=\boldsymbol{\alpha}(\omega)$, since $\boldsymbol{\alpha}$ is real, and $\boldsymbol{\Gamma}'(-\omega; \mathbf{r}_0', \mathbf{r}_1')=\boldsymbol{\Gamma}'{}^*(\omega; \mathbf{r}_0', \mathbf{r}_1')$.

Before proceeding to differentiation of these point-separated $\mathcal F'$ expressions, let us note that, from the symmetry of Eq.~\eqref{4.3}, there are two possible identifications of the field point and the dipole point, before separation, which introduces a multiplicity factor of 2. Also, although we could equally well differentiate with respect to the dipole point, corresponding to the location of the particle, we will find it more convenient to differentiate with respect to the field point, as in the proof of Appendix E.

So, differentiating the expression in Eq.~\eqref{4.10a} with respect to the time coordinate of the field point, $t_1'$, and then taking the limit as the field point approaches the dipole point, we obtain
\begin{equation}\label{4.11}
P_0'= -\int\frac{d\omega}{2\pi}\int\frac{dk_x}{2\pi}\,i\omega\tr\left[\boldsymbol{\alpha}(\omega)\operatorname{Im} \mathbf{G}'(\omega, k_x)\right]\coth\left(\frac{\beta\gamma}{2}(\omega+vk_x)\right)=0,
\end{equation}
since the integrand is odd under combined reflection in $\omega$ and $k_x$. Likewise, keeping only the part of the integrand that is even under this combined reflection, Eq.~\eqref{4.10b} yields
\begin{align}\label{4.12}
P_{\text{I}}'&=\int\frac{d\omega}{2\pi}\int\frac{dk_x}{2\pi}\,\omega\tr\left[\boldsymbol{\alpha}(\omega)\operatorname{Im}\boldsymbol{\Gamma}'(\omega; \mathbf{r}_0',\mathbf{r}_0')\,\boldsymbol{\alpha}(\omega)\operatorname{Im} \mathbf{G}'(\omega, k_x)\right]\coth\left(\frac{\beta\gamma}{2}(\omega+vk_x)\right)\nonumber\\
&=\frac1{6\pi} \int \frac{d\omega}{2\pi}\int \frac{dk_x}{2\pi}\,\omega^4\tr\left[\boldsymbol{\alpha}^2(\omega)\operatorname{Im} \mathbf{G}'(\omega, k_x)\right]\coth\left(\frac{\beta\gamma}{2}(\omega+vk_x)\right)\nonumber\\
&=\frac{1}{36\pi^3}\int_0^{\infty}d\omega\,\omega^7\int_{y_-}^{y_+}d \bar y \,\tr\left[\boldsymbol{\alpha}^2(\omega) \operatorname{diag}\left(f^X(\bar y), f^Y(\bar y), f^Y(\bar y)\right)\right]\coth\left(\frac{\beta}{2}\omega\bar y\right)\nonumber\\
&=\frac{1}{36\pi^3}\int_0^{\infty}d\omega\,\omega^7\int_{y_-}^{y_+}d \bar y \left[(\boldsymbol{\alpha}^2)_{xx}(\omega) \,f^X(\bar y) +\left((\boldsymbol{\alpha}^2)_{yy}(\omega)+(\boldsymbol{\alpha}^2)_{zz}(\omega)\right)f^Y(\bar y)\right]\coth\left(\frac{\beta}{2}\omega\bar y\right),
\end{align}
where we have used the properties of $\operatorname{Im} \mathbf{G}'(\omega, k_x)$ detailed in Appendix A to deduce that
\begin{equation}\label{4.13}
\operatorname{Im}\boldsymbol{\Gamma}'(\omega; \mathbf{r}_0', \mathbf{r}_0')=\int\frac{dk_x}{2\pi}\operatorname{Im}\mathbf{G}'(\omega,k_x)=\frac{\omega^3}{6\pi} \mathbf{1},
\end{equation}
and
\begin{equation}\label{4.14}
\int\frac{dk_x}{2\pi} \operatorname{Im} \mathbf{G}'(\omega,k_x) \,\coth\left(\frac{\beta\gamma}{2}\left(\omega+vk_x\right)\right)=\frac{\omega^3}{6\pi}\int_{y_-}^{y_+}d\bar y \,\operatorname{diag}\left(f^X(\bar y), f^Y(\bar y), f^Y(\bar y)\right)\coth\left(\frac{\beta}{2}\omega \bar y\right),
\end{equation}
and, in the latter expression, have employed the change of variable from $k_x$ to $\bar y$, defined by $\omega \bar y \equiv \gamma(\omega +vk_x)$. Since $f^X(\bar y)\ge 0$ and $f^Y(\bar y) \ge 0$ on the $\bar y$ integration interval $[y_-, y_+]$, it is clear from Eq.~\eqref{4.12} that $P'_{\text{I}}>0$ if the diagonal elements of $\boldsymbol{\alpha}^2(\omega)$ are non-negative. That this is indeed the case follows from the fact that these diagonal elements are sums of squares: for example, $(\boldsymbol{\alpha}^2)_{xx}=(\alpha_{xx})^2+(\alpha_{xy})^2+(\alpha_{xz})^2$.

For $P_{\text{II}}'$, the only change is the replacement of $\boldsymbol{\Gamma}'$ in Eq~\eqref{4.10b} by $\boldsymbol{\Gamma}'^*$ in Eq.~ \eqref{4.10c}, so $P'_{\text{II}}=-P'_{\text{I}}<0$. Thus, $P'=0$ to second order in $\boldsymbol{\alpha}$.

In fact, it is not difficult to show that $P'=0$ to all orders in $\boldsymbol{\alpha}$. Applying the above approach to Eq.~\eqref{4.3}, we obtain
\begin{equation}\label{4.15}
P'=-\int\frac{d\omega}{2\pi}\int\frac{dk_x}{2\pi}\,i\omega \tr\left[\frac{1}{\mathbf{1}-\boldsymbol{\alpha}(-\omega)\,\boldsymbol{\Gamma}'(-\omega;\mathbf{r}_0',\mathbf{r}_0')}\,\boldsymbol{\alpha}(\omega)\,\frac{1}{\mathbf{1}-\boldsymbol{\Gamma}'(\omega;\mathbf{r}_0',\mathbf{r}_0')\,\boldsymbol{\alpha}(\omega)}\operatorname{Im}\mathbf{G}'(\omega,k_x)\right]\coth\left(\frac{\beta\gamma}{2}(\omega+vk_x)\right).
\end{equation}
Since the matrix
\begin{equation}\label{4.16}
\bar{\boldsymbol{\alpha}}(\omega)\equiv\frac{1}{\mathbf{1}-\boldsymbol{\alpha}(-\omega)\,\boldsymbol{\Gamma}'(-\omega;\mathbf{r}_0',\mathbf{r}_0')}\,\boldsymbol{\alpha}(\omega)\,\frac{1}{\mathbf{1}-\boldsymbol{\Gamma}'(\omega;\mathbf{r}_0',\mathbf{r}_0')\,\boldsymbol{\alpha}(\omega)}
\end{equation}
is contracted with the symmetric matrix $\operatorname{Im}\mathbf{G}'(\omega,k_x)$, it may be replaced in Eq.~\eqref{4.15} by its transpose, 
\begin{equation}\label{4.17}
\bar{\boldsymbol{\alpha}}^T(\omega)=\frac{1}{\mathbf{1}-\boldsymbol{\alpha}(\omega)\,\boldsymbol{\Gamma}'(\omega;\mathbf{r}_0',\mathbf{r}_0')}\,\boldsymbol{\alpha}(\omega)\,\frac{1}{\mathbf{1}-\boldsymbol{\Gamma}'(-\omega;\mathbf{r}_0',\mathbf{r}_0')\,\boldsymbol{\alpha}(-\omega)}=\bar{\boldsymbol{\alpha}}(-\omega),
\end{equation}
or, indeed, by the average,
\begin{equation}\label{4.18}
\frac12\left(\bar{\boldsymbol{\alpha}}(\omega)+\bar{\boldsymbol{\alpha}}^T(\omega)\right)=\frac12\left(\bar{\boldsymbol{\alpha}}(\omega)+\bar{\boldsymbol{\alpha}}(-\omega)\right),
\end{equation}
which is manifestly reflection symmetric in $\omega$. It is then easily seen that the integrand in Eq.~\eqref{4.15} is odd under combined reflection in $\omega$ and $k_x$, and therefore that $P'=0$ to all orders in $\boldsymbol{\alpha}$.

Differentiating the expression in Eq.~\eqref{4.10a} with respect to the spatial coordinate of the field point, $x_1'$, and then taking the limit as the field point approaches the dipole point, we obtain
\begin{equation}\label{4.19}
F_0'=-\int\frac{d\omega}{2\pi}\int\frac{dk_x}{2\pi}\,ik_x\tr\left[\boldsymbol{\alpha}(\omega)\operatorname{Im} \mathbf{G}'(\omega, k_x)\right]\coth\left(\frac{\beta\gamma}{2}(\omega+vk_x)\right)=0,
\end{equation}
since the integrand is odd under combined reflection in $\omega$ and $k_x$. Likewise, keeping only the part of the integrand that is even under this combined reflection, Eq.~\eqref{4.10b} yields
\begin{align}\label{4.20}
F_{\text{I}}'&=\int\frac{d\omega}{2\pi}\int\frac{dk_x}{2\pi}\,k_x\tr\left[\boldsymbol{\alpha}(\omega)\operatorname{Im}\boldsymbol{\Gamma}'(\omega; \mathbf{r}_0',\mathbf{r}_0')\,\boldsymbol{\alpha}(\omega)\operatorname{Im} \mathbf{G}'(\omega, k_x)\right]\coth\left(\frac{\beta\gamma}{2}(\omega+vk_x)\right)\nonumber\\
&=\frac1{6\pi} \int \frac{d\omega}{2\pi}\int \frac{dk_x}{2\pi}\,\omega^3 \,k_x\tr\left[\boldsymbol{\alpha}^2(\omega)\operatorname{Im} \mathbf{G}'(\omega, k_x)\right]\coth\left(\frac{\beta\gamma}{2}(\omega+vk_x)\right)\nonumber\\
&=\frac{1}{36\pi^3\gamma v}\int_0^{\infty}d\omega\,\omega^7\int_{y_-}^{y_+}d \bar y \,(\bar y-\gamma)\tr\left[\boldsymbol{\alpha}^2(\omega) \operatorname{diag}\left(f^X(\bar y), f^Y(\bar y), f^Y(\bar y)\right)\right]\coth\left(\frac{\beta}{2}\omega\bar y\right)\nonumber\\
&=\frac{1}{36\pi^3\gamma v}\int_0^{\infty}d\omega\,\omega^7\int_{y_-}^{y_+}d \bar y \,(\bar y-\gamma)\left[(\boldsymbol{\alpha}^2)_{xx}(\omega) \,f^X(\bar y) +\left((\boldsymbol{\alpha}^2)_{yy}(\omega)+(\boldsymbol{\alpha}^2)_{zz}(\omega)\right)f^Y(\bar y)\right]\coth\left(\frac{\beta}{2}\omega\bar y\right)\nonumber\\
&=\frac{1}{18\pi^3\gamma v}\int_0^{\infty}d\omega\,\omega^7\int_{y_-}^{y_+}d \bar y \,(\bar y-\gamma)\left[(\boldsymbol{\alpha}^2)_{xx}(\omega)\, f^X(\bar y) +\left((\boldsymbol{\alpha}^2)_{yy}(\omega)+(\boldsymbol{\alpha}^2)_{zz}(\omega)\right)f^Y(\bar y)\right]\frac{1}{e^{\beta\omega\bar y}-1},
\end{align}
where we have invoked Eq.~\eqref{4.13} and have used the properties of $\operatorname{Im} \mathbf{G}'(\omega, k_x)$ detailed in Appendix A to deduce that
\begin{equation}\label{4.21}
\int\frac{dk_x}{2\pi}\,k_x \operatorname{Im} \mathbf{G}'(\omega,k_x) \,\coth\left(\frac{\beta\gamma}{2}\left(\omega+vk_x\right)\right)=\frac{\omega^4}{6\pi\gamma v}\int_{y_-}^{y_+}d\bar y \,(\bar y-\gamma)\operatorname{diag}\left(f^X(\bar y), f^Y(\bar y), f^Y(\bar y)\right)\coth\left(\frac{\beta}{2}\omega \bar y\right).
\end{equation}
Apart from the thermal factor, $\left(e^{\beta\omega\bar y}-1\right)^{-1}$, the $\bar y$ integrand in Eq.~\eqref{4.20} is odd about the midpoint, $\gamma$, of the integration interval. However, the thermal factor is a decreasing function of $\bar y$ on this interval, and so gives less weight to the positive $\bar y >\gamma$ contributions compared to the negative $\bar y < \gamma$ contributions. Thus, $F'_{\text{I}}<0$, that is, the frictional force is indeed a drag.

For $F_{\text{II}}'$, note that, in differentiating the expression in Eq.~\eqref{4.10c}, we encounter
\begin{equation}\label{4.22}
\left.\frac{\partial}{\partial x_1'} \boldsymbol{\Gamma}'^*(\omega; \mathbf{r}_0', \mathbf{r}_1')\right|_{\mathbf{r}_1'\to \mathbf{r}_0'}=\left.\frac{\partial}{\partial x_1'} \boldsymbol{\Gamma}'^*(\omega; \mathbf{r}_1', \mathbf{r}_0')\right|_{\mathbf{r}_1'\to \mathbf{r}_0'}=-\left.\frac{\partial}{\partial x_0'} \boldsymbol{\Gamma}'^*(\omega; \mathbf{r}_1', \mathbf{r}_0')\right|_{\mathbf{r}_1'\to \mathbf{r}_0'}=-\left.\frac{\partial}{\partial x_1'} \boldsymbol{\Gamma}'^*(\omega; \mathbf{r}_0', \mathbf{r}_1')\right|_{\mathbf{r}_1'\to \mathbf{r}_0'}=\mathbf{0}.
\end{equation}
This also follows directly from the property, inherited from the symmetry of the spatial configuration, that the Green's dyadic is reflection symmetric in the $x$-direction: $\boldsymbol{\Gamma}'(\omega; \mathbf{r}_0'+\delta\mathbf{\hat x}, \mathbf{r}'_0)=\boldsymbol{\Gamma}'(\omega; \mathbf{r}_0'-\delta\mathbf{\hat x}, \mathbf{r}'_0)$, for $\delta\ge 0$. Thus, $F_{\text{II}}'=0$, which simply expresses the fact that, because of the isotropic nature of the (induced) dipole radiation emitted by the particle in its rest frame, there is no radiation reaction force on the particle in this frame. 

It is easily verified that the formulas for quantum vacuum frictional force Eqs.~\eqref{eq3-16}, \eqref{eq3-17}, \eqref{eq3-17.5}, \eqref{eq3-18} obtained by quantization in frame $ \mathcal{R} $ are now systematically summarized in Eq.~\eqref{4.20}.

\subsection{Extension to all orders in the intrinsic polarizability}
Again, it is not difficult to extend these results to all orders in $\boldsymbol{\alpha}$. Using $\left(\mathbf{1}-\boldsymbol{\Gamma}'\boldsymbol{\alpha}\right)^{-1}-\mathbf{1}=\boldsymbol{\Gamma}'\boldsymbol{\alpha} \left(\mathbf{1}-\boldsymbol{\Gamma}'\boldsymbol{\alpha}\right)^{-1}$ in Eq.~\eqref{4.3} to generalise  Eq.~\eqref{4.6} and Eq.~\eqref{4.10} to all orders in $\boldsymbol{\alpha}$, we obtain
\begin{subequations}\label{4.23}
\begin{align}\label{4.23a}
\mathcal F_{\text{I}}'\equiv-\frac12 \int\frac{d\omega}{2\pi}\int\frac{dk_x}{2\pi}&e^{-i\omega(t_0'-t_1')}e^{ik_x(x_0'-x_1')}\tr\left[\boldsymbol{\alpha}(\omega)\,\boldsymbol{\Gamma}'(\omega; \mathbf{r}_0',\mathbf{r}_0')\,\boldsymbol{\alpha}(\omega)\phantom{\frac{1}{\boldsymbol{\Gamma}'}}\right.\nonumber\\
&\times\left.\left.\frac{1}{\mathbf{1}-\boldsymbol{\Gamma}'(\omega; \mathbf{r}_0', \mathbf{r}_0')\,\boldsymbol{\alpha}(\omega)}
\operatorname{Im} \mathbf{G}'(\omega, k_x)\right]\right|_{t_1'\to t_0', \mathbf{r}_1'\to \mathbf{r}_0'}
\coth\left(\frac{\beta\gamma}{2}(\omega+vk_x)\right)
\end{align}
and, likewise, 
\begin{align}\label{4.23b}
\mathcal F_{\text{II}}'\equiv-\frac12 \int\frac{d\omega}{2\pi}\int\frac{dk_x}{2\pi}&e^{-i\omega(t_0'-t_1')}e^{ik_x(x_0'-x_0')}\tr\left[\frac{1}{\mathbf{1}-\boldsymbol{\alpha}(\omega)\,\boldsymbol{\Gamma}'{}^*(\omega;\mathbf{r}_0',\mathbf{r}_0')}\,\boldsymbol{\alpha}(\omega)\,\boldsymbol{\Gamma}'{}^*(\omega; \mathbf{r}_0', \mathbf{r}_1')\,\boldsymbol{\alpha}(\omega)\right.\nonumber\\
&\times \left.\left.\frac{1}{\mathbf{1}-\boldsymbol{\Gamma}'(\omega;\mathbf{r}_0',\mathbf{r}_0')\,\boldsymbol{\alpha}(\omega)}\operatorname{Im} \mathbf{G}'(\omega, k_x)\right]\right|_{t_1'\to t_0', \mathbf{r}_1'\to \mathbf{r}_0'}
\coth\left(\frac{\beta\gamma}{2}(\omega+vk_x)\right), 
\end{align}
\end{subequations}
where, in the second of these, we have again employed the reflection properties $\boldsymbol{\alpha}(-\omega)=\boldsymbol{\alpha}(\omega)$, recalling that $\boldsymbol{\alpha}$ is real, and $\boldsymbol{\Gamma}'(-\omega; \mathbf{r}_0', \mathbf{r}_1')=\boldsymbol{\Gamma}'{}^*(\omega; \mathbf{r}_0', \mathbf{r}_1')$.

Differentiating the expression in Eq.~\eqref{4.23a} with respect to the spatial coordinate of the field point, $x_1'$, and then taking the limit as the field point approaches the dipole point, we obtain, using Eq.~\eqref{4.21},
\begin{align}\label{4.24}
F_{\text{I}}'&=\int\frac{d\omega}{2\pi}\int\frac{dk_x}{2\pi}\,k_x\tr\left[\operatorname{Im}\hat{\boldsymbol{\alpha}}(\omega)
\operatorname{Im} \mathbf{G}'(\omega, k_x)\right]
\coth\left(\frac{\beta\gamma}{2}(\omega+vk_x)\right)\nonumber\\
&=\frac{1}{6\pi^2\gamma v} \int_0^{\infty}d\omega\,\omega^4\int_{y_-}^{y_+} d\bar y\,(\bar y-\gamma)\tr\left[\operatorname{Im}\hat{\boldsymbol{\alpha}}(\omega)
\operatorname{diag}\left(f^X(\bar y), f^Y(\bar y), f^Y(\bar y)\right)\right]\coth\left(\frac{\beta}{2}\omega\bar y\right)\nonumber\\
&=\frac{1}{6\pi^2\gamma v} \int_0^{\infty}d\omega\,\omega^4\int_{y_-}^{y_+} d\bar y\,(\bar y-\gamma)\left[\operatorname{Im}\hat{\alpha}_{xx}(\omega)\, f^X(\bar y)+\left(\operatorname{Im}\hat{\alpha}_{yy}(\omega)+\operatorname{Im}\hat{\alpha}_{zz}(\omega)\right) f^Y(\bar y)\right]\coth\left(\frac{\beta}{2}\omega\bar y\right)\nonumber\\
&=\frac{1}{3\pi^2\gamma v} \int_0^{\infty}d\omega\,\omega^4\int_{y_-}^{y_+} d\bar y\,(\bar y-\gamma)\left[\operatorname{Im}\hat{\alpha}_{xx}(\omega) \,f^X(\bar y)+\left(\operatorname{Im}\hat{\alpha}_{yy}(\omega)+\operatorname{Im}\hat{\alpha}_{zz}(\omega)\right) f^Y(\bar y)\right]\frac{1}{e^{\beta\omega\bar y}-1},
\end{align}
where
\begin{equation}\label{4.25}
\hat{\boldsymbol{\alpha}}(\omega)\equiv\boldsymbol{\alpha}(\omega)\,\frac{1}{\vb{1}-\boldsymbol{\Gamma}'(\omega; \mathbf{r}_0', \mathbf{r}_0') \,\boldsymbol{\alpha}(\omega)}
\end{equation}
denotes the effective, or dressed, polarizability, which is clearly symmetric. For $F_{\text{II}}'$,  it is immediate from Eq.~\eqref{4.22} that Eq.~\eqref{4.23b} yields 
\begin{equation}\label{4.26}
F_{\text{II}}'=0,
\end{equation}
to all orders in $\boldsymbol{\alpha}$.

Likewise, differentiating the expression in \eqref{4.23a} with respect to the time coordinate of the field point, $t'_1$, and then taking the limit as the field point approaches the dipole point, we obtain, using \eqref{4.14},
\begin{align}\label{4.27}
P_{\text{I}}'&=\int\frac{d\omega}{2\pi}\int\frac{dk_x}{2\pi}\,\omega\tr\left[\operatorname{Im}\hat{\boldsymbol{\alpha}}(\omega)
\operatorname{Im} \mathbf{G}'(\omega, k_x)\right]
\coth\left(\frac{\beta\gamma}{2}(\omega+vk_x)\right)\nonumber\\
&=\frac{1}{6\pi^2} \int_0^{\infty}d\omega\,\omega^4\int_{y_-}^{y_+} d\bar y\,\tr\left[\operatorname{Im}\hat{\boldsymbol{\alpha}}(\omega)
\operatorname{diag}\left(f^X(\bar y), f^Y(\bar y), f^Y(\bar y)\right)\right]\coth\left(\frac{\beta}{2}\omega\bar y\right)\nonumber\\
&=\frac{1}{6\pi^2} \int_0^{\infty}d\omega\,\omega^4\int_{y_-}^{y_+} d\bar y\,\left[\operatorname{Im}\hat{\alpha}_{xx}(\omega)\, f^X(\bar y)+\left(\operatorname{Im}\hat{\alpha}_{yy}(\omega)+\operatorname{Im}\hat{\alpha}_{zz}(\omega)\right) f^Y(\bar y)\right]\coth\left(\frac{\beta}{2}\omega\bar y\right).
\end{align}
For $P'_{\text{II}}$, it similarly follows from Eq.~\eqref{4.23b}, or immediately from the earlier result  that $P'=0$, that 
\begin{equation}\label{4.28}
P'_{\text{II}} =-P'_{\text{I}}=-\frac{1}{6\pi^2} \int_0^{\infty}d\omega\,\omega^4\int_{y_-}^{y_+} d\bar y\,\left[\operatorname{Im}\hat{\alpha}_{xx}(\omega)\, f^X(\bar y)+\left(\operatorname{Im}\hat{\alpha}_{yy}(\omega)+\operatorname{Im}\hat{\alpha}_{zz}(\omega)\right) f^Y(\bar y)\right]\coth\left(\frac{\beta}{2}\omega\bar y\right),
\end{equation}
to all orders in $\boldsymbol{\alpha}$. Of course, in each of Eq.~\eqref{4.27} and Eq.~\eqref{4.28}, the $\coth$ factor includes the corresponding zero-point energy of the electromagnetic field, and so each integral over $\omega$ is formally divergent. However, these two contributions to $P'$ come in with opposite signs, so the divergences due to the zero-point energy exactly cancel.

It is easily verified that Eq.~\eqref{4.24} and Eq.~\eqref{4.27} agree with Eq.~\eqref{4.20} and Eq.~\eqref{4.12}, respectively, to second order in $\boldsymbol{\alpha}$. To this order, Eq.~\eqref{4.25} becomes
\begin{equation}\label{4.29}
\hat{\boldsymbol{\alpha}}(\omega)=\boldsymbol{\alpha}(\omega)+\boldsymbol{\alpha}(\omega)\,\boldsymbol{\Gamma}'(\omega;\mathbf{r}_0', \mathbf{r}_0')\,\boldsymbol{\alpha}(\omega),
\end{equation}
whence, using the vacuum Green's dyadic in Eq.~\eqref{4.13},
\begin{equation}\label{4.30}
\operatorname{Im}\hat{\boldsymbol{\alpha}}(\omega)= \boldsymbol{\alpha}(\omega)\operatorname{Im}\boldsymbol{\Gamma}'(\omega; \mathbf{r}_0', \mathbf{r}_0')\,\boldsymbol{\alpha}(\omega)=\frac{\omega^3}{6\pi}\,\boldsymbol{\alpha}^2(\omega).
\end{equation}
When Eq.~\eqref{4.30} is inserted into Eq.~\eqref{4.24} and Eq.~\eqref{4.27}, Eq.~\eqref{4.20} and Eq.~\eqref{4.12}, respectively, are reproduced. 

\subsection{Renormalization of the intrinsic polarizability}
However, beyond second order in $\boldsymbol{\alpha}$, Eq.~\eqref{4.24} and Eq.~\eqref{4.27} are formally divergent, because, beyond this order, $\operatorname{Im} \hat{\boldsymbol{\alpha}}(\omega)$ involves $\operatorname{Re} \boldsymbol{\Gamma}'(\omega; \mathbf{r}_0', \mathbf{r}_0')$, which is itself divergent. In fact, this divergence is inherent in all of our work thus far, but we have been able to eliminate or ignore it by making use of certain symmetry properties in our analyses of $P_{\text{I}}'$, $P_{\text{II}}'$,  $F_{\text{I}}'$, and $F_{\text{II}}'$ to second order in $\boldsymbol{\alpha}$, and of $P'$ and $F_{\text{II}}'$ to all orders in $\boldsymbol{\alpha}$. To analyze $F_{\text{I}}'$ and $P'_{\text{I}}$, or $P'_{\text{II}}$,  beyond second order in $\boldsymbol{\alpha}$, these symmetry properties alone are insufficient, and we need to properly recognize and address this divergence.

There are two contributions to $\boldsymbol{\Gamma}'(\omega; \mathbf{r}_0', \mathbf{r}_0')$: $\operatorname{Re} \boldsymbol{\Gamma}'(\omega; \mathbf{r}_0', \mathbf{r}_0')$ is purely real and divergent; 
$i \operatorname{Im} \boldsymbol{\Gamma}'(\omega; \mathbf{r}_0', \mathbf{r}_0')$ is purely imaginary and finite. The first contribution gives rise to an infinite particle-field interaction free energy. In order to obtain finite results, we effect an infinite renormalization of the bare intrinsic polarizability by replacing $\boldsymbol{\Gamma}'(\omega; \mathbf{r}_0', \mathbf{r}_0')$ in Eq.~\eqref{4.25} by the reduced Green's dyadic $ i\operatorname{Im} \boldsymbol{\Gamma}'(\omega; \mathbf{r}_0', \mathbf{r}_0')$. We now proceed to formalize this prescription.\footnote{In similar contexts, where the interaction occurs with a surface, these two contributions are associated with evanescent electromagnetic waves and with propagating electromagnetic waves, respectively \cite{Volokitin:book}. The propagating wave is the solution that picks up the contribution from the imaginary part of the wavenumber $ \kappa $. In this view, the renormalization prescription may be thought of as discarding the contribution of the evanescent waves and retaining only the contribution of the propagating waves.}

Let us relabel the bare intrinsic polarizability by $\boldsymbol{\alpha}_0(\omega)$. Then Eq.~\eqref{4.25} may be written as
\begin{align}\label{4.31}
\hat{\boldsymbol{\alpha}}&=\boldsymbol{\alpha}_0\left[\mathbf{1}-\boldsymbol{\Gamma}'\,\boldsymbol{\alpha}_0\right]^{-1}
=\boldsymbol{\alpha}_0\left[\mathbf{1}-\operatorname{Re} \boldsymbol{\Gamma}'\,\boldsymbol{\alpha}_0-i\operatorname{Im}\boldsymbol{\Gamma}'\, \boldsymbol{\alpha}_0\right]^{-1}\nonumber\\
&=\boldsymbol{\alpha}_0\left[\left(\mathbf{1}-i\operatorname{Im}\boldsymbol{\Gamma}'\,\boldsymbol{\alpha}_0\left(\mathbf{1}-\operatorname{Re}\boldsymbol{\Gamma}'\,\boldsymbol{\alpha}_0\right)^{-1}\right)\left(\mathbf{1}-\operatorname{Re}\boldsymbol{\Gamma}'\,\boldsymbol{\alpha}_0\right)\right]^{-1}\nonumber\\
&=\boldsymbol{\alpha}_0\left(\mathbf{1}-\operatorname{Re}\boldsymbol{\Gamma}'\,\boldsymbol{\alpha}_0\right)^{-1}\left[\mathbf{1}-i\operatorname{Im}\boldsymbol{\Gamma}'\,\boldsymbol{\alpha}_0\left(\mathbf{1}-\operatorname{Re}\boldsymbol{\Gamma}'\,\boldsymbol{\alpha}_0\right)^{-1}\right]^{-1}\nonumber\\
&=\boldsymbol{\alpha}\left[\mathbf{1}-i\operatorname{Im}\boldsymbol{\Gamma}'\,\boldsymbol{\alpha}\right]^{-1},
\end{align}
where we have dropped all of the obvious arguments and 
\begin{equation}\label{4.32}
\boldsymbol{\alpha}(\omega)\equiv\boldsymbol{\alpha}_0(\omega)\,\frac{1}{\mathbf{1}-\operatorname{Re}\boldsymbol{\Gamma}'(\omega; \mathbf{r}'_0, \mathbf{r}'_0)\,\boldsymbol{\alpha}_0(\omega)}
\end{equation}
is the renormalized intrinsic polarizability, which clearly inherits from the bare intrinsic polarizability the properties of being real and symmetric. In vacuum, the effective polarizability may therefore be written, using Eq.~\eqref{4.13}, as
\begin{equation}\label{4.33}
\hat{\boldsymbol{\alpha}}(\omega)
=\boldsymbol{\alpha}(\omega)\,\frac{1}{\mathbf{1}-i\operatorname{Im}\boldsymbol{\Gamma}{}'(\omega; \mathbf{r}_0', \mathbf{r}_0') \,\boldsymbol{\alpha}(\omega)}=\boldsymbol{\alpha}(\omega)\,\frac{1}{\mathbf{1}-i\,\frac{\omega^3}{6\pi} \,\boldsymbol{\alpha}(\omega)},
\end{equation}
which agrees with the functional form of the atomic polarizability obtained in Ref.~\cite{Jentschura:EPJD} by a perturbative analysis of the energy shift; see also Refs.~\cite{Albaladejo:Radiative, Ru:Radiative, Wokaun:Radiation}. Whence,\footnote{This is a version of the optical theorem \citep{Berman:OptTrm}.} the imaginary part of the effective polarizability reads
\begin{equation}\label{4.34}
\operatorname{Im}\hat{\boldsymbol{\alpha}}(\omega)=\frac{\omega^3}{6\pi}\boldsymbol{\alpha}^2(\omega)\,\frac{1}{\mathbf{1}+\left(\frac{\omega^3}{6\pi}\right)^2\boldsymbol{\alpha}^2(\omega)}.
\end{equation}

The diagonal elements of $\operatorname{Im}\hat{\boldsymbol{\alpha}}(\omega)$, constructed as in Eq.~\eqref{4.34}, are non-negative. This may be obvious, given its structure and the fact that, as noted earlier,  the diagonal elements of $\boldsymbol{\alpha}^2(\omega)$ are non-negative, but we provide a formal proof in Appendix F. We may therefore conclude, as for our earlier second-order expressions, that $F'_{\text{I}}<0$, $F'_{\text{II}}=0$, $P'_{\text{I}}>0$, and $P'_{\text{II}}=-P'_{\text{I}}<0$ to all orders in the renormalized intrinsic polarizability $\boldsymbol{\alpha}$. Following the above renormalization, those earlier second-order expressions themselves should be re-interpreted in terms of the renormalized, rather than bare, intrinsic polarizability. 

\subsection{Asymptotic behavior of the frictional force}
The structure of Eq.~\eqref{4.34} has implications for the low- and high-temperature behavior of the frictional force, as we now illustrate in the case of isotropic polarizability, $\boldsymbol{\alpha}(\omega) = \alpha(\omega) \,\mathbf{1}$, where, using Eq.~\eqref{fiso}, Eq.~\eqref{4.24} simplifies to 
\begin{align}\label{4.35}
F'_{\text{I}}&=\frac{1}{2\pi^2\gamma^2 v^2} \int_{y_-}^{y_+} d\bar y\,(\bar y-\gamma)\int_0^{\infty}d\omega\,\omega^4\operatorname{Im}\hat{\alpha}(\omega)\, \frac{1}{e^{\beta\omega\bar y}-1}\nonumber\\
&=\frac{1}{2\pi^2\gamma^2 v^2\beta^5} \int_{y_-}^{y_+} d\bar y\,\frac{(\bar y-\gamma)}{{\bar y}^5}\int_0^{\infty}du\,u^4\operatorname{Im}\hat{\alpha}\left(\frac{u}{\beta \bar y}\right) \frac{1}{e^{u}-1}.
\end{align}
It is therefore clear that the low-frequency modes dominate in the force integral at low temperatures while the high-frequency modes dominate at higher temperatures.

Let us assume the asymptotic behavior of the renormalized intrinsic polarizability at low and high frequencies are described by the different power laws
\begin{equation}\label{4.36}
\alpha(\omega) \sim A\,\omega^a \qquad (\omega \to 0^+),\qquad \alpha(\omega) \sim B\,\omega^b \qquad (\omega \to \infty)
\end{equation}
with $A\ne 0$ and  $B\ne 0$. Then the low-temperature and high-temperature limits of the quantum vacuum friction can be readily obtained as follow,
\begin{equation}\label{lowhigh}
F'_{\text{I, LowT}}\sim
\begin{cases}
\frac{A^2\,\Gamma(8+2a)\,\zeta(8+2a)}{6\pi^3\beta^{8+2a}}\,\eta(8+2a, v),\qquad&a>-3,\\\\
\frac{A^2}{36\pi\left[1+\left(\frac{A}{6\pi}\right)^2\right]\beta^2}\,\eta(2, v),\qquad&a=-3,\\\\
\frac{\pi}{\beta^2}\,\eta(2, v),\qquad&a<-3.
\end{cases}
\quad
F'_{\text{I, HighT}}\sim
\begin{cases}
\frac{B^2\,\Gamma(8+2b)\,\zeta(8+2b)}{6\pi^3\beta^{8+2b}}\,\eta(8+2b, v),\qquad&b<-3,\\\\
\frac{B^2}{36\pi\left[1+\left(\frac{B}{6\pi}\right)^2\right]\beta^2}\,\eta(2, v),\qquad&b=-3,\\\\
\frac{\pi}{\beta^2}\,\eta(2, v),\qquad&b>-3.
\end{cases}
\end{equation}
where we have introduced 
\begin{equation}\label{4.38}
\eta(n, v)\equiv \frac{1}{2\gamma^2 v^2}\int_{y_-}^{y_+}d\bar y\,\frac{(\bar y-\gamma)}{{\bar y}^{n}}.
\end{equation}


The obvious symmetry between the two limits of the force in Eq.~\eqref{lowhigh} can be easily understood. First, in the low-temperature limit with $ a>-3 $ and in the high-temperature limit with $ b<-3 $, $ \Im\alpha(\omega) $ in Eq.~\eqref{4.34} reduces to $ \omega^{3}\alpha^{2}(\omega)/6\pi $. Then for the special cases with critical power $a=-3 $ or $ b=-3 $, both terms in the denominator for $ \Im \hat{\alpha} $ should be kept, contributing an additional numerical factor multiplying $ \omega^{3}\alpha^{2}(\omega)/6\pi $.
Finally, in the low-temperature limit with $ a<-3 $ and in the high-temperature limit with $ b>-3 $, $ \Im\alpha(\omega) $ reduces to $ 6\pi/\omega^{3}$. As a result, the asymptotic behavior of the force becomes independent of the details (power and coefficients) of the model for the renormalized intrinsic polarizability $ \alpha(\omega) $, and is always proportional to $ T^{2} $:
\begin{equation}\label{4.42}
F'_{\text{I}}\sim \frac{\pi\,T^2}{2\gamma^2v^2}\left[\ln \left(\frac{1+v}{1-v}\right)-2\gamma^2v\right].
\end{equation}


\subsection{Interpretation}
It will be clear from the above that working in frame $ \mathcal{P} $, rather than in frame $ \mathcal{R} $, enables more efficient derivation of expressions for the power and for the frictional force. In particular, we have been able to handle all polarization states together, and to extend the earlier analysis to all orders in the renormalized intrinsic polarizability. Working in $ \mathcal{P} $ also aids in the elucidation and interpretation of the underlying physics, to which we now turn.

The NESS condition is simply stated as $P'=0$ in frame $ \mathcal{P} $, and we have shown that it holds to all orders in the renormalized intrinsic polarizability. This means that the particle absorbs ($P'_{\text{I}}>0$) and emits ($P'_{\text{II}}<0$) energy at the same rate. (We explicitly show in Appendix G that $ P_{\rm{II}}' $ is the power radiated due to the induced dipole radiation.) In frame $ \mathcal{P} $, the blackbody radiation carries a momentum bias, oriented opposite to the direction of motion of the particle in frame $ \mathcal{R} $; blackbody radiation is isotropic only in its own rest frame. As the blackbody radiation is absorbed, this momentum bias is transferred to the particle, and gives rise to the frictional force ($F'_{\text{I}}<0$) that the particle experiences; indeed, the rate and direction of this momentum transfer are equal to the magnitude and sense of the frictional force.

\begin{figure*}[htbp]
\subfloat[]{\label{point}%
\includegraphics[width=0.45\linewidth]{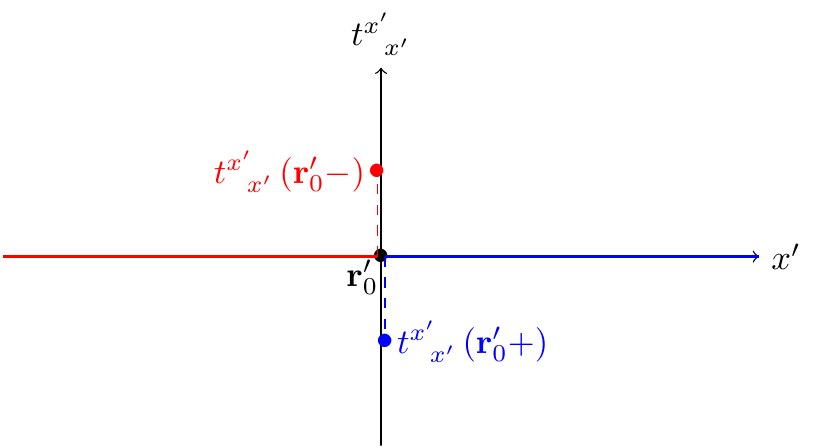}%
}\,
\subfloat[]{\label{extended}%
\includegraphics[width=0.45\linewidth]{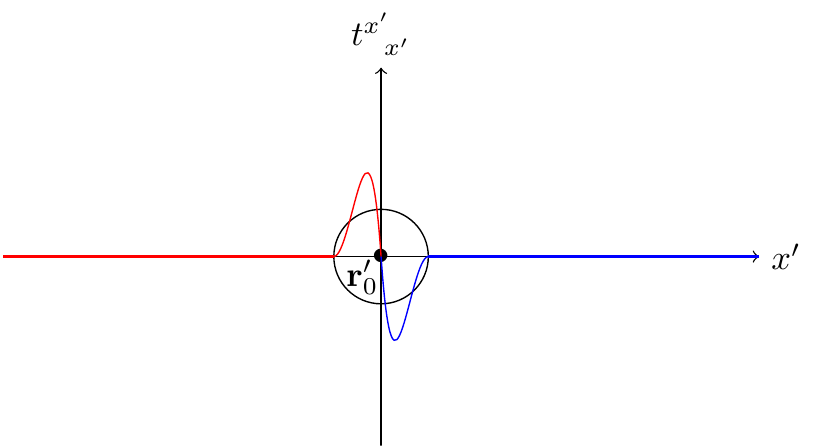}%
}
\caption{Stress tensor for (a) a point particle, and (b) an extended particle.} 
\label{}
\end{figure*}

The absorption and extraction of momentum from the blackbody radiation are accomplished through a process that is entirely internal to the particle. Since we consider only a point interaction, we cannot probe the precise mechanism further here. However, in the similar realistic case of a nanoparticle made of some dielectric material, one might imagine that this material may suffer mechanical deformation as it absorbs and extracts the momentum from the blackbody radiation, such that it is compressed at one end and stretched at the other,  resulting in a mechanical stress tensor gradient that produces an internal force equal and opposite to that due to the momentum transfer from the blackbody radiation. Likewise, in the case of an atom, one might imagine that the wave function is similarly distorted from its symmetric at-rest configuration, in response to the absorption of the momentum imparted by the blackbody radiation. If this is the case, so that the process of momentum absorption is indeed entirely internal, one might wonder how, when viewed externally, the particle can experience a frictional force. The resolution of this seeming paradox is that the internal stress tensor gradient also produces a gradient at the boundary of the particle, between that boundary and the outside, and it is {\it this\/} stress tensor gradient, or difference, that gives rise to the frictional force that the particle experiences. Even in our simple point-interaction model, we see evidence for this interpretation in the discontinuity in the stress tensor from one side of the point to the other that emerges from the proof in Appendix E of the principle of virtual work applicable to the current context. Figure~\ref{point} illustrates the ($ y $-$ z $ cross-sectional integral of the) stress tensor for a point particle, and Fig.~\ref{extended} that for an extended particle.


So, the particle functions like a shock absorber, or momentum converter: it extracts momentum from the absorbed blackbody radiation, and, in doing so, gives rise to a difference in the stress tensor between the boundary of the particle and the outside, which, in turn, results in the frictional force that it experiences. This interpretation only holds water if the radiation emitted by the particle in its rest frame, which arises from the induced dipole fluctuations, has no momentum bias, that is, is isotropic. That this is indeed the case is evidenced by the fact that $F'_{\text{II}}=0$, which, as we have shown, holds to all orders in the renormalized intrinsic polarizability. In spite of this difference between the absorbed and emitted radiation --- that absorbed carries a momentum bias, whereas that emitted does not --- they transfer energy into and out of the particle, respectively, at the same rate. The energy, or spectral, distributions of the absorbed and emitted radiation are therefore identical. Because the Lorentz transformation between frame $ \mathcal{R} $ and frame $ \mathcal{P} $ mixes frequency and momentum, the spectral density of the blackbody radiation in frame $ \mathcal{P} $ is not purely Planckian but is instead a mixture of Planckian spectral densities, with an appropriate mixing  distribution over inverse temperature. This is true of both the absorbed and the emitted radiation.

Indeed, we may write Eq.~\eqref{4.27} as
\begin{equation}\label{4.43}
P'_{\text{I}}=\int_0^{\infty}d\omega\,\omega\left[\operatorname{Im}\hat \alpha_{xx}(\omega)\,\rho^X(\omega)+\left(\operatorname{Im}\hat \alpha_{yy}(\omega)+\operatorname{Im}\hat \alpha_{zz}(\omega)\right)\rho^Y(\omega)\right],
\end{equation}
where 
\begin{equation}\label{4.44}
\rho^{X,Y}(\omega)\equiv\int_{y_-}^{y_+}d\bar y\,\frac{\omega^3}{6\pi^2}\, f^{X,Y}(\bar y)\coth\left(\frac{\beta_{\bar y}\,\omega}{2}\right)
\end{equation}
denote the spectral densities appropriate to $X$ and $Y$ polarizations and $\beta_{\bar y}\equiv \beta \bar y$ denotes the $\bar y$-dependent inverse temperature.  In fact, we could have started from Eq.~\eqref{4.43}, recognising both that the spectral density of the field is the sum of the diagonal elements of Eq.~\eqref{4.14}, which, itself, results from the FDT, Eq.~\eqref{4.8}, and that the diagonal elements of the imaginary part of the effective polarizability in Eq.~\eqref{4.43} serve merely to reflect the extent to which the corresponding components of the field energy are absorbed by the particle through the particle-field interaction.

It is easily verified from Eq.~\eqref{4.44} that, for $v=0$, $\rho^X(\omega)$ and $\rho^Y(\omega)$ each reduce, as they should, to one-third of the Planckian spectral density,
\begin{equation}\label{4.45}
\rho^{\text{PL}}(\omega)\equiv\frac{\omega^3}{2\pi^2}\,\coth\left(\frac{\beta\, \omega}{2}\right).
\end{equation}
More generally, for $v\ne 0$, Eq.~\eqref{4.44} indicates how the spectral density deviates from purely Planckian form: one of the $\omega$ factors represents the energy, the remaining $\frac{\omega^2}{6\pi^2} f^{X,Y}(\bar y)$ factor represents the density of states for that energy, which is $x$-momentum dependent, here represented through the corresponding transformed variable $\bar y$, and the $\coth$ factor governs the occupation of any such specific energy and $x$-momentum state, through the corresponding transformed energy, $\omega \bar y=\gamma(\omega+vk_x)$, in frame $ \mathcal{R} $, where the purely Planckian distribution appropriate to inverse temperature $\beta$ pertains; finally, the expression is integrated over the permitted $x$-momentum values, as represented through $\bar y$. Because of its structure, Eq.~\eqref{4.44} may, equivalently, be thought of as the energy factor, $\omega$, multiplied by the ($v=0$ and therefore $x$-momentum-independent) Planckian density of states factor, $\frac{\omega^2}{6\pi^2}$, multiplied by the $\coth$ thermal occupation factor for that energy, but at transformed inverse temperature $\beta_{\bar y}$, and finally integrated over the mixture of such transformed inverse temperatures governed by the appropriate probability density function $f^{X,Y}(\bar y)$ for the mixing factor $\bar y$. So, in this view, Eq.~\eqref{4.44} is a mixture of Planckian spectral densities, with $f^{X,Y}(\bar y)$ serving as the appropriate probability density function that mixes the corresponding inverse temperatures.

In the case of isotropic polarizability, Eq.~\eqref{4.43} simplifies to 
\begin{equation}\label{4.46}
P'_{\text{I}}=\int_0^{\infty}d\omega\,\omega\,\operatorname{Im}\hat \alpha(\omega)\,\rho(\omega),
\end{equation}
where
\begin{equation}\label{4.47}
\rho(\omega)\equiv \rho^{X}(\omega)+2\,\rho^Y(\omega)=\frac{\omega^3}{4\pi^2\gamma v}
\int_{y_-}^{y_+} d\bar y\,\coth \left(\frac{\beta_{\bar y}\,\omega }{2}\right)=
\frac{\omega^2}{2\pi^2\gamma v\beta}\ln\left[\frac{\sinh \left(\frac{\beta\omega y_+}{2}\right)}{\sinh \left(\frac{\beta\omega y_-}{2}\right)}\right],
\end{equation}
which explicitly exhibits the non-Planckian nature of the corresponding spectral density.

It is clear from its construction that Eq.~\eqref{4.43} represents the rate of absorption of energy per unit time by the particle from the field: the leading $\omega$ factor represents the rate of change per unit time, the spectral density represents the energy distribution of the field, and the effective polarizability represents the extent to which this field energy is transferred to the particle through the particle-field interaction. It is also clear that the corresponding expression for the rate of absorption of $x$-momentum per unit time by the particle from the field, that is, the frictional force experienced by the particle, should be obtained simply by replacing the energy factor, $\omega$, in Eq.~\eqref{4.44} for the spectral density, by the corresponding $x$-momentum factor, $k_x=\omega\frac{(\bar y-\gamma)}{\gamma v}$:
\begin{equation}\label{4.48}
F'_{\text{I}}=\int_0^{\infty}d\omega\,\omega\left[\operatorname{Im}\hat \alpha_{xx}(\omega)\,p^X(\omega)+\left(\operatorname{Im}\hat \alpha_{yy}(\omega)+\operatorname{Im}\hat \alpha_{zz}(\omega)\right)p^Y(\omega)\right],
\end{equation}
where
\begin{equation}\label{4.49}
p^{X,Y}(\omega)\equiv\int_{y_-}^{y_+}d\bar y\,\frac{\omega^3}{6\pi^2}\,\frac{(\bar y -\gamma)}{\gamma v}\,f^{X,Y}(\bar y)\coth\left(\frac{\beta_{\bar y}\,\omega}{2}\right).
\end{equation}
It is easily verified that this approach reproduces Eq.~\eqref{4.24}.

Finally, let us relate the NESS condition to the optical theorem. Using Eq.~\eqref{4.31} to rewrite Eq.~\eqref{4.3} in terms of the effective polarizability, $\hat{\boldsymbol{\alpha}}$, we easily obtain the decomposition
\begin{equation}\label{4.50}
\mathcal F'=\mathcal F'_{\mathbf{E}'^f\mathbf{E}'^f}+\mathcal F'_{\hat{\mathbf{d}}'\hat{\mathbf{d}}'}\equiv-\frac12  \left\langle \mathbf{E}'^{fT} \hat{\boldsymbol{\alpha}}\, \mathbf{E}'^f\right\rangle 
-\frac12  \left\langle \hat{\mathbf{d}}'^T \boldsymbol{\Gamma}'\, \hat{\mathbf{d}}'\right\rangle,
\end{equation}
where $\hat{\mathbf{d}}'\equiv \hat{\boldsymbol{\alpha}}\,\mathbf{E}'^f$ is the corresponding induced dipole moment. In fact, this decomposition is simply a restatement of that met earlier in Eq.~\eqref{4.5},
with 
\begin{subequations}
\begin{equation}\label{4.51a}
\mathcal F'_{\mathbf{E}'^f\mathbf{E}'^f}=\mathcal F'_{0}+\mathcal F'_{\text{I}}
\end{equation}
and 
\begin{equation}\label{4.51b}
\mathcal F'_{\hat{\mathbf{d}}'\hat{\mathbf{d}}'}=\mathcal F'_{\text{II}}.
\end{equation}
\end{subequations}
The correlation function in Eq.~\eqref{4.7} is symmetrized, so $\mathbf{C}'(\omega, \omega'; \mathbf{r}_0', \mathbf{r}_0')$ is symmetric on interchange of $\omega$ and $\omega'$, and may be expressed as
\begin{equation}\label{4.52}
\mathbf{C}'(\omega, \omega'; \mathbf{r}_0', \mathbf{r}_0')=2\pi \delta(\omega+\omega')\,\hat{\mathbf{C}}'(\omega; \mathbf{r}_0'),
\end{equation}
where the symmetric matrix $\hat{\mathbf{C}}'(\omega; \mathbf{r}_0')$ is even in $\omega$. It follows that $P'$ may be correspondingly decomposed as
\begin{equation}\label{4.53}
P'=P'_{\mathbf{E}'^f\mathbf{E}'^f}+P'_{\hat{\mathbf{d}}'\hat{\mathbf{d}}'},
\end{equation}
where
\begin{subequations}
\begin{equation}\label{4.54a}
P'_{\mathbf{E}'^f\mathbf{E}'^f}=\int \frac{d\omega}{2\pi}\,\omega\tr\left[\operatorname{Im}\hat{\boldsymbol{\alpha}}(\omega) \,\hat{\mathbf{C}}'(\omega; \mathbf{r}'_0)\right]
\end{equation}
and
\begin{equation}\label{4.54b}
P'_{\hat{\mathbf{d}}'\hat{\mathbf{d}}'}=\int \frac{d\omega}{2\pi}\,\omega\tr\left[\operatorname{Im}\left\{\hat{\boldsymbol{\alpha}}^*(\omega)\,\boldsymbol{\Gamma}'^{*}(\omega; \mathbf{r}'_0, \mathbf{r}'_0)\,\hat{\boldsymbol{\alpha}}(\omega)\right\} \hat{\mathbf{C}}'(\omega; \mathbf{r}'_0)\right].
\end{equation}
\end{subequations}
Since the matrix in braces in Eq.~\eqref{4.54b} is contracted with the symmetric matrix 
$\hat{\mathbf{C}}'(\omega; \mathbf{r}'_0)$, it may be replaced by its transpose, or, indeed, by the average of itself and its transpose, 
\begin{equation}\label{4.55}
\frac12\left\{\hat{\boldsymbol{\alpha}}^*\,\boldsymbol{\Gamma}'^{*}\,\hat{\boldsymbol{\alpha}}+\hat{\boldsymbol{\alpha}}\,\boldsymbol{\Gamma}'^{*}\,\hat{\boldsymbol{\alpha}}^*\right\}
=\frac12\left\{\hat{\boldsymbol{\alpha}}^*\operatorname{Re}\boldsymbol{\Gamma}'\,\hat{\boldsymbol{\alpha}}
-i\,\hat{\boldsymbol{\alpha}}^*\operatorname{Im}\boldsymbol{\Gamma}'\,\hat{\boldsymbol{\alpha}}\right\}
+\frac12\left\{\hat{\boldsymbol{\alpha}}\operatorname{Re}\boldsymbol{\Gamma}'\,\hat{\boldsymbol{\alpha}}^*
-i\,\hat{\boldsymbol{\alpha}}\operatorname{Im}\boldsymbol{\Gamma}'\,\hat{\boldsymbol{\alpha}}^*\right\}.
\end{equation}
The sum of the terms in Eq.~\eqref{4.55} that involve $\operatorname{Re} \boldsymbol{\Gamma}'$ is real, and therefore even in $\omega$, so does not contribute to the integral in Eq.~\eqref{4.54b}; the sum of the terms in Eq.~\eqref{4.55} that involve $\operatorname{Im} \boldsymbol{\Gamma}'$ is imaginary, and therefore odd in $\omega$, so does contribute to the integral in Eq.~\eqref{4.54b}. We may therefore rewrite Eq.~\eqref{4.54b} as
\begin{equation}\label{4.56}
P'_{\hat{\mathbf{d}}'\hat{\mathbf{d}}'}=-\int \frac{d\omega}{2\pi}\,\omega\tr\left[\frac12\left\{\hat{\boldsymbol{\alpha}}^*(\omega)\operatorname{Im}\boldsymbol{\Gamma}'(\omega; \mathbf{r}'_0, \mathbf{r}'_0)\,\hat{\boldsymbol{\alpha}}(\omega)+\hat{\boldsymbol{\alpha}}(\omega)\operatorname{Im}\boldsymbol{\Gamma}'(\omega; \mathbf{r}'_0, \mathbf{r}'_0)\,\hat{\boldsymbol{\alpha}}^*(\omega)\right\} \hat{\mathbf{C}}'(\omega; \mathbf{r}'_0)\right].
\end{equation}
However, from Eq.~\eqref{4.31}, it follows that 
\begin{equation}\label{4.57}
\operatorname{Im}\hat{\boldsymbol{\alpha}}(\omega)=\frac12\left\{\hat{\boldsymbol{\alpha}}^*(\omega)\operatorname{Im}\boldsymbol{\Gamma}'(\omega; \mathbf{r}'_0, \mathbf{r}'_0)\,\hat{\boldsymbol{\alpha}}(\omega)+\hat{\boldsymbol{\alpha}}(\omega)\operatorname{Im}\boldsymbol{\Gamma}'(\omega; \mathbf{r}'_0, \mathbf{r}'_0)\,\hat{\boldsymbol{\alpha}}^*(\omega)\right\},
\end{equation}
which is a statement of the optical theorem in the current context~\citep{Berman:OptTrm}. Thus, we immediately find from Eq.~\eqref{4.54a} and \eqref{4.56} that 
\begin{equation}\label{4.58}
P'=P'_{\mathbf{E}'^f\mathbf{E}'^f}+P'_{\hat{\mathbf{d}}'\hat{\mathbf{d}}'}=0.
\end{equation}

We may conclude, therefore, that, in the current context, satisfaction of the NESS condition is an immediate consequence of the optical theorem. In the second paper in this series, we will again consider the decomposition in Eq.~\eqref{4.50}, but will allow for dipole fluctuations that are uncorrelated with field fluctuations, and satisfy a separate FDT at a corresponding dipole temperature. In this case, as one might expect, the NESS condition is not automatically satisfied, since the optical theorem does not account for the intrinsic (rather than induced) dipole radiation emitted by the particle, but this condition may be engineered to hold by suitably relating the temperatures of the absorbed blackbody radiation and of the emitted dipole radiation.

The radiative corrections included in the effective polarizability, $ \hat{\bm{\alpha}} $, discussed here
are reminiscent of radiative corrections in quantum electrodynamics.  The
radiative corrections to the intrinsic polarizability $ \bm{\alpha}$ are already accounted
for in the phenomenological value---see Refs.~\cite{Lach:radiative,Piszczatowski:Frequency} for an example of the helium atom. 
However, the correction to the photon propagator (vacuum
polarization) should be considered. The imaginary part of the modified photon propagator will 
indeed give a correction, but only when the frequency $\omega>2m_e$, where
$m_e$ is the mass of the electron.  Because, in the present context, the
typical frequency modes that contribute to the quantum vacuum friction are of the same order as the temperature $T$ of the
blackbody radiation, this effect would only be expected to be significant
if $T> 10^{10}$ K, which should be far beyond the range of applicability
of our considerations.  Even in that extreme regime, the effect of vacuum polarization would be
small: relative to the imaginary part of the vacuum Green's function in 
Eq.~(4.13), the one loop effect is calculated to be only
\be
\frac{\Im \mathbf{\Gamma}^{\prime (1)}(\omega;\mathbf{0,0})}
{\Im \mathbf{\Gamma}^{\prime (0)}(\omega;\mathbf{0,0})}
\sim \frac{2}{3\pi}\left(\ln\frac{2\omega}{m_{e}}-\frac{5}{3}\right)\alpha,\quad \omega\gg 2m_e,
\ee
where $\alpha\approx1/137$ is the fine structure constant.

\section{numerical estimate for quantum vacuum friction of a gold atom}\label{numerics}
How big is quantum vacuum friction? Will it be accessible to experiments? To answer these questions, in this section let us obtain an estimate for the quantum friction on a gold atom moving uniformly in vacuum.  We will assume the intrinsic (renormalized, but not dressed) polarizability is isotropic and static, $ \bm{\alpha}(\omega)=\alpha(0) \vb{1} $. Then the effective polarizability in Eq.~\eqref{4.34} becomes
\begin{equation}\label{alphahat1}
\Im\hat{\alpha}(\omega)=\frac{\omega^{3}}{6\pi}\alpha^{2}(0)\frac{1}{1+(\frac{\omega^{3}}{6\pi})^{2}\alpha^{2}(0)},
\end{equation}
where $ \alpha(0) $ is the static polarizability of the gold atom, being $5.33\cross 10^{-24} \rm{cm}^{3}$ according to Ref.~\cite{polarizability-table}. In the low-frequency limit, $ \Im\hat{\alpha}(\omega) $ reduces to the well-known radiation reaction model~\cite{Jentschura:PRL}, $ \frac{\omega^{3}}{6\pi}\alpha_{0}^{2} $. In the high-frequency limit, $ \Im\hat{\alpha}(\omega) $ becomes  $ \frac{6\pi}{\omega^{3}} $, which is independent of the value for the static polarizability. 

Plugging Eq.~\eqref{alphahat1} into the formula for quantum friction Eq.~\thetag{4.35}, we obtain
\begin{equation}\label{F1}
F^{\rm{ISO}}=\frac{\alpha_{0}^{2}}{12\pi^{3}}\int_{0}^{\infty} d\omega \frac{\omega^{7}}{1+(\frac{\omega^{3}}{6\pi})^{2}\alpha_{0}^{2}}\int_{y_{-}}^{y_{+}} \frac{dy}{\gamma^{2}v^{2}} \frac{y-\gamma}{e^{\beta\omega y}-1}.
\end{equation}
For ease of numerical evaluation, let us introduce the dimensionless frequency $ x=\frac{\beta\omega}{2} $ and temperature $ \lambda=\left(\frac{\alpha_{0}}{6\pi}\right)^{1/3}\frac{2}{\beta} $.
Then Eq.~\eqref{F1} can be rewritten as 
\begin{equation}\label{F2}
F^{\rm{ISO}}= C\int_{0}^{\infty} dx \frac{\lambda^{8}x^{7}}{1+\lambda^{6}x^{6}} \int_{y_{-}}^{y_{+}} \frac{dy}{\gamma^{2}v^{2}} \frac{y-\gamma}{e^{2x y}-1}, \qquad C=\frac{(6\pi)^{8/3}}{12\pi^{3}}\alpha_{0}^{-2/3}.
\end{equation}
For a gold atom, the dimensional factor $ C $ in Eq.~\eqref{F2}, which is independent of temperature and velocity, evaluates to $ C=6.99\cross 10^{-6} \,\rm{N} $, after converting to \rm{SI} units. The remaining factor in Eq.~\eqref{F2} is a dimensionless function of velocity $ v $ and rescaled temperature $ \lambda $.
The integral in Eq.~\eqref{F2} is dominated by the low-frequency contributions in the low-temperature limit ($ \lambda\ll 1 $) and the high-frequency contributions in the high-temperature limit ($ \lambda\gg 1 $). Effectively, the model for $ \Im\hat{\alpha}(\omega) $ reduces to the radiation reaction model at low temperatures  and $ 6\pi/\omega^{3} $ at high temperatures.
The more general expressions of the two limits have been worked out in Eq.~\eqref{lowhigh}. Since the gold atom considered has a static polarizability before being dressed by the radiation, we just need to set $ a=0 $ and $ b=0 $ together with $ A=B=\alpha_{0} $ in these equations to obtain

\begin{subequations}\label{Flowhigh}
\begin{equation}\label{Flow}
F^{\rm{ISO}}_{\rm{Low T}}=\frac{4\pi^{5}\alpha_{0}^{2}}{45\beta^{8}}\eta(8,v)=-\frac{4\pi^{5}\alpha_{0}^{2}\gamma^{6}}{45\beta^{8}}\left(\frac{8}{3}v+\frac{16}{3}v^{3}+\frac{8}{7}v^{5}\right),
\end{equation}
\begin{equation}\label{Fhigh}
F^{\rm{ISO}}_{\rm{High T}}=\frac{\pi}{\beta^{2}}\eta(2,v)=\frac{\pi}{2\gamma^{2}v^{2}\beta^{2}}\left[\ln(\frac{1+v}{1-v})-2\gamma^{2}v\right],
\end{equation}
\end{subequations}
where Eq.~\eqref{Fhigh} is identical to Eq.~\eqref{4.42} and the force is independent of the actual value of the intrinsic polarizability of the gold atom. Here, we note the nonrelativistic limit of the low-temperature blackbody friction agrees exactly with Eq.~(15) in Ref.~\cite{Jentschura:PRL}. But our Eqs.~\eqref{Flowhigh} also give the high-temperature and relativistic behavior of the blackbody friction. 

\begin{figure*}[h!]
\subfloat[]{\label{fig1a}%
\includegraphics[width=0.48\linewidth]{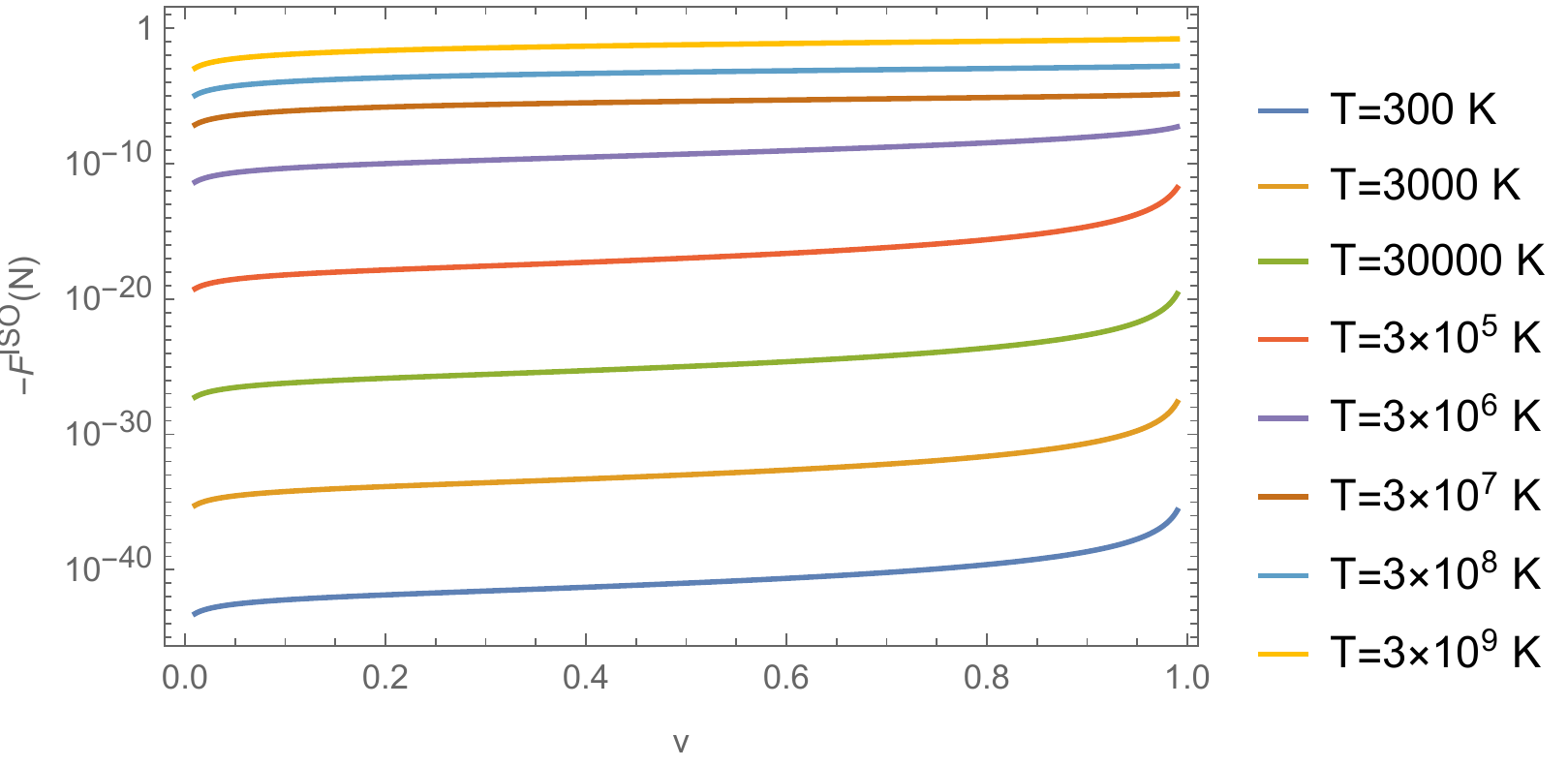}%
}\,
\subfloat[]{\label{fig1b}%
\includegraphics[width=0.45\linewidth]{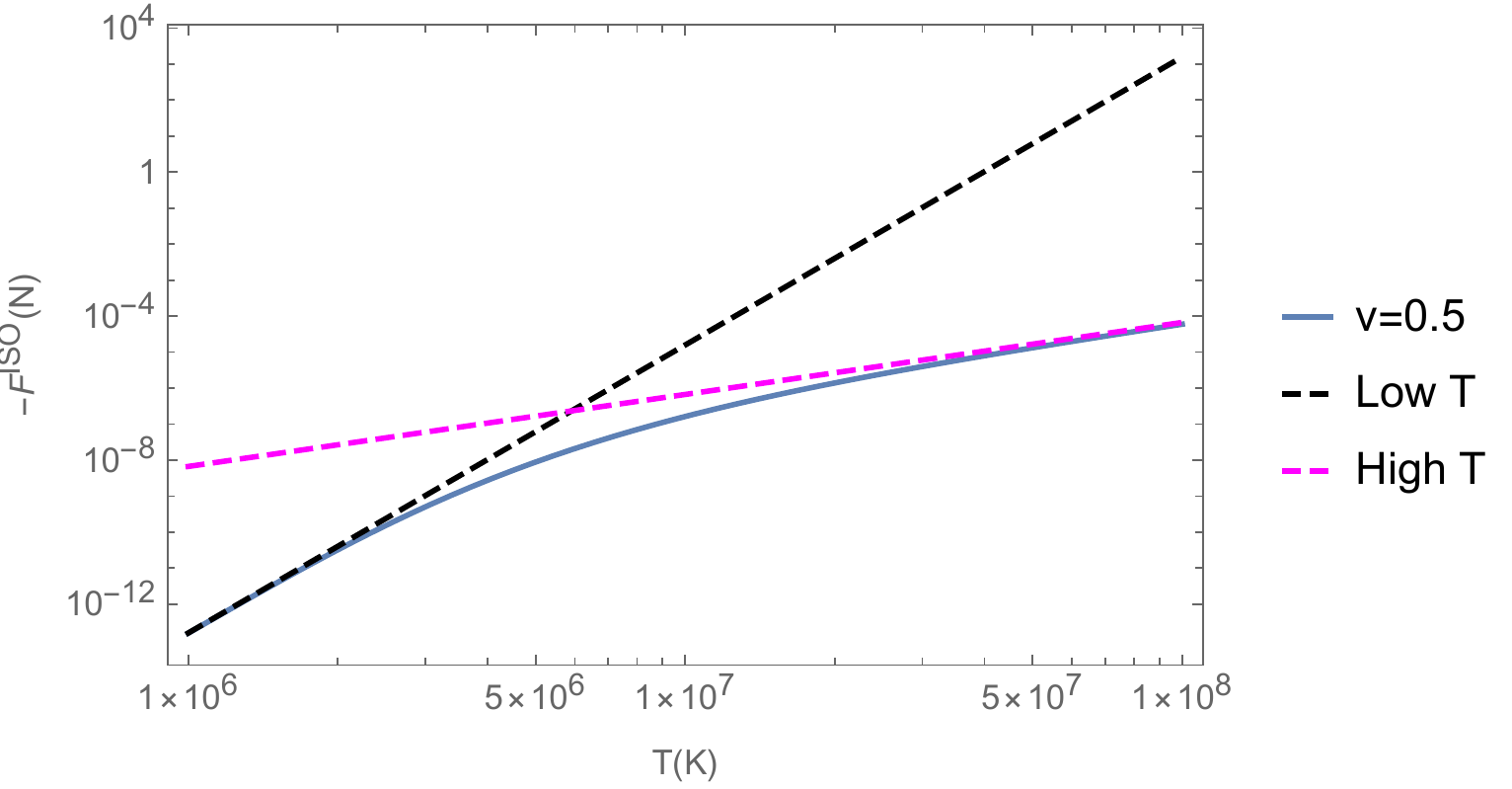}%
}
\caption{The magnitude of the quantum vacuum frictional force $ -F^{\rm{ISO}} $ on a gold atom moving uniformly with velocity $ v $ through the blackbody raditation at temperature $ T $ is illustrated in Newtons. (a)\, At various temperatures of the blackbody radiation, $ -F^{\rm{ISO}} $ is plotted as a function of velocity of the gold atom for $ v\in [0.01,0.99] $.  (b)\, The solid blue curve plots $ -F^{\rm{ISO}} $ as a function of temperature of the blackbody radiation $ T $ at a fixed velocity $ v=0.5 $. The dashed black curve and the dashed mangenta curve plot the low and high temperature limits of the quantum vacuum friction shown in Eq.~\eqref{Flow} and Eq.~\eqref{Fhigh}, respectively, as a function of the blackbody temperature $ T $ for $ v=0.5 $. It is seen the transition occurs at an incredibily high temperature around $ T=10^{7}\, \rm{K} $.} 
\label{Fig1}
\end{figure*}

In Fig.~\ref{Fig1}, we illustrate the velocity dependence and temperature dependence of the quantum vacuum friction on the moving gold atom. It is seen from Fig.~\ref{fig1a} that the magnitude of the quantum vacuum friction monotonically increases with velocity. For a fixed velocity, raising the temperature in general enhances the quantum friction. But the temperature effect is more pronounced for the lower temperatures than the higher temperatures because the frictional force is proportional to $ T^{8} $ in the low-temperature limit but only $ T^{2} $ in the high-temperature limit as predicted by Eq.~\eqref{Flow} and Eq.~\eqref{Fhigh}. In Fig.~\ref{fig1b}, the transition of different temperature behaviors is seen to occur roughly in between $10^{6}\,\rm{K}$ and $ 10^{8}\,\rm{K} $. That is, the low-frequency radiation model is a good approximation so long as the temperature does not get above $ 10^{6} \,\rm{K} $.

For the sake of attracting the attention of experimentalists, let us comment on the possibility of detecting the effects of quantum vacuum friction. It will cause the gold atom to decelerate when the external driving force is removed. To make a rough estimate of the time taken for the atom to decelerate by a noticeable amount,  we assume the gold atom would be in a ``quasi nonequilibrium steady state'' where the friction on it could still be calculated using the NESS formulas. Since it is hard, experimentally,  to accelerate a neutral particle to relativistic velocities or raise the temperature of the vacuum above $ 10^{6}\,\rm{K} $, we restrict our calculation in the low-temperature and nonrelativistic regime, where we can safely apply Newton's second law together with the lowest order (in $ v $) approximation of the frictional force shown in Eq.~\eqref{Flow}:
\begin{equation}\label{Flowv}
F(v)=-\frac{32\pi^{5}\alpha_{0}^{2}}{135\beta^{8}}v=m\frac{dv}{dt}.
\end{equation}
The time taken for the gold atom to decelerate from an initial velocity $ v_{i} $ to a final velocity $ v_{f} $ is then found to be 
\begin{equation}\label{Deltat}
\Delta t=-\tau\ln\frac{v_{f}}{v_{i}}, \quad \tau=\frac{135m\beta^{8}}{32\pi^{5}\alpha_{0}^{2}},
\end{equation}
where $ \tau$ is evaluated to be $1.72 \cross 10^{25} \,\rm{s} $ at room temperature $ T=300 \, \rm{K} $.\footnote{The mass of a gold atom is $ 197 u = 1.84\cross 10^{11} \, \rm{eV} $. The conversion factors used in the estimate are $k_{B}=8.62\cross 10^{-5} \rm{eV/K}  $, $ \hbar c=1.97\cross 10^{-5}\, \rm{eV}\cdot\rm{cm} $ and $ c=3.00\cross 10^{10}\,\rm{cm/s} $.} 

For example, the time taken for the velocity of to be reducing by 10\% is $ \Delta t = 1.81\cross 10^{24}\, \rm{s} $. It then seems hopeless to detect the quantum vacuum frictional effect at room temperature. However, if the experiment could be performed at $ T=30,000\, \rm{K} $, $ \Delta t $ would be $ 16 $ orders of magnitude shorter, being $ 1.81\cross 10^{8} \,\rm{s}=5.91\, yrs $. Coincidentally, this is close to the average time graduate students spend in a US institution!

%
%

\section{conclusions}\label{conclusions}
In this paper, we provide fully relativistic and finite temperature formulations for calculating the quantum frictional power and force on a neutral particle with  real intrinsic polarizability $ \bm{\alpha}(\omega) $. The focus of our exploration is on the quantum vacuum frictional phenomenon for a particle maintained in the nonequilibrium steady state (NESS). That is, the particle is assumed to be moving with constant velocity, relative to a background filled only with blackbody radiation, and its energy is conserved.

We perform calculations both in the rest frame of the blackbody radiation (frame $ \mathcal{R} $) and in the rest frame of the particle (frame $ \mathcal{P} $). In both frames, we obtain explicit expressions for the quantum vacuum frictional power and force for different polarization states to second order in $ \bm{\alpha}(\omega) $. It is easily seen from our results that the quantum vacuum friction is always a true drag, independent of the assumed model for polarizability of the particle and its polarization state. The frictional power and force in each of these two frames are simply related, through Eqs.~\eqref{4.1}. It is a special property of NESS that the quantum vacuum frictional force is the same in the two frames. As a consequence, the external force that maintains the motion of the particle at constant velocity in frame $ \mathcal{R} $ must equal that which keeps the particle at rest in frame $ \mathcal{P} $. The energetics, however, are different in the two frames. In frame $ \mathcal{R} $, this frictional force does negative work on the particle, which is compensated by the positive work done by the external driving force. In frame $ \mathcal{P} $, neither force does any work on the particle as it is not moving. 


At a microscopic level, the particle acts as both a momentum converter and an energy bookkeeper. Due to the relative motion between the particle and the blackbody radiation, in frame $ \mathcal{P} $, this radiation carries a momentum bias oriented opposite to the velocity of the particle in frame $ \mathcal{R} $. It is this momentum bias that is transferred to the particle and gives rise to the quantum vacuum frictional force. But the particle can never absorb nor emit net energy because its intrinsic polarizability is purely real, $ \Im\bm{\alpha}(\omega)=0 $, or, equivalently, its effective polarizability $ \hat{\bm{\alpha}}(\omega) $ satisfies the optical theorem, $ \Im\hat{\bm{\alpha}}(\omega)=\Im\Gamma'(\omega)|\hat{\bm{\alpha}}(\omega)|^{2} $. In both $ \mathcal{R} $ and $ \mathcal{P} $, the net power gained by the particle can be broken into an absorbed part, $ P_{\rm{I}} $ and $ P'_{\rm{I}} $, respectively, and an emitted part, $ -P_{\rm{II}} $ and $-P'_{\rm{II}}$, respectively. The emitted power is entirely a quantum version of the classical dipole radiation with the dipole being induced by the field fluctuations, which is clearly illustrated in Appendix \ref{apG}. It must be combined with the absorbed power to account for the total rate of work done on the particle by the quantum vacuum friction, $ P_{1}+P_{2}=P=Fv $ in frame $ \mathcal{R} $ and $ P_{1}'+P_{2}'=P'=0 $ in frame $ \mathcal{P} $.

Replacing $\bm{\alpha}\bm{\Gamma}'\bm{\alpha}  $ in the second-order expressions for power and force with the effective polarizability written in terms of the intrinsic polarizability, $\bm{\hat{\alpha}}=\bm{\alpha}\left(1-\bm{\Gamma}'\cdot \bm{\alpha}\right)^{-1} $, we are able to extend our expressions to all orders in $ \bm{\alpha} $. Since the real part of the vacuum Green's dyadic $ \Re\bm{\Gamma}' $ is divergent, it must be absorbed through a renormalization of the bare intrinsic polarizability.  A numerical estimate of the quantum vacuum friction on a moving gold atom is obtained using the full order expression of the force; it is seen that the higher order (in $\bm{\alpha}$) corrections only become important at rather high temperatures, above $ 10^{6}\, \rm{K} $ for a gold atom. The numerical results also show that the effect of quantum vacuum friction is too tiny to be observed at room temperature but it may become observable when the temperature is raised by at least two orders of magnitude. 

The exclusion of any intrinsic dissipation may be too idealistic. Susceptibilities which are consistent with causality requirements must develop both real and imaginary parts in order to respect the Kramers-Kronig relation.\footnote{Atoms, however, acquire dissipation only through interaction with the radiation field.} In view of this, we will further study in a subsequent paper the quantum frictional phenomenon associated with a neutral particle that does possess intrinsic dissipation. In that case, NESS is no longer automatically satisfied because the particle itself has the ability to act as a net absorber or emitter of energy. Indeed, satisfaction of the NESS requirement imposes a relation between the temperature of the particle and the temperature of the blackbody radiation.

In the course of this investigation on quantum vacuum friction, we have also derived formulas for the quantum frictional power and force applicable to a more general backgound. We hope, in the near future, to revisit the classic situation where a neutral particle is passing above a dielectric surface. Of course, many other authors have already studied this problem but have obtained disparate answers. Oelschl\"{a}ger's thesis~\cite{Marty:thesis} contains a very useful summary of the different results. Our relativistic and finite temperature formulation may prove to be advantageous in resolving this controversy.

\begin{acknowledgments}
We thank the US National Science Foundation, grants Nos. 1707511, 2008417, for partial support of this work. We thank Steve Fulling, John Joseph Marchetta and Prachi Parashar for insightful comments. This paper reflects solely the authors' personal opinions and does not represent the opinions of the authors' employers, present and past, in any way.
\end{acknowledgments}

\appendix
\section{THE GREEN'S DYADIC}\label{apA}
The Green's dyadic $\vb{\Gamma}(\vb{r},\vb{r'};\omega)$  satisfies the following differential equation derived from Maxwell-Heaviside equations,
\begin{equation}\label{eqA-1}
\left[\frac{1}{\omega^{2}}\curl\curl - \vb{\varepsilon}(\vb{r};\omega)\right]\vb{\Gamma}(\vb{r},\vb{r'};\omega)=\delta(\vb{r}-\vb{r'}),
\end{equation}
where $ \vb{\varepsilon}(\vb{r};\omega) $ is the permittivity of the dielectric, if any. 
The source of the field could be some polarization field or any current,
\begin{equation}\label{eqA-2}
\vb{E}(\vb{r};\omega)=\int d\vb{r'} \,\vb{\Gamma}(\vb{r},\vb{r'};\omega) \cdot \vb{P}(\vb{r'};\omega)=\int d\vb{r'} \,\left(-\frac{1}{i\omega}\right)\vb{\Gamma}(\vb{r},\vb{r'};\omega) \cdot \vb{j}(\vb{r'};\omega).
\end{equation}
The geometry of the problem we consider always possesses a translational symmetry in the $ x$-$y $ plane, which permits us to Fourier transform the Green's dyadic in these spatial directions,
\begin{equation}\label{eqA-3}
\vb{\Gamma}(\vb{r},\vb{r'};\omega)=\int \frac{d^{2}\vb{k}_{\perp}}{(2\pi)^{2}} e^{i\vb{k}_{\perp}\cdot (\vb{r}_{\perp}-\vb{r'}_{\perp})} \vb{g}(z,z';\omega,\vb{k}_{\perp}).
\end{equation}
Solving the differential equation Eq.~\eqref{eqA-1} amounts to finding the reduced Green's dyadic $ \vb{g}(z,z';\omega,\vb{k}_{\perp}) $.

Imagining a uniform dielectric occupying the semi-space $ z<0 $, we define wave numbers
\begin{equation}
k=\sqrt{k_{x}^{2}+k_{y}^{2}}, \quad
\kappa=\sqrt{k^{2}-\omega^{2}}, \quad
\kappa'=\sqrt{k^{2}-\omega^{2}\varepsilon}
\end{equation}
and the reflection coefficients  for the transverse electric and transverse magnetic modes
\begin{equation}
r^{E}=\frac{\kappa-\kappa'}{\kappa+\kappa'}, \qquad
r^{H}=\frac{\kappa-\kappa'/\varepsilon}{\kappa+\kappa'/\varepsilon}.
\end{equation}
The scalar Green's functions for the transverse electric and transverse magnetic modes are
\begin{equation}\label{gEH}
g^{E,H}(z,z';\omega,k)=\frac{1}{2\kappa}e^{-\kappa|z-z'|}+\frac{r^{E,H}}{2\kappa}e^{-\kappa(z+z')}.
\end{equation}
Then the reduced Green's dyadic can be conveniently written in terms of these scalar Green's functions,
\begin{align}\label{eqA-4}
\vb{g}(z,z';\omega,\vb{k}_{\perp})=&\mqty(\frac{k_{x}^{2}}{k^{2}}\frac{1}{\varepsilon}\partial_{z}\frac{1}{\varepsilon'}\partial_{z'}g^{H}+\frac{k_{y}^{2}}{k^{2}}\omega^{2}g^{E}
&
\frac{k_{x}k_{y}}{k^{2}}\frac{1}{\varepsilon}\partial_{z}\frac{1}{\varepsilon'}\partial_{z'}g^{H}-\frac{k_{x}k_{y}}{k^{2}}\omega^{2}g^{E}
&
\frac{ik_{x}}{\varepsilon\varepsilon'}\partial_{z}g^{H}
\\
\frac{k_{x}k_{y}}{k^{2}}\frac{1}{\varepsilon}\partial_{z}\frac{1}{\varepsilon'}\partial_{z'}g^{H}-\frac{k_{x}k_{y}}{k^{2}}\omega^{2}g^{E}
&
\frac{k_{y}^{2}}{k^{2}}\frac{1}{\varepsilon}\partial_{z}\frac{1}{\varepsilon'}\partial_{z'}g^{H}+\frac{k_{x}^{2}}{k^{2}}\omega^{2}g^{E}
&
\frac{ik_{y}}{\varepsilon\varepsilon'}\partial_{z}g^{H}
\\
\frac{-ik_{x}}{\varepsilon\varepsilon'}\partial_{z'}g^{H}
&
\frac{-ik_{y}}{\varepsilon\varepsilon'}\partial_{z'}g^{H}
&
\frac{k^{2}}{\varepsilon\varepsilon'}g^{H}).
\end{align}
Here $ \varepsilon $ and $ \varepsilon' $ are the permittivities evaluated at $ z $ and $ z' $, respectively.

In this paper, we are entirely concerned with the vacuum situation with $ \varepsilon=\varepsilon'=1 $. As a result, the reflection coefficients vanish and the scalar Green's functions coincide, 
\begin{equation}\label{eqA-4.5}
g^{E}(z,z';\omega,k)=g^{H}(z,z';\omega,k)=\frac{1}{2\kappa}e^{-\kappa|z-z'|}.
\end{equation}
The vacuum Green's dyadic therefore reads
\begin{align}\label{eqA-7}
\vb{g}(z,z';\omega,\vb{k}_{\perp})=\frac{1}{2\kappa}e^{-\kappa|z-z'|}
&\mqty(\omega^{2}-k_{x}^{2}
&
-k_{x}k_{y}
&
-ik_{x}\kappa\sgn(z-z')
\\
-k_{x}k_{y}
&
\omega^{2}-k_{y}^{2}
&
-ik_{y}\kappa\sgn(z-z')
\\
ik_{x}\kappa\sgn(z'-z)
&
\quad ik_{y}\kappa\sgn(z'-z)
&
k^{2}).
\end{align}
Obviously, the vacuum Green's dyadic is a symmetric matrix. 

The symmetries of the vacuum Green's functions in frequency and wave number are frequently taken advantage of in various calculations throughout the paper.
The symmetries in $ k_{x} $ and $ k_{y} $ are obvious from Eq.~\eqref{eqA-7}. All the diagonal Green's functions are even in both $ k_{x} $ and $ k_{y} $ but the off-diagonal Green's functions are odd  either in $ k_{x} $ or $ k_{y} $.

The Fourier transform of $ g_{ij} $ in $ y $ evaluated at coincident $ y $ coordinates is
\begin{equation}\label{eqA-7.6}
G_{ij}(\omega,k_{x})=\int\frac{dk_{y}}{2\pi}  g_{ij}(\omega,k_{x},k_{y}).
\end{equation}
The real parts of $ G_{ij} $ are all even in $ \omega $ and actually divergent.
Only if $ \omega^{2}>k^{2} $ does the wavenumber $ \kappa $ become imaginary,
\begin{equation}\label{eqA-7.4}
\kappa=-i\sgn(\omega)\sqrt{\omega^{2}-k^{2}}.
\end{equation}
The branch is chosen so that the Green's functions are retarded. The imaginary part of $ G_{ij} $ are always odd in $\omega $ because of the additional factor of $ \sgn(\omega)$ in $ \kappa $. As an example, the imaginary part of $ G_{xx}(\omega,k_{x}) $ is worked out in detail in Eq.~\eqref{eq2-17.6}. Below we give the explicit forms for the imaginary parts of all different $ G_{ij} $:
\begin{subequations}
\begin{equation}\label{Imgxx}
\addtolength{\arraycolsep}{-3pt}
 \Im G_{xx}(\omega,k_{x})=\left\{%
 \begin{array}{lcrcl}
\frac{1}{4}(\omega^{2}-k_{x}^{2})\sgn(\omega),& \qquad k_{x}^{2}<\omega^{2},\\
0, & \qquad k_{x}^{2}>\omega^{2};
\end{array}\right.
\end{equation}
\begin{equation}\label{Imgyy}
\addtolength{\arraycolsep}{-3pt}
 \Im G_{yy}(\omega,k_{x})=\left\{%
 \begin{array}{lcrcl}
\frac{1}{8}(\omega^{2}+k_{x}^{2})\sgn(\omega),& \qquad k_{x}^{2}<\omega^{2},\\
0, & \qquad k_{x}^{2}>\omega^{2};
\end{array}\right.
\end{equation}
 \begin{equation}\label{Imgzz}
\addtolength{\arraycolsep}{-3pt}
 \Im G_{zz}(\omega,k_{x})=\left\{%
 \begin{array}{lcrcl}
\frac{1}{8}(\omega^{2}+k_{x}^{2})\sgn(\omega),& \qquad k_{x}^{2}<\omega^{2},\\
0, & \qquad k_{x}^{2}>\omega^{2};
\end{array}\right.
\end{equation}
\begin{equation}\label{Imgxy}
\Im G_{xy}(\omega,k_{x})=\Im G_{yz}(\omega,k_{x})=\Im G_{xz}(\omega,k_{x})=0\,.
\end{equation}

We note $ \Im G_{zz} $ has exactly the same form as $ \Im G_{yy} $, reflecting the symmetry of the geometry of the vacuum problem. The imaginary part of all off-diagonal Green's functions vanish, but for different reasons: $ g_{xy} $ is odd in $ k_{y} $ so that $ \Im G_{xy} $ evaluates to zero when taking the $ k_{y} $ integration over an even interval;  $ g_{yz} $ and $ g_{xz} $ each contains a factor of $ \sgn(z-z') $, so $ \Im G_{yz} $ and $ \Im G_{xz} $ both evaluate to zero because we are taking the limit of the  coincident $ z $ coordinates.
 \end{subequations}
 
\section{LORENTZ TRANSFORMATION PROPERTIES OF DIPOLE AND FIELD}\label{apB}
Here, we describe how both the dipole and the field transform under a Lorentz boost in the $ x $ direction with speed $ v $.

The dipole is often transformed from its moving frame to its rest frame. In the frequency domain, the transformations of $ x $ and $ y $ components of the dipole are
\begin{equation}\label{eqB-1}
d_{x}(\omega)=d_{x}'(\gamma \omega )
\qquad
d_{y}(\omega)=\gamma d_{y}'(\gamma\omega ).
\end{equation}
In the time domain, the transformations for $ d_{x} $ and $ d_{y} $ are
\begin{equation}\label{eqB-2}
d_{x}(t)=\frac{1}{\gamma}d_{x}'\left(\frac{t}{\gamma} \right)
\qquad
d_{y}(t)=d_{y}'\left(\frac{t}{\gamma}\right ).
\end{equation}

However, the field is often transformed from frame $ \mathcal{P} $ to frame $ \mathcal{R} $.
In spacetime coordinates, the transformations of different components of the electric field read
\begin{align}\label{eqB-3}
E_{x}'(\vb{r'},t')&=E_{x}(\vb{r},t),\nonumber\\
E_{y}'(\vb{r'},t')&=\gamma[E_{y}(\vb{r},t)-vB_{z}(\vb{r},t)],\\\nonumber
E_{z}'(\vb{r'},t')&=\gamma[E_{z}(\vb{r},t)+vB_{y}(\vb{r},t)],
\end{align} 
where the transformation of the coordinates are
\begin{equation}
t=\gamma(t'+vx'), \quad x=\gamma(x'+vt'), \quad y=y', \quad z=z'.
\end{equation}
In momentum space, these transformations become
\begin{align}\label{eqB-4}
E_{x}'(\omega,k_{x},k_{y})&=E_{x}\left[\gamma(\omega+k_{x}v),\gamma(k_{x}+\omega v),k_{y}\right],\nonumber\\
E_{y}'(\omega,k_{x},k_{y})&=\gamma(E_{y}-vB_{z})[\gamma(\omega+k_{x}v),\gamma(k_{x}+\omega v),k_{y}],\\
E_{z}'(\omega,k_{x},k_{y})&=\gamma(E_{z}+vB_{y})[\gamma(\omega+k_{x}v),\gamma(k_{x}+\omega v),k_{y}]\nonumber.
\end{align}

\section{THE MOMENTUM DISTRIBUTION FUNCTIONS FOR DIFFERENT POLARIZATIONS}\label{apC}

The momentum distribution functions for the $ x $ polarization and $ y $ polarization 
are defined respectively as
\begin{align}\label{C1}
f^{X}(y)&=\frac{3}{4\gamma v} \left[y^{2}-\left(y-\frac{1}{\gamma}\right)^{2} \frac{1}{v^{2}}\right]=\frac{3}{4\gamma v} \left[1-\frac{1}{\gamma^{2}v^{2}}\left(y-\gamma\right)^{2}\right],\nonumber\\
f^{Y}(y)&=\frac{3}{4\gamma v}\left\lbrace 1-\frac{1}{2}\left[y^{2}-\left(y-\frac{1}{\gamma}\right)^{2} \frac{1}{v^{2}}\right]\right\rbrace=\frac{3}{4\gamma v}\left\lbrace1-\frac{1}{2}\left[1-\frac{1}{\gamma^{2}v^{2}}\left(y-\gamma\right)^{2}\right]\right\rbrace.
\end{align}
These functions are normalized to $ 1 $ with respect to the integral on $ y $,
\begin{equation}\label{C2}
\int_{y_{-}}^{y_{+}} dy \, f^{X,Y}(y)=1,
\end{equation}
with $ y_{+}=\sqrt{\frac{1+v}{1-v}} $ and $ y_{-}=\sqrt{\frac{1-v}{1+v}} $.
Other integrals of these distribution functions used in the formulas for quantum vacuum frictional power and force are 
\begin{equation}\label{C3}
\int_{y_{-}}^{y_{+}} dy\, y f^{X,Y}(y)=\gamma, \qquad 
\int_{y_{-}}^{y_{+}} dy\, \frac{1}{v}\left(y-\frac{1}{\gamma}\right) f^{X,Y}(y)=\gamma v.
\end{equation}

The momentum distribution function for the isotropic polarization is defined as
\begin{equation}\label{C4}
f^{\rm{ISO}}(y)=f^{X}(y)+2f^{Y}(y)=\frac{3}{2\gamma v},
\end{equation}
which is normalized to $ 3 $, reflecting the contributions from the three different diagonal polarizations,
\begin{equation}\label{C5}
\int_{y_{-}}^{y_{+}} dy \, f^{\rm{ISO}}(y)=3.
\end{equation}
Other integrals of $ f^{\rm{ISO}} $ used in the formulas for quantum vacuum frictional power and force are 
\begin{equation}\label{C6}
\int_{y_{-}}^{y_{+}} dy\, y f^{\rm{ISO}}(y)=3\gamma, \qquad 
\int_{y_{-}}^{y_{+}} dy\, \frac{1}{v}\left(y-\frac{1}{\gamma}\right) f^{\rm{ISO}}(y)=3\gamma v.
\end{equation}

\section{THE QUANTUM FRICTIONAL FORCE IN A GENERAL BACKGROUND}\label{apD}
The following formulas for the quantum frictional force are written in terms of components of the general Green's dyadic. Therefore, they can be applied to situations where the neutral particle is passing through a more complicated background, for example, a surface that is translational invariant in the $ x $-$ y $ plane.
\begin{align}\label{Y1}
F_{\rm{I}}^{Y}=\gamma &\int\frac{d\tilde{\omega}}{2\pi}\int\frac{d^{2}k_{\perp}}{(2\pi)^{2}}\int\frac{d^{2}\bar{k}_{\perp}}{(2\pi)^{2}} \,\alpha_{yy}^{2}(\tilde{\omega})\,\bar{k}_{x} \,\coth\left[\frac{\beta}{2}\left(\frac{\tilde{\omega}}{\gamma}+\bar{k}_{x}v\right)\right]\nonumber\\
\cross&\frac{1}{(\frac{\tilde{\omega}}{\gamma}+k_{x}v)^{2}}\left\lbrace k_{y}^{2}v^{2}\Im g_{xx}+\frac{\tilde{\omega}^{2}}{\gamma^{2}}\Im g_{yy}+\frac{\tilde{\omega}}{\gamma}k_{y}v\Im g_{xy}+\frac{\tilde{\omega}}{\gamma}k_{y}v\Im g_{yx}\right\rbrace\left(\frac{\tilde{\omega}}{\gamma}+k_{x}v,\vb{k}_{\perp};z,\tilde{z}\right) \nonumber\\
\cross &\frac{1}{(\frac{\tilde{\omega}}{\gamma}+\bar{k}_{x}v)^{2}}\left\lbrace \bar{k}_{y}^{2}v^{2}\Im g_{xx}+\frac{\tilde{\omega}^{2}}{\gamma^{2}}\Im g_{yy}+\frac{\tilde{\omega}}{\gamma}\bar{k}_{y}v\Im g_{xy}+\frac{\tilde{\omega}}{\gamma}\bar{k}_{y}v\Im g_{yx}\right\rbrace\left(\frac{\tilde{\omega}}{\gamma}+\bar{k}_{x}v,\bar{\vb{k}}_{\perp};z,\tilde{z}\right)
.
\end{align}

\begin{align}\label{Y2}
F_{\rm{II}}^{Y}=-\gamma & \int\frac{d\tilde{\omega}}{2\pi}\int\frac{d^{2}k_{\perp}}{(2\pi)^{2}}\int\frac{d^{2}\bar{k}_{\perp}}{(2\pi)^{2}} \,\alpha_{yy}^{2}(\tilde{\omega})\,k_{x} \,\coth\left[\frac{\beta}{2}\left(\frac{\tilde{\omega}}{\gamma}+\bar{k}_{x}v\right)\right]\nonumber\\
\cross&\frac{1}{(\frac{\tilde{\omega}}{\gamma}+k_{x}v)^{2}}\left\lbrace k_{y}^{2}v^{2}\Im g_{xx}+\frac{\tilde{\omega}^{2}}{\gamma^{2}}\Im g_{yy}+\frac{\tilde{\omega}}{\gamma}k_{y}v\Im g_{xy}+\frac{\tilde{\omega}}{\gamma}k_{y}v\Im g_{yx}\right\rbrace\left(\frac{\tilde{\omega}}{\gamma}+k_{x}v,\vb{k}_{\perp};z,\tilde{z}\right) \nonumber\\
\cross &\frac{1}{(\frac{\tilde{\omega}}{\gamma}+\bar{k}_{x}v)^{2}}\left\lbrace \bar{k}_{y}^{2}v^{2}\Im g_{xx}+\frac{\tilde{\omega}^{2}}{\gamma^{2}}\Im g_{yy}+\frac{\tilde{\omega}}{\gamma}\bar{k}_{y}v\Im g_{xy}+\frac{\tilde{\omega}}{\gamma}\bar{k}_{y}v\Im g_{yx}\right\rbrace\left(\frac{\tilde{\omega}}{\gamma}+\bar{k}_{x}v,\bar{\vb{k}}_{\perp};z,\tilde{z}\right)
.
\end{align}

\begin{align}\label{Z1}
F_{\rm{I}}^{Z}=\gamma &\int\frac{d\tilde{\omega}}{(2\pi)^{2}}\int\frac{d^{2}k_{\perp}}{(2\pi)^{2}}\int\frac{d^{2}\bar{k}_{\perp}}{(2\pi)^{2}} \,\alpha_{zz}^{2}(\tilde{\omega})\,\bar{k}_{x} \,\coth\left[\frac{\beta}{2}\left(\frac{\tilde{\omega}}{\gamma}+\bar{k}_{x}v\right)\right]\nonumber\\
\cross&\frac{1}{(\frac{\tilde{\omega}}{\gamma}+k_{x}v)^{2}}\left\lbrace v^{2}\partial_{z}\partial_{\tilde{z}}\Im g_{xx}+\frac{\tilde{\omega}^{2}}{\gamma^{2}}\Im g_{zz}-i\frac{\tilde{\omega}}{\gamma}v\partial_{z}\Im g_{xz}+i\frac{\tilde{\omega}}{\gamma}v\partial_{\tilde{z}}\Im g_{zx}\right\rbrace\left(\frac{\tilde{\omega}}{\gamma}+k_{x}v,\vb{k}_{\perp};z,\tilde{z}\right) \nonumber\\
\cross &\frac{1}{(\frac{\tilde{\omega}}{\gamma}+\bar{k}_{x}v)^{2}}\left\lbrace v^{2}\partial_{z}\partial_{\tilde{z}}\Im g_{xx}+\frac{\tilde{\omega}^{2}}{\gamma^{2}}\Im g_{zz}-i\frac{\tilde{\omega}}{\gamma}v\partial_{z}\Im g_{xz}+i\frac{\tilde{\omega}}{\gamma}v\partial_{\tilde{z}}\Im g_{zx}\right\rbrace\left(\frac{\tilde{\omega}}{\gamma}+\bar{k}_{x}v,\bar{\vb{k}}_{\perp};z,\tilde{z}\right)
.
\end{align}

\begin{align}\label{Z2}
F_{\rm{II}}^{Z}=-\gamma &\int\frac{d\tilde{\omega}}{(2\pi)^{2}}\int\frac{d^{2}k_{\perp}}{(2\pi)^{2}}\int\frac{d^{2}\bar{k}_{\perp}}{(2\pi)^{2}} \,\alpha_{zz}^{2}(\tilde{\omega})\,k_{x} \,\coth\left[\frac{\beta}{2}\left(\frac{\tilde{\omega}}{\gamma}+\bar{k}_{x}v\right)\right]\nonumber\\
\cross&\frac{1}{(\frac{\tilde{\omega}}{\gamma}+k_{x}v)^{2}}\left\lbrace v^{2}\partial_{z}\partial_{\tilde{z}}\Im g_{xx}+\frac{\tilde{\omega}^{2}}{\gamma^{2}}\Im g_{zz}-i\frac{\tilde{\omega}}{\gamma}v\partial_{z}\Im g_{xz}+i\frac{\tilde{\omega}}{\gamma}v\partial_{\tilde{z}}\Im g_{zx}\right\rbrace\left(\frac{\tilde{\omega}}{\gamma}+k_{x}v,\vb{k}_{\perp};z,\tilde{z}\right) \nonumber\\
\cross &\frac{1}{(\frac{\tilde{\omega}}{\gamma}+\bar{k}_{x}v)^{2}}\left\lbrace v^{2}\partial_{z}\partial_{\tilde{z}}\Im g_{xx}+\frac{\tilde{\omega}^{2}}{\gamma^{2}}\Im g_{zz}-i\frac{\tilde{\omega}}{\gamma}v\partial_{z}\Im g_{xz}+i\frac{\tilde{\omega}}{\gamma}v\partial_{\tilde{z}}\Im g_{zx}\right\rbrace\left(\frac{\tilde{\omega}}{\gamma}+\bar{k}_{x}v,\bar{\vb{k}}_{\perp};z,\tilde{z}\right)
.
\end{align}

We notice the symmetry between the $ \rm{I} $ and $ \rm{II} $ contributions for the quantum vacuum friction formula in the main text are still at work for a general background: $ F_{\rm{II}} $ can be obtained directly by changing the overall sign and trading the momentum factor $ \bar{k}_{x} $ for $ k_{x} $ in $  F_{\rm{I}} $. In addition, there exists an obvious reflection symmetry between the formulas for $ F^{Y} $ and $ F^{Z} $. To be precise, $ F^{Z} $ can be obtained by the following replacement rules: $ k_{y}^{2}g_{xx} \to \partial_{z}\partial_{\bar{z}}g_{xx} $, $ g_{yy}\to g_{zz} $, $ k_{y}g_{xy} \to -i\partial_{z}g_{xz} $, $ k_{y}g_{yx} \to i\partial_{\bar{z}}g_{zx} $. At a first sight, the distinct replacements for the two off-diagonal contributions seem less aesthetic. But when the explicit Green's functions in Eq.~\eqref{eqA-4} is considered, we see that the replacement rules render the two off-diagonal terms in $ F^{Z} $ contribute the same to the frictional force.

The quantum friction in a general background from the $ x $ polarization has already been written down in Eq.~\eqref{eq3-12} and Eq.~\eqref{eq3-14}. But these formulas could indeed be recast into more complicated forms to make the symmetry between polarizations more obvious ($ k_{y}^{2}g_{xx} \to k_{x}^{2}g_{xx} $, $ g_{yy}\to g_{xx} $, $ k_{y}g_{xy} \to k_{x}g_{xx} $, $ k_{y}g_{yx} \to k_{x}g_{xx} $):
\begin{align}\label{X1}
F_{\rm{I}}^{X}=\frac{1}{\gamma^{3}} &\int\frac{d\tilde{\omega}}{(2\pi)^{2}}\int\frac{d^{2}k_{\perp}}{(2\pi)^{2}}\int\frac{d^{2}\bar{k}_{\perp}}{(2\pi)^{2}} \,\alpha_{xx}^{2}(\tilde{\omega})\,\bar{k}_{x} \,\coth\left[\frac{\beta}{2}\left(\frac{\tilde{\omega}}{\gamma}+\bar{k}_{x}v\right)\right]\nonumber\\
\cross&\frac{1}{(\frac{\tilde{\omega}}{\gamma}+k_{x}v)^{2}}\left\lbrace k_{x}^{2}v^{2}\Im g_{xx}+\frac{\tilde{\omega}^{2}}{\gamma^{2}}\Im g_{xx}+\frac{\tilde{\omega}}{\gamma}k_{x}v\Im g_{xx}+\frac{\tilde{\omega}}{\gamma}k_{x}v\Im g_{xx}\right\rbrace\left(\frac{\tilde{\omega}}{\gamma}+k_{x}v,\vb{k}_{\perp};z,\tilde{z}\right) \nonumber\\
\cross &\frac{1}{(\frac{\tilde{\omega}}{\gamma}+\bar{k}_{x}v)^{2}}\left\lbrace \bar{k}_{x}^{2}v^{2}\Im g_{xx}+\frac{\tilde{\omega}^{2}}{\gamma^{2}}\Im g_{xx}+\frac{\tilde{\omega}}{\gamma}\bar{k}_{x}v\Im g_{xx}+\frac{\tilde{\omega}}{\gamma}\bar{k}_{x}v\Im g_{xx}\right\rbrace\left(\frac{\tilde{\omega}}{\gamma}+\bar{k}_{x}v,\bar{\vb{k}}_{\perp};z,\tilde{z}\right)
.
\end{align}

\begin{align}\label{X2}
F_{\rm{II}}^{X}=-\frac{1}{\gamma^{3}}& \int\frac{d\tilde{\omega}}{(2\pi)^{2}}\int\frac{d^{2}k_{\perp}}{(2\pi)^{2}}\int\frac{d^{2}\bar{k}_{\perp}}{(2\pi)^{2}} \,\alpha_{xx}^{2}(\tilde{\omega})\,k_{x} \,\coth\left[\frac{\beta}{2}\left(\frac{\tilde{\omega}}{\gamma}+\bar{k}_{x}v\right)\right]\nonumber\\
\cross&\frac{1}{(\frac{\tilde{\omega}}{\gamma}+k_{x}v)^{2}}\left\lbrace k_{x}^{2}v^{2}\Im g_{xx}+\frac{\tilde{\omega}^{2}}{\gamma^{2}}\Im g_{xx}+\frac{\tilde{\omega}}{\gamma}k_{x}v\Im g_{xx}+\frac{\tilde{\omega}}{\gamma}k_{x}v\Im g_{xx}\right\rbrace\left(\frac{\tilde{\omega}}{\gamma}+k_{x}v,\vb{k}_{\perp};z,\tilde{z}\right) \nonumber\\
\cross &\frac{1}{(\frac{\tilde{\omega}}{\gamma}+\bar{k}_{x}v)^{2}}\left\lbrace \bar{k}_{x}^{2}v^{2}\Im g_{xx}+\frac{\tilde{\omega}^{2}}{\gamma^{2}}\Im g_{xx}+\frac{\tilde{\omega}}{\gamma}\bar{k}_{x}v\Im g_{xx}+\frac{\tilde{\omega}}{\gamma}\bar{k}_{x}v\Im g_{xx}\right\rbrace\left(\frac{\tilde{\omega}}{\gamma}+\bar{k}_{x}v,\bar{\vb{k}}_{\perp};z,\tilde{z}\right)
.
\end{align}
The overall factor has different powers of $ \gamma $ for $ F^{X} $ than $ F^{Y} $ and $ F^{Z} $. This can be understood as a result of the different $ \gamma $ factors in the Lorentz transformation of dipole and field, i.e., Eq.~\eqref{eqB-1} and Eq.~\eqref{eqB-3}, for parallel polarization and perpendicular polarizations. 

\section{THE PRINCIPLE OF VIRTUAL WORK}\label{apE}
Here, we provide a proof of the principle of virtual work, applicable to the current context. A similar proof, applicable to a different context, may be found in Ref.~\cite{LY:ihm}.

Let $\mathcal F'(\mathbf{r}_0', \mathbf{r}_1')$ denote the point-separated particle-field interaction free energy in frame $ \mathcal{P} $, where we have identified and separated the dipole point, $\mathbf{r}_0'=(x_0', 0, 0)$, and the field point, $\mathbf{r}_1'=(x_1', 0, 0)$. Here, for simplicity of exposition, we ignore the temporal coordinates of these two points, which are set equal. Under the combined coordinate scaling
\begin{equation}\tag{E1a}
x' \to x'^{\lambda}=x'_0+\frac{1}{\lambda} (x'-x_0'),
\end{equation}
where $\lambda >0$, and dual metric scaling
\begin{equation}\tag{E1b}
g_{x'x'}\to g_{x'x'}^{\lambda}=\lambda^2 g_{x'x'},
\end{equation}
localized to a neighborhood of the line segment between the two points, $\{x'_0\le x'\le x'_1, y'=0, z'=0\}$, $\mathcal F'$ is invariant, that is, in an obvious notation, $\mathcal F'^{\lambda}=\mathcal F'$. Thus, to first order in $\delta \lambda$,
\begin{equation}\tag{E2}
\delta \mathcal F'^{\lambda}=\frac{\partial \mathcal F'^{\lambda}}{\partial x'^{\lambda}_1} \,\delta x'^{\lambda}_1 + \int d\mathbf{r}'^{\lambda} \frac{\delta \mathcal F'^{\lambda}}{\delta g_{x'x'}^{\lambda}(\mathbf{r}'^{\lambda})}\,\delta g_{x'x'}^{\lambda}(\mathbf{r}'^{\lambda})=0.
\end{equation}
Setting $\lambda=1$, Eq. \thetag{E2} becomes
\begin{equation}\tag{E3}
-\frac{\partial \mathcal F'}{\partial x'_1} (x'_1-x'_0)\, \delta \lambda+\int d\mathbf{r}'\frac{\delta \mathcal F'}{\delta g_{x'x'}(\mathbf{r}')}\,2 \,g_{x'x'}(\mathbf{r}')\,\delta \lambda=0,
\end{equation}
that is, 
\begin{equation}\tag{E4}
-\frac{\partial \mathcal F'}{\partial x'_1}=\frac{1}{\lvert x'_1-x'_0\rvert}\int_{x'_0}^{x'_1} dx'\int dy'dz' \,\hat T^{x'}_{\phantom{x'}x'}(x',y', z')=\frac{1}{\lvert x'_1-x'_0\rvert}\int_{x'_0}^{x'_1} dx' \,\hat t^{x'}_{\phantom{x'}x'}(x',0,0),
\end{equation}
where we have used the fact that the free energy is the negative of the effective Lagrangian of the particle-field interaction, and 
\begin{equation}\tag{E5}
\hat t^{x'}_{\phantom{x'}x'}(x', 0, 0)\equiv\int dy'dz' \,\hat T^{x'}_{\phantom{x'}x'}(x',y', z')
\end{equation}
is the transverse integral of the $x'x'$ component of the corresponding stress tensor, which is localized to the line segment between the dipole point and the field point.  
In the limit as the field point approaches the dipole point, Eq. \thetag{E4} yields
\begin{equation}\tag{E6}
\left.-\frac{\partial}{\partial x'_1}\mathcal F' (\mathbf{r}'_0, \mathbf{r}'_1)\right\rvert_{\mathbf{r}'_1\to\mathbf{r}'_0\pm}=\pm \,\hat t^{x'}_{\phantom{x'}x'}(\mathbf{r}'_0\pm).
\end{equation}
However, since changing the argument of differentiation in the left-hand side of Eq. \thetag{E6} simply changes the sign, the limit in the left-hand side must be independent of the sense of approach, that is, 
\begin{equation}\tag{E7}
\left.\frac{\partial}{\partial x'_0}\mathcal F' (\mathbf{r}'_0, \mathbf{r}'_1)\right\rvert_{\mathbf{r}'_1\to\mathbf{r}'_0}=
\left.\frac{\partial}{\partial x'_0}\mathcal F' (\mathbf{r}'_0, \mathbf{r}'_1)\right\rvert_{\mathbf{r}'_1\to\mathbf{r}'_0\pm}=
\left.-\frac{\partial}{\partial x'_1}\mathcal F' (\mathbf{r}'_0, \mathbf{r}'_1)\right\rvert_{\mathbf{r}'_1\to\mathbf{r}'_0\pm}=
\pm \,\hat t^{x'}_{\phantom{x'}x'}(\mathbf{r}'_0\pm).
\end{equation}
From Eq. \thetag{E7}, we deduce that 
\begin{equation}\tag{E8}
\left.\frac{\partial}{\partial x'_0}\mathcal F' (\mathbf{r}'_0, \mathbf{r}'_1)\right\rvert_{\mathbf{r}'_1\to\mathbf{r}'_0}=
\left.-\frac{\partial}{\partial x'_1}\mathcal F' (\mathbf{r}'_0, \mathbf{r}'_1)\right\rvert_{\mathbf{r}'_1\to\mathbf{r}'_0}=
\hat t^{x'}_{\phantom{x'}x'}(\mathbf{r}'_0+)=-\hat t^{x'}_{\phantom{x'}x'}(\mathbf{r}'_0-),
\end{equation}
that is, the transverse integral of the $x'x'$ component of the stress tensor changes sign as we pass from one side of the interaction point to the other. The corresponding force acting on either spatial side of the interaction point is therefore the same:
\begin{equation}\tag{E9}
\left.\frac{\partial}{\partial x'_0}\mathcal F' (\mathbf{r}'_0, \mathbf{r}'_1)\right\rvert_{\mathbf{r}'_1\to\mathbf{r}'_0}=
\left.-\frac{\partial}{\partial x'_1}\mathcal F' (\mathbf{r}'_0, \mathbf{r}'_1)\right\rvert_{\mathbf{r}'_1\to\mathbf{r}'_0}=
\hat F'(\mathbf{r}'_0+)=\hat F'(\mathbf{r}'_0-)=\hat F'(\mathbf{r}'_0),
\end{equation}
where $\hat F'(\mathbf{r}'_0+)\equiv \hat t^{x'}_{\phantom{x'}x'}(\mathbf{r}'_0+)-0$ and $\hat F'(\mathbf{r}'_0-) \equiv 0-\hat t^{x'}_{\phantom{x'}x'}(\mathbf{r}'_0-)$. 
Finally, we must account for the fact that, because of the symmetry of $\mathcal F'$, there are two ways to identify and separate the dipole point and the field point, each of which results in a contribution identical to the above. Thus, including the corresponding multiplicity factor of 2, the principle of virtual work in the current context may be stated as
\begin{equation}\tag{E10}
-\frac{\partial \mathcal F'}{\partial x'}=
\left.-2\frac{\partial}{\partial x'_1}\mathcal F' (\mathbf{r}'_0, \mathbf{r}'_1)\right\rvert_{\mathbf{r}'_1\to\mathbf{r}'_0}=
F'(\mathbf{r}'_0)=F',
\end{equation}
where $F'=2\hat F'$.

A variant of the above approach may be used to establish the corresponding relationship for the power. In this case, we let $\mathcal F'(t_0', t_1')$ denote the point-separated particle-field interaction free energy in frame $ \mathcal{P} $, where we have identified and separated the dipole point, $t_0'$, and the field point, $t_1'$. Here, for simplicity of exposition, we ignore the spatial coordinates of these two points, which are set equal. We also define
\begin{equation}\tag{E11}
\mathcal G'(t'_0, t'_1)\equiv \mathcal F'(t'_0, t'_1)-\mathcal F'(t'_0,t'_0)=\int_{t'_0}^{t'_1}dt'\,\frac{\partial}{\partial t'}\mathcal F'(t'_0, t'). 
\end{equation}
Under the combined coordinate scaling
\begin{equation}\tag{E12a}
t' \to t'^{\lambda}=t'_0+\frac{1}{\lambda} (t'-t_0'),
\end{equation}
where $\lambda >0$, and dual metric scaling
\begin{equation}\tag{E12b}
g_{t't'}\to g_{t't'}^{\lambda}=\lambda^2 g_{t't'},
\end{equation}
localized to a neighborhood of the line segment between the two points, $\{t'_0\le t'\le t'_1,x'=x'_0, y'=0, z'=0\}$, $\mathcal G'$ is invariant, that is, in an obvious notation, $\mathcal G'^{\lambda}=\mathcal G'$. Thus, to first order in $\delta \lambda$,
\begin{equation}\tag{E13}
\delta \mathcal G'^{\lambda}=\frac{\partial \mathcal F'^{\lambda}}{\partial t'^{\lambda}_1} \,\delta t'^{\lambda}_1 + \int_{t'_0}^{t'^{\lambda}_1}dt'^{\lambda}\frac{\partial}{\partial t'^{\lambda}}\int d\mathbf{r}'\frac{\delta \mathcal F'^{\lambda}}{\delta g_{t't'}^{\lambda}(t'^{\lambda})}\,\delta g_{t't'}^{\lambda}(t'^{\lambda})=0.
\end{equation}
Setting $\lambda=1$, Eq. \thetag{E13} becomes
\begin{equation}\tag{E14}
-\frac{\partial \mathcal F'}{\partial t'_1} (t'_1-t'_0)\, \delta \lambda+\int_{t'_0}^{t'_1}dt'\,\frac{\partial}{\partial t'}\int d\mathbf{r}'\frac{\delta \mathcal F'}{\delta g_{t't'}(t')}\,2 \,g_{t't'}(t')\,\delta \lambda=0,
\end{equation}
that is, 
\begin{equation}\tag{E15}
\frac{\partial \mathcal F'}{\partial t'_1}=-\frac{1}{t'_1-t'_0}\int_{t'_0}^{t'_1} dt'\,\frac{\partial}{\partial t'}\int d\mathbf{r}' \,\hat T^{t'}_{\phantom{t'}t'}(t')=\frac{1}{t'_1-t'_0}\int_{t'_0}^{t'_1} dt' \,\frac{\partial}{\partial t'}\hat E'(t')=\frac{1}{t'_1-t'_0}\int_{t'_0}^{t'_1} dt' \,\hat P'(t'),
\end{equation}
where we have used the fact that the free energy is the negative of the effective Lagrangian of the particle-field interaction, and the energy 
\begin{equation}\tag{E16}
\hat E'(t') \equiv -\int d\mathbf{r}' \,\hat T^{t'}_{\phantom{t'}t'}(t')
\end{equation}
is the spatial integral of the $t't'$ component of the corresponding stress tensor, which is localized to the line segment between the dipole point and the field point.  
In the limit as the field point approaches the dipole point, Eq. \thetag{E15} yields
\begin{equation}\tag{E17}
\left.\frac{\partial}{\partial t'_1}\mathcal F' (t'_0, t'_1)\right\rvert_{t'_1\to t'_0\pm}=\hat P'(t'_0\pm).
\end{equation}
However, since changing the argument of differentiation in the left-hand side of Eq. \thetag{E17} simply changes the sign, the limit in the left-hand side must be independent of the sense of approach, that is, 
\begin{equation}\tag{E18}
\left.-\frac{\partial}{\partial t'_0}\mathcal F' (t'_0, t'_1)\right\rvert_{t'_1\to t'_0}=
\left.-\frac{\partial}{\partial t'_0}\mathcal F' (t'_0, t'_1)\right\rvert_{t'_1\to t'_0\pm}=
\left.\frac{\partial}{\partial t'_1}\mathcal F' (t'_0, t'_1)\right\rvert_{t'_1\to t'_0\pm}=
\hat P'(t'_0\pm).
\end{equation}
From Eq. \thetag{E18}, we deduce that 
\begin{equation}\tag{E19}
\left.-\frac{\partial}{\partial t'_0}\mathcal F' (t'_0, t'_1)\right\rvert_{t'_1\to t'_0}=
\left.\frac{\partial}{\partial t'_1}\mathcal F' (t'_0, t'_1)\right\rvert_{t'_1\to t'_0}=
\hat P'(t'_0+)=\hat P'(t'_0-)=\hat P'(t'_0),
\end{equation}
that is, the power on either temporal side of the interaction point is the same. 
Finally, we must account for the fact that, because of the symmetry of $\mathcal F'$, there are two ways to identify and separate the dipole point and the field point, each of which results in a contribution identical to the above. Thus, including the corresponding multiplicity factor of 2, the relationship for the power in the current context may be stated as
\begin{equation}\tag{E20}
\frac{\partial \mathcal F'}{\partial t'}=
\left.2\frac{\partial}{\partial t'_1}\mathcal F' (t'_0, t'_1)\right\rvert_{t'_1\to t'_0}=
P'(t'_0)=P',
\end{equation}
where $P'=2\hat P'$.

\section{SIGN OF THE IMAGINARY PART OF THE EFFECTIVE POLARIZABILITY}\label{apF}
In this appendix, we prove that the diagonal elements of the imaginary part of the effective, or dressed, polarizability, $\operatorname{Im} \hat{\boldsymbol{\alpha}}(\omega)$, are non-negative. 

Since the (renormalized) intrinsic polarizability, $\boldsymbol{\alpha}\equiv\boldsymbol{\alpha}(\omega)$, is a real symmetric matrix, it may be diagonalized by the  transformation
\begin{equation}\tag{F1}
\mathbf{U}^{-1} \boldsymbol{\alpha} \,\mathbf{U}=\boldsymbol{\lambda}, 
\end{equation}
where the column vectors, $\mathbf{u}_1$, $\mathbf{u}_2$, and $\mathbf{u}_3$, of the orthogonal matrix $\mathbf{U}=(\mathbf{u}_1, \mathbf{u}_2, \mathbf{u}_3)$ are orthonormal eigenvectors of $\boldsymbol{\alpha}$, and the diagonal elements of $\boldsymbol{\lambda}=\operatorname{diag}(\lambda_1, \lambda_2, \lambda_3)$ are the corresponding eigenvalues. Then
\begin{equation}\tag{F2}
\boldsymbol{\alpha}=\mathbf{U}\,\boldsymbol{\lambda}\,\mathbf{U}^{-1}
\end{equation}
and
\begin{equation}\tag{F3}
\boldsymbol{\alpha}^2=\mathbf{U}\,\boldsymbol{\lambda}\,\mathbf{U}^{-1}\mathbf{U}\,\boldsymbol{\lambda}\,\mathbf{U}^{-1}=\mathbf{U}\,\boldsymbol{\lambda}^2\,\mathbf{U}^{-1}.
\end{equation}
We may therefore write
\begin{equation}\tag{F4}
\left[\mathbf{1}+\left(\frac{\omega^3}{6\pi}\right)^{\!2} \boldsymbol{\alpha}^2\right]\left[\mathbf{1}+\left(\frac{\omega^3}{6\pi}\right)^{\!2} \boldsymbol{\alpha}^2\right]^{-1}=\mathbf{1}
\end{equation}
as
\begin{equation}\tag{F5}
\mathbf{U}\left[\mathbf{1}+\left(\frac{\omega^3}{6\pi}\right)^{\!2} \boldsymbol{\lambda}^2\right]\mathbf{U}^{-1}\left[\mathbf{1}+\left(\frac{\omega^3}{6\pi}\right)^{\!2} \boldsymbol{\alpha}^2\right]^{-1}=\mathbf{1},
\end{equation}
whence, 
\begin{equation}\tag{F6}
\left[\mathbf{1}+\left(\frac{\omega^3}{6\pi}\right)^{\!2} \boldsymbol{\alpha}^2\right]^{-1}=\mathbf{U}\left[\mathbf{1}+\left(\frac{\omega^3}{6\pi}\right)^{\!2} \boldsymbol{\lambda}^2\right]^{-1}\mathbf{U}^{-1}.
\end{equation}

From Eq. \thetag{F3} and Eq. \thetag{F6}, it follows that
\begin{equation}\tag{F7}
\operatorname{Im} \hat{\boldsymbol{\alpha}}=\frac{\omega^3}{6\pi} \,\boldsymbol{\alpha}^2 \left[\mathbf{1}+\left(\frac{\omega^3}{6\pi}\right)^{\!2} \boldsymbol{\alpha}^2\right]^{-1}=\frac{\omega^3}{6\pi}\, \mathbf{U}\,\boldsymbol{\lambda}^2\left[\mathbf{1}+\left(\frac{\omega^3}{6\pi}\right)^{\!2}\boldsymbol{\lambda}^2\right]^{-1}\mathbf{U}^{-1},
\end{equation}
that is, 
\begin{equation}\tag{F8}
\operatorname{Im}\hat{\boldsymbol{\alpha}}=\frac{\omega^3}{6\pi}\, \mathbf{U}\operatorname{diag}\left(\frac{\lambda_1^2}{1+\left(\frac{\omega^3}{6\pi}\right)^{\!2}\lambda_1^2},\frac{\lambda_2^2}{1+\left(\frac{\omega^3}{6\pi}\right)^{\!2}\lambda_2^2}, \frac{\lambda_3^2}{1+\left(\frac{\omega^3}{6\pi}\right)^{\!2}\lambda_3^2}\right) \mathbf{U}^{-1}.
\end{equation}
Since $\mathbf{U}$ is orthogonal, $\mathbf{U}^{-1}=\mathbf{U}^T$, so Eq. \thetag{F8} may be written as
\begin{equation}\tag{F9}
\operatorname{Im}\hat{\boldsymbol{\alpha}}= \frac{\omega^3}{6\pi} \,\sum_{k=1}^{3} \frac{\lambda_k^2}{1+\left(\frac{\omega^3}{6\pi}\right)^{\!2}\lambda_k^2}\,\mathbf{u}_k\,\mathbf{u}^T_k.
\end{equation}

The elements of $\operatorname{Im}\hat{\bm{\alpha}}$ are therefore given by 
\begin{equation}\tag{F10}
\operatorname{Im} \hat{\alpha}_{ij}= \frac{\omega^3}{6\pi}\,\sum_{k=1}^3\frac{\lambda_k^2}{1+\left(\frac{\omega^3}{6\pi}\right)^{\!2}\lambda_k^2}\,U_{ik}\,U_{jk},
\end{equation}
where $i, j \in \{x, y, z\}$. In particular, the diagonal elements of $\operatorname{Im}\hat{\boldsymbol{\alpha}}$ are manifestly non-negative:
\begin{equation}\tag{F11}
\operatorname{Im} \hat{\alpha}_{ii}= \frac{\omega^3}{6\pi}\,\sum_{k=1}^3\frac{\lambda_k^2}{1+\left(\frac{\omega^3}{6\pi}\right)^{\!2}\lambda_k^2}\,U_{ik}^2\ge 0.
\end{equation}

\section{INDUCED DIPOLE RADIATION}\label{apG}
In Sec.~IV we provided a physical intepretation of the power emitted
in frame $ \mathcal{P} $, -$P'_{\rm II}$, and the power absorbed, 
$P'_{\rm I}$.
But is this interpretation correct?  Indeed it is, as a simple calculation
based on classical dipole radiation, supplemented by the 
fluctuation-dissipation theorem, shows.

We start from the formula for the energy emitted per unit frequency interval
by dipole radiation ($\omega>0$) (see Ref.~\cite{Schwinger:ce}, Eq.~(35.36), except
we are now using rationalized Heaviside-Lorentz units)
\be
\frac{dE'_{\rm rad}}{d\omega}=\frac1{6\pi^2}\omega^4|\mathbf{d}'(\omega)|^2,
\ee
in frame $ \mathcal{P} $.
Here, we envisage that there is no intrinsic dipole moment, but rather,
the dipole moment is induced by the fluctuating electromagnetic field,
\be
\mathbf{d}'(\omega)=\bm{\alpha}(\omega)\cdot \mathbf{E}'(\omega),
\ee
where the product of $\mathbf{E}'$ fields is given by the fluctuation-%
dissipation theorem (4.7--8).  The imaginary parts of the Green's functions
there are given by Eqs.~(A12).  Then, with the interpretation that the delta
function at coincident frequencies is interpreted as $\mathcal{T}'=2\pi \delta(0)$, where
$\mathcal{T}'$ is the time the configuration exists, we have
\be
P'_{\rm rad}=\frac{E'_{\rm rad}}{\mathcal{T'}}=\frac1{6\pi^2}\int_{0}^\infty d\omega \,\omega^{4}
\int_{-\omega}^{\omega} \frac{dk_x}{2\pi}\left\{
(\bm{\alpha}^2)_{xx}\frac14(\omega^2-k_x^2)+\left[(\bm{\alpha}^2)_{yy}
+(\bm{\alpha}^2)_{zz}\right]\frac18(\omega^2+k_x^2)\right\}\coth
\frac{\beta\gamma}2 (\omega+k_xv).
\ee
When the same variable change is made as in the main text $\omega y=\gamma(\omega+k_{x}v)$, we 
recover exactly the formula (4.12), thereby proving $P'_{\rm rad}=
-P'_{\rm II}=P'_{\rm I}$.
\bibliography{eqfcite}

\begin{thebibliography}{33}%
\makeatletter
\providecommand \@ifxundefined [1]{%
 \@ifx{#1\undefined}
}%
\providecommand \@ifnum [1]{%
 \ifnum #1\expandafter \@firstoftwo
 \else \expandafter \@secondoftwo
 \fi
}%
\providecommand \@ifx [1]{%
 \ifx #1\expandafter \@firstoftwo
 \else \expandafter \@secondoftwo
 \fi
}%
\providecommand \natexlab [1]{#1}%
\providecommand \enquote  [1]{``#1''}%
\providecommand \bibnamefont  [1]{#1}%
\providecommand \bibfnamefont [1]{#1}%
\providecommand \citenamefont [1]{#1}%
\providecommand \href@noop [0]{\@secondoftwo}%
\providecommand \href [0]{\begingroup \@sanitize@url \@href}%
\providecommand \@href[1]{\@@startlink{#1}\@@href}%
\providecommand \@@href[1]{\endgroup#1\@@endlink}%
\providecommand \@sanitize@url [0]{\catcode `\\12\catcode `\$12\catcode
  `\&12\catcode `\#12\catcode `\^12\catcode `\_12\catcode `\%12\relax}%
\providecommand \@@startlink[1]{}%
\providecommand \@@endlink[0]{}%
\providecommand \url  [0]{\begingroup\@sanitize@url \@url }%
\providecommand \@url [1]{\endgroup\@href {#1}{\urlprefix }}%
\providecommand \urlprefix  [0]{URL }%
\providecommand \Eprint [0]{\href }%
\providecommand \doibase [0]{https://doi.org/}%
\providecommand \selectlanguage [0]{\@gobble}%
\providecommand \bibinfo  [0]{\@secondoftwo}%
\providecommand \bibfield  [0]{\@secondoftwo}%
\providecommand \translation [1]{[#1]}%
\providecommand \BibitemOpen [0]{}%
\providecommand \bibitemStop [0]{}%
\providecommand \bibitemNoStop [0]{.\EOS\space}%
\providecommand \EOS [0]{\spacefactor3000\relax}%
\providecommand \BibitemShut  [1]{\csname bibitem#1\endcsname}%
\let\auto@bib@innerbib\@empty
\bibitem [{\citenamefont {Milton}\ \emph
  {et~al.}(2020{\natexlab{a}})\citenamefont {Milton}, \citenamefont {Li},
  \citenamefont {Guo},\ and\ \citenamefont {Kennedy}}]{Kim:charged}%
  \BibitemOpen
  \bibfield  {author} {\bibinfo {author} {\bibfnamefont {K.~A.}\ \bibnamefont
  {Milton}}, \bibinfo {author} {\bibfnamefont {Y.}~\bibnamefont {Li}}, \bibinfo
  {author} {\bibfnamefont {X.}~\bibnamefont {Guo}},\ and\ \bibinfo {author}
  {\bibfnamefont {G.}~\bibnamefont {Kennedy}},\ }\bibfield  {title} {\bibinfo
  {title} {Electrodynamic friction of a charged particle passing a conducting
  plate},\ }\href@noop {} {\bibfield  {journal} {\bibinfo  {journal} {Phys.
  Rev. Research}\ }\textbf {\bibinfo {volume} {2}},\ \bibinfo {pages} {023114}
  (\bibinfo {year} {2020}{\natexlab{a}})}\BibitemShut {NoStop}%
\bibitem [{\citenamefont {Milton}\ \emph
  {et~al.}(2020{\natexlab{b}})\citenamefont {Milton}, \citenamefont {Day},
  \citenamefont {Li}, \citenamefont {Guo},\ and\ \citenamefont
  {Kennedy}}]{Kim:dipole}%
  \BibitemOpen
  \bibfield  {author} {\bibinfo {author} {\bibfnamefont {K.~A.}\ \bibnamefont
  {Milton}}, \bibinfo {author} {\bibfnamefont {H.}~\bibnamefont {Day}},
  \bibinfo {author} {\bibfnamefont {Y.}~\bibnamefont {Li}}, \bibinfo {author}
  {\bibfnamefont {X.}~\bibnamefont {Guo}},\ and\ \bibinfo {author}
  {\bibfnamefont {G.}~\bibnamefont {Kennedy}},\ }\bibfield  {title} {\bibinfo
  {title} {Self-force on moving electric and magnetic dipoles: Dipole
  radiation, {V}avilov-\v{C}erenkov radiation, friction with a conducting
  surface, and the {E}instein-{H}opf effect},\ }\href@noop {} {\bibfield
  {journal} {\bibinfo  {journal} {Phys. Rev. Research}\ }\textbf {\bibinfo
  {volume} {2}},\ \bibinfo {pages} {043347} (\bibinfo {year}
  {2020}{\natexlab{b}})}\BibitemShut {NoStop}%
\bibitem [{\citenamefont {Pendry}(1997)}]{Pendry:1997}%
  \BibitemOpen
  \bibfield  {author} {\bibinfo {author} {\bibfnamefont {J.~B.}\ \bibnamefont
  {Pendry}},\ }\bibfield  {title} {\bibinfo {title} {Shearing the vacuum --
  quantum friction},\ }\href@noop {} {\bibfield  {journal} {\bibinfo  {journal}
  {J. Phys.: Condens. Matter}\ }\textbf {\bibinfo {volume} {9}},\ \bibinfo
  {pages} {10301} (\bibinfo {year} {1997})}\BibitemShut {NoStop}%
\bibitem [{\citenamefont {Teodorovich}(1978)}]{Teodorovich:1978}%
  \BibitemOpen
  \bibfield  {author} {\bibinfo {author} {\bibfnamefont {E.~V.}\ \bibnamefont
  {Teodorovich}},\ }\bibfield  {title} {\bibinfo {title} {Contribution of
  macroscopic van der {W}aals interactions to frictional force},\ }\href@noop
  {} {\bibfield  {journal} {\bibinfo  {journal} {Proc. R. Soc. Lond. A}\
  }\textbf {\bibinfo {volume} {362}},\ \bibinfo {pages} {71} (\bibinfo {year}
  {1978})}\BibitemShut {NoStop}%
\bibitem [{\citenamefont {Levitov}(1989)}]{Levitov:1989}%
  \BibitemOpen
  \bibfield  {author} {\bibinfo {author} {\bibfnamefont {L.~S.}\ \bibnamefont
  {Levitov}},\ }\bibfield  {title} {\bibinfo {title} {Van der {W}aals
  friction},\ }\href@noop {} {\bibfield  {journal} {\bibinfo  {journal}
  {Europhys. Lett.}\ }\textbf {\bibinfo {volume} {8}},\ \bibinfo {pages} {499}
  (\bibinfo {year} {1989})}\BibitemShut {NoStop}%
\bibitem [{\citenamefont {H\o{}ye}\ and\ \citenamefont
  {Brevik}(1992)}]{Hoye:1992}%
  \BibitemOpen
  \bibfield  {author} {\bibinfo {author} {\bibfnamefont {J.~S.}\ \bibnamefont
  {H\o{}ye}}\ and\ \bibinfo {author} {\bibfnamefont {I.}~\bibnamefont
  {Brevik}},\ }\bibfield  {title} {\bibinfo {title} {Friction force between
  moving harmonic oscillators},\ }\href@noop {} {\bibfield  {journal} {\bibinfo
   {journal} {Physica A}\ }\textbf {\bibinfo {volume} {181}},\ \bibinfo {pages}
  {413} (\bibinfo {year} {1992})}\BibitemShut {NoStop}%
\bibitem [{\citenamefont {H\o{}ye}\ and\ \citenamefont
  {Brevik}(1993)}]{Hoye:1993}%
  \BibitemOpen
  \bibfield  {author} {\bibinfo {author} {\bibfnamefont {J.~S.}\ \bibnamefont
  {H\o{}ye}}\ and\ \bibinfo {author} {\bibfnamefont {I.}~\bibnamefont
  {Brevik}},\ }\bibfield  {title} {\bibinfo {title} {Friction force with
  non-instantaneous interaction between moving harmonic oscillators},\
  }\href@noop {} {\bibfield  {journal} {\bibinfo  {journal} {Physica A}\
  }\textbf {\bibinfo {volume} {196}},\ \bibinfo {pages} {241} (\bibinfo {year}
  {1993})}\BibitemShut {NoStop}%
\bibitem [{\citenamefont {Milton}\ \emph {et~al.}(2016)\citenamefont {Milton},
  \citenamefont {H\o{}ye},\ and\ \citenamefont {Brevik}}]{Kim:reality}%
  \BibitemOpen
  \bibfield  {author} {\bibinfo {author} {\bibfnamefont {K.~A.}\ \bibnamefont
  {Milton}}, \bibinfo {author} {\bibfnamefont {J.~S.}\ \bibnamefont
  {H\o{}ye}},\ and\ \bibinfo {author} {\bibfnamefont {I.}~\bibnamefont
  {Brevik}},\ }\bibfield  {title} {\bibinfo {title} {The reality of {C}asimir
  friction},\ }\href@noop {} {\bibfield  {journal} {\bibinfo  {journal}
  {Symmetry}\ }\textbf {\bibinfo {volume} {8}},\ \bibinfo {pages} {29}
  (\bibinfo {year} {2016})}\BibitemShut {NoStop}%
\bibitem [{\citenamefont {Mkrtchian}\ \emph {et~al.}(2003)\citenamefont
  {Mkrtchian}, \citenamefont {Parsegian}, \citenamefont {Podgornik},\ and\
  \citenamefont {Saslow}}]{Mkrtchian:universal}%
  \BibitemOpen
  \bibfield  {author} {\bibinfo {author} {\bibfnamefont {V.}~\bibnamefont
  {Mkrtchian}}, \bibinfo {author} {\bibfnamefont {V.~A.}\ \bibnamefont
  {Parsegian}}, \bibinfo {author} {\bibfnamefont {R.}~\bibnamefont
  {Podgornik}},\ and\ \bibinfo {author} {\bibfnamefont {W.~M.}\ \bibnamefont
  {Saslow}},\ }\bibfield  {title} {\bibinfo {title} {Universal thermal
  radiation drag on neutral objects},\ }\href@noop {} {\bibfield  {journal}
  {\bibinfo  {journal} {Phys. Rev. Lett.}\ }\textbf {\bibinfo {volume} {91}},\
  \bibinfo {pages} {220801} (\bibinfo {year} {2003})}\BibitemShut {NoStop}%
\bibitem [{\citenamefont {Volokitin}\ and\ \citenamefont
  {Persson}(2017)}]{Volokitin:book}%
  \BibitemOpen
  \bibfield  {author} {\bibinfo {author} {\bibfnamefont {A.~I.}\ \bibnamefont
  {Volokitin}}\ and\ \bibinfo {author} {\bibfnamefont {B.~N.~J.}\ \bibnamefont
  {Persson}},\ }\href@noop {} {\emph {\bibinfo {title} {Electromagnetic
  Fluctuations at the Nanoscale}}}\ (\bibinfo  {publisher} {Springer},\
  \bibinfo {address} {Berlin},\ \bibinfo {year} {2017})\BibitemShut {NoStop}%
\bibitem [{\citenamefont {Dedkov}\ and\ \citenamefont
  {Kyasov}(2020)}]{Dedkov:friction}%
  \BibitemOpen
  \bibfield  {author} {\bibinfo {author} {\bibfnamefont {G.~V.}\ \bibnamefont
  {Dedkov}}\ and\ \bibinfo {author} {\bibfnamefont {A.~A.}\ \bibnamefont
  {Kyasov}},\ }\bibfield  {title} {\bibinfo {title} {Nonlocal friction forces
  in the particle-plate and plate-plate configurations: Nonretarded
  approximation},\ }\href@noop {} {\bibfield  {journal} {\bibinfo  {journal}
  {Surface Science}\ }\textbf {\bibinfo {volume} {700}},\ \bibinfo {pages}
  {121681} (\bibinfo {year} {2020})}\BibitemShut {NoStop}%
\bibitem [{\citenamefont {Intravaia}\ \emph {et~al.}(2014)\citenamefont
  {Intravaia}, \citenamefont {Behunin},\ and\ \citenamefont
  {Dalvit}}]{Intravaia:2014PRA}%
  \BibitemOpen
  \bibfield  {author} {\bibinfo {author} {\bibfnamefont {F.}~\bibnamefont
  {Intravaia}}, \bibinfo {author} {\bibfnamefont {R.~O.}\ \bibnamefont
  {Behunin}},\ and\ \bibinfo {author} {\bibfnamefont {D.~A.~R.}\ \bibnamefont
  {Dalvit}},\ }\bibfield  {title} {\bibinfo {title} {Quantum friction and
  fluctuation theorems},\ }\href@noop {} {\bibfield  {journal} {\bibinfo
  {journal} {Phys. Rev. A}\ }\textbf {\bibinfo {volume} {89}},\ \bibinfo
  {pages} {050101(R)} (\bibinfo {year} {2014})}\BibitemShut {NoStop}%
\bibitem [{\citenamefont {Intravaia}\ \emph
  {et~al.}(2016{\natexlab{a}})\citenamefont {Intravaia}, \citenamefont
  {Behunin}, \citenamefont {Henkel}, \citenamefont {Busch},\ and\ \citenamefont
  {Dalvit}}]{Intravaia:2016PRL}%
  \BibitemOpen
  \bibfield  {author} {\bibinfo {author} {\bibfnamefont {F.}~\bibnamefont
  {Intravaia}}, \bibinfo {author} {\bibfnamefont {R.~O.}\ \bibnamefont
  {Behunin}}, \bibinfo {author} {\bibfnamefont {C.}~\bibnamefont {Henkel}},
  \bibinfo {author} {\bibfnamefont {K.}~\bibnamefont {Busch}},\ and\ \bibinfo
  {author} {\bibfnamefont {D.~A.~R.}\ \bibnamefont {Dalvit}},\ }\bibfield
  {title} {\bibinfo {title} {Failure of local thermal equilibrium in quantum
  friction},\ }\href@noop {} {\bibfield  {journal} {\bibinfo  {journal} {Phys.
  Rev. Lett.}\ }\textbf {\bibinfo {volume} {117}},\ \bibinfo {pages} {100402}
  (\bibinfo {year} {2016}{\natexlab{a}})}\BibitemShut {NoStop}%
\bibitem [{\citenamefont {Intravaia}\ \emph
  {et~al.}(2016{\natexlab{b}})\citenamefont {Intravaia}, \citenamefont
  {Behunin}, \citenamefont {Henkel}, \citenamefont {Busch},\ and\ \citenamefont
  {Dalvit}}]{Intravaia:2016PRA}%
  \BibitemOpen
  \bibfield  {author} {\bibinfo {author} {\bibfnamefont {F.}~\bibnamefont
  {Intravaia}}, \bibinfo {author} {\bibfnamefont {R.~O.}\ \bibnamefont
  {Behunin}}, \bibinfo {author} {\bibfnamefont {C.}~\bibnamefont {Henkel}},
  \bibinfo {author} {\bibfnamefont {K.}~\bibnamefont {Busch}},\ and\ \bibinfo
  {author} {\bibfnamefont {D.~A.~R.}\ \bibnamefont {Dalvit}},\ }\bibfield
  {title} {\bibinfo {title} {Non-{M}arkovianity in atom-surface dispersion
  forces},\ }\href@noop {} {\bibfield  {journal} {\bibinfo  {journal} {Phys.
  Rev. A}\ }\textbf {\bibinfo {volume} {94}},\ \bibinfo {pages} {042114}
  (\bibinfo {year} {2016}{\natexlab{b}})}\BibitemShut {NoStop}%
\bibitem [{\citenamefont {Reiche}\ \emph {et~al.}(2020)\citenamefont {Reiche},
  \citenamefont {Intravaia}, \citenamefont {Hsiang}, \citenamefont {Busch},\
  and\ \citenamefont {Hu}}]{Intravaia:NTQF}%
  \BibitemOpen
  \bibfield  {author} {\bibinfo {author} {\bibfnamefont {D.}~\bibnamefont
  {Reiche}}, \bibinfo {author} {\bibfnamefont {F.}~\bibnamefont {Intravaia}},
  \bibinfo {author} {\bibfnamefont {J.-T.}\ \bibnamefont {Hsiang}}, \bibinfo
  {author} {\bibfnamefont {K.}~\bibnamefont {Busch}},\ and\ \bibinfo {author}
  {\bibfnamefont {B.-L.}\ \bibnamefont {Hu}},\ }\bibfield  {title} {\bibinfo
  {title} {Nonequilibrium thermodynamics of quantum friction},\ }\href@noop {}
  {\bibfield  {journal} {\bibinfo  {journal} {Phys. Rev. A}\ }\textbf {\bibinfo
  {volume} {102}},\ \bibinfo {pages} {050203(R)} (\bibinfo {year}
  {2020})}\BibitemShut {NoStop}%
\bibitem [{\citenamefont {{A. O. Caldeira and A. J.
  Leggett}}(1983)}]{Caldeira:Quantum}%
  \BibitemOpen
  \bibfield  {author} {\bibinfo {author} {\bibnamefont {{A. O. Caldeira and A.
  J. Leggett}}},\ }\bibfield  {title} {\bibinfo {title} {Quantum tunnelling in
  a dissipative system},\ }\href@noop {} {\bibfield  {journal} {\bibinfo
  {journal} {Ann. Phys.}\ }\textbf {\bibinfo {volume} {149}},\ \bibinfo {pages}
  {374} (\bibinfo {year} {1983})}\BibitemShut {NoStop}%
\bibitem [{\citenamefont {{B. L. Hu, J. P. Paz, and Y. Zhang}}(1992)}]{HuPaz}%
  \BibitemOpen
  \bibfield  {author} {\bibinfo {author} {\bibnamefont {{B. L. Hu, J. P. Paz,
  and Y. Zhang}}},\ }\bibfield  {title} {\bibinfo {title} {Quantum {B}rownian
  motion in a general environment: Exact master equation with nonlocal
  dissipation and colored noise},\ }\href@noop {} {\bibfield  {journal}
  {\bibinfo  {journal} {Phys. Rev. D}\ }\textbf {\bibinfo {volume} {45}},\
  \bibinfo {pages} {2843} (\bibinfo {year} {1992})}\BibitemShut {NoStop}%
\bibitem [{\citenamefont {{B. L. Hu and A. Matacz}}(1994)}]{HuMatacz}%
  \BibitemOpen
  \bibfield  {author} {\bibinfo {author} {\bibnamefont {{B. L. Hu and A.
  Matacz}}},\ }\bibfield  {title} {\bibinfo {title} {Quantum {B}rownian motion
  in a bath of parametric oscillators: A model for system-field interactions},\
  }\href@noop {} {\bibfield  {journal} {\bibinfo  {journal} {Phys. Rev. D}\
  }\textbf {\bibinfo {volume} {49}},\ \bibinfo {pages} {6612} (\bibinfo {year}
  {1994})}\BibitemShut {NoStop}%
\bibitem [{\citenamefont {Polonyi}(2018)}]{Polonyi:Dissipation}%
  \BibitemOpen
  \bibfield  {author} {\bibinfo {author} {\bibfnamefont {J.}~\bibnamefont
  {Polonyi}},\ }\bibfield  {title} {\bibinfo {title} {Instantaneous and
  dynamical decoherence},\ }\href@noop {} {\bibfield  {journal} {\bibinfo
  {journal} {{J. Phys. A: Math. Theor.}}\ }\textbf {\bibinfo {volume} {51}},\
  \bibinfo {pages} {145302} (\bibinfo {year} {2018})}\BibitemShut {NoStop}%
\bibitem [{\citenamefont {Calzetta}\ and\ \citenamefont {{B.-L.
  Hu}}(2008)}]{Hu:Nonequilibrium}%
  \BibitemOpen
  \bibfield  {author} {\bibinfo {author} {\bibfnamefont {E.~A.}\ \bibnamefont
  {Calzetta}}\ and\ \bibinfo {author} {\bibnamefont {{B.-L. Hu}}},\ }\href@noop
  {} {\emph {\bibinfo {title} {Nonequilibrium quantum field theory}}}\
  (\bibinfo  {publisher} {Cambridge Univ. Press},\ \bibinfo {year}
  {2008})\BibitemShut {NoStop}%
\bibitem [{\citenamefont {Einstein}\ and\ \citenamefont
  {Hopf}(1910)}]{Einstein:Hopf}%
  \BibitemOpen
  \bibfield  {author} {\bibinfo {author} {\bibfnamefont {A.}~\bibnamefont
  {Einstein}}\ and\ \bibinfo {author} {\bibfnamefont {L.}~\bibnamefont
  {Hopf}},\ }\bibfield  {title} {\bibinfo {title} {Statistische {U}ntersuchung
  der {B}ewegung eines {R}esonators in einem {S}trahlungsfeld},\ }\href@noop {}
  {\bibfield  {journal} {\bibinfo  {journal} {Ann. Phys.}\ }\textbf {\bibinfo
  {volume} {338}},\ \bibinfo {pages} {1105} (\bibinfo {year}
  {1910})}\BibitemShut {NoStop}%
\bibitem [{\citenamefont {Jentschura}\ and\ \citenamefont
  {Pachucki}(2015)}]{Jentschura:EPJD}%
  \BibitemOpen
  \bibfield  {author} {\bibinfo {author} {\bibfnamefont {U.~D.}\ \bibnamefont
  {Jentschura}}\ and\ \bibinfo {author} {\bibfnamefont {K.}~\bibnamefont
  {Pachucki}},\ }\bibfield  {title} {\bibinfo {title} {Functional form of the
  imaginary part of the atomic polarizability},\ }\href@noop {} {\bibfield
  {journal} {\bibinfo  {journal} {Eur. Phys. J. D}\ }\textbf {\bibinfo {volume}
  {69}},\ \bibinfo {pages} {118} (\bibinfo {year} {2015})}\BibitemShut
  {NoStop}%
\bibitem [{\citenamefont {{S. Albaladejo et
  al.}}(2010)}]{Albaladejo:Radiative}%
  \BibitemOpen
  \bibfield  {author} {\bibinfo {author} {\bibnamefont {{S. Albaladejo et
  al.}}},\ }\bibfield  {title} {\bibinfo {title} {Radiative corrections to the
  polarizability tensor of an electrically small anisotropic dielectric
  particle},\ }\href@noop {} {\bibfield  {journal} {\bibinfo  {journal} {Opt.
  Exp.}\ }\textbf {\bibinfo {volume} {18}},\ \bibinfo {pages} {3556} (\bibinfo
  {year} {2010})}\BibitemShut {NoStop}%
\bibitem [{\citenamefont {{Eric C. Le Ru, W. R. C. Somerville, and B.
  Augui\'e}}(2013)}]{Ru:Radiative}%
  \BibitemOpen
  \bibfield  {author} {\bibinfo {author} {\bibnamefont {{Eric C. Le Ru, W. R.
  C. Somerville, and B. Augui\'e}}},\ }\bibfield  {title} {\bibinfo {title}
  {Radiative correction in approximate treatments of electromagnetic scattering
  by point and body scattering},\ }\href@noop {} {\bibfield  {journal}
  {\bibinfo  {journal} {Phys. Rev. A}\ }\textbf {\bibinfo {volume} {87}},\
  \bibinfo {pages} {012504} (\bibinfo {year} {2013})}\BibitemShut {NoStop}%
\bibitem [{\citenamefont {{A. Wokaun, J. P. Gordon, and P. F.
  Liao}}(1982)}]{Wokaun:Radiation}%
  \BibitemOpen
  \bibfield  {author} {\bibinfo {author} {\bibnamefont {{A. Wokaun, J. P.
  Gordon, and P. F. Liao}}},\ }\bibfield  {title} {\bibinfo {title} {Radiation
  damping in surface-enhanced {R}aman scattering},\ }\href@noop {} {\bibfield
  {journal} {\bibinfo  {journal} {Phys. Rev. Lett.}\ }\textbf {\bibinfo
  {volume} {48}},\ \bibinfo {pages} {957} (\bibinfo {year} {1982})}\BibitemShut
  {NoStop}%
\bibitem [{\citenamefont {Berman}\ \emph {et~al.}(2006)\citenamefont {Berman},
  \citenamefont {Boyd},\ and\ \citenamefont {Milonni}}]{Berman:OptTrm}%
  \BibitemOpen
  \bibfield  {author} {\bibinfo {author} {\bibfnamefont {P.~R.}\ \bibnamefont
  {Berman}}, \bibinfo {author} {\bibfnamefont {R.~W.}\ \bibnamefont {Boyd}},\
  and\ \bibinfo {author} {\bibfnamefont {P.~W.}\ \bibnamefont {Milonni}},\
  }\bibfield  {title} {\bibinfo {title} {Polarizability and the optical theorem
  for a two-level atom with radiative broadening},\ }\href@noop {} {\bibfield
  {journal} {\bibinfo  {journal} {Phys. Rev. A}\ }\textbf {\bibinfo {volume}
  {74}},\ \bibinfo {pages} {053816} (\bibinfo {year} {2006})}\BibitemShut
  {NoStop}%
\bibitem [{\citenamefont {\L{}ach}\ \emph {et~al.}(2004)\citenamefont
  {\L{}ach}, \citenamefont {Jeziorski},\ and\ \citenamefont
  {Szalewicz}}]{Lach:radiative}%
  \BibitemOpen
  \bibfield  {author} {\bibinfo {author} {\bibfnamefont {G.}~\bibnamefont
  {\L{}ach}}, \bibinfo {author} {\bibfnamefont {B.}~\bibnamefont {Jeziorski}},\
  and\ \bibinfo {author} {\bibfnamefont {K.}~\bibnamefont {Szalewicz}},\
  }\bibfield  {title} {\bibinfo {title} {Radiative corrections to the
  polarizability of helium},\ }\href@noop {} {\bibfield  {journal} {\bibinfo
  {journal} {Phys. Rev. Lett.}\ }\textbf {\bibinfo {volume} {92}},\ \bibinfo
  {pages} {233001} (\bibinfo {year} {2004})}\BibitemShut {NoStop}%
\bibitem [{\citenamefont {Piszczatowski}\ \emph {et~al.}(2015)\citenamefont
  {Piszczatowski}, \citenamefont {Puchalski}, \citenamefont {Komasa},
  \citenamefont {Jeziorski},\ and\ \citenamefont
  {Szalewicz}}]{Piszczatowski:Frequency}%
  \BibitemOpen
  \bibfield  {author} {\bibinfo {author} {\bibfnamefont {K.}~\bibnamefont
  {Piszczatowski}}, \bibinfo {author} {\bibfnamefont {M.}~\bibnamefont
  {Puchalski}}, \bibinfo {author} {\bibfnamefont {J.}~\bibnamefont {Komasa}},
  \bibinfo {author} {\bibfnamefont {B.}~\bibnamefont {Jeziorski}},\ and\
  \bibinfo {author} {\bibfnamefont {K.}~\bibnamefont {Szalewicz}},\ }\bibfield
  {title} {\bibinfo {title} {Frequency{-}dependent polarizability of helium
  including relativistic effects with nuclear recoil terms},\ }\href@noop {}
  {\bibfield  {journal} {\bibinfo  {journal} {Phys. Rev. Lett.}\ }\textbf
  {\bibinfo {volume} {114}},\ \bibinfo {pages} {173004} (\bibinfo {year}
  {2015})}\BibitemShut {NoStop}%
\bibitem [{\citenamefont {Schwerdtfeger}\ and\ \citenamefont
  {Nagle}(2019)}]{polarizability-table}%
  \BibitemOpen
  \bibfield  {author} {\bibinfo {author} {\bibfnamefont {P.}~\bibnamefont
  {Schwerdtfeger}}\ and\ \bibinfo {author} {\bibfnamefont {J.~K.}\ \bibnamefont
  {Nagle}},\ }\bibfield  {title} {\bibinfo {title} {2018 table of static dipole
  polarizabilities of the neutral elements in the periodic table},\ }\href@noop
  {} {\bibfield  {journal} {\bibinfo  {journal} {Molecular Physics}\ }\textbf
  {\bibinfo {volume} {117}},\ \bibinfo {pages} {1200} (\bibinfo {year}
  {2019})}\BibitemShut {NoStop}%
\bibitem [{\citenamefont {Jentschura}\ \emph {et~al.}(2015)\citenamefont
  {Jentschura}, \citenamefont {{\L}ach}, \citenamefont {DeKieviet},\ and\
  \citenamefont {Pachucki}}]{Jentschura:PRL}%
  \BibitemOpen
  \bibfield  {author} {\bibinfo {author} {\bibfnamefont {U.~D.}\ \bibnamefont
  {Jentschura}}, \bibinfo {author} {\bibfnamefont {G.}~\bibnamefont {{\L}ach}},
  \bibinfo {author} {\bibfnamefont {M.}~\bibnamefont {DeKieviet}},\ and\
  \bibinfo {author} {\bibfnamefont {K.}~\bibnamefont {Pachucki}},\ }\bibfield
  {title} {\bibinfo {title} {One-loop dominance in the imaginary part of the
  polarizability: application to blackbody and noncontact van der {W}aals
  friction},\ }\href@noop {} {\bibfield  {journal} {\bibinfo  {journal} {Phys.
  Rev. Lett.}\ }\textbf {\bibinfo {volume} {114}},\ \bibinfo {pages} {043001}
  (\bibinfo {year} {2015})}\BibitemShut {NoStop}%
\bibitem [{\citenamefont {Oelschl\"{a}ger}(2020)}]{Marty:thesis}%
  \BibitemOpen
  \bibfield  {author} {\bibinfo {author} {\bibfnamefont {M.}~\bibnamefont
  {Oelschl\"{a}ger}},\ }\emph {\bibinfo {title} {Fluctuation-induced phenomena
  in nanophotonic systems}},\ \href@noop {} {Ph.D. thesis},\ \bibinfo  {school}
  {Institut f\"{u}r Physik, Humboldt-Universit\"{a}t zu Berlin} (\bibinfo
  {year} {2020})\BibitemShut {NoStop}%
\bibitem [{\citenamefont {Li}\ \emph {et~al.}(2019)\citenamefont {Li},
  \citenamefont {Milton}, \citenamefont {Guo}, \citenamefont {Kennedy},\ and\
  \citenamefont {Fulling}}]{LY:ihm}%
  \BibitemOpen
  \bibfield  {author} {\bibinfo {author} {\bibfnamefont {Y.}~\bibnamefont
  {Li}}, \bibinfo {author} {\bibfnamefont {K.~A.}\ \bibnamefont {Milton}},
  \bibinfo {author} {\bibfnamefont {X.}~\bibnamefont {Guo}}, \bibinfo {author}
  {\bibfnamefont {G.}~\bibnamefont {Kennedy}},\ and\ \bibinfo {author}
  {\bibfnamefont {S.~A.}\ \bibnamefont {Fulling}},\ }\bibfield  {title}
  {\bibinfo {title} {Casimir forces in inhomogeneous media: Renormalization and
  the principle of virtual work},\ }\href@noop {} {\bibfield  {journal}
  {\bibinfo  {journal} {Phys. Rev. D}\ }\textbf {\bibinfo {volume} {99}},\
  \bibinfo {pages} {125004} (\bibinfo {year} {2019})}\BibitemShut {NoStop}%
\bibitem [{\citenamefont {Schwinger}\ \emph {et~al.}(1998)\citenamefont
  {Schwinger}, \citenamefont {{L. L. DeRaad, Jr.}}, \citenamefont {Milton},\
  and\ \citenamefont {{W.-y. Tsai}}}]{Schwinger:ce}%
  \BibitemOpen
  \bibfield  {author} {\bibinfo {author} {\bibfnamefont {J.}~\bibnamefont
  {Schwinger}}, \bibinfo {author} {\bibnamefont {{L. L. DeRaad, Jr.}}},
  \bibinfo {author} {\bibfnamefont {K.~A.}\ \bibnamefont {Milton}},\ and\
  \bibinfo {author} {\bibnamefont {{W.-y. Tsai}}},\ }\href@noop {} {\emph
  {\bibinfo {title} {Classical Electrodynamics}}}\ (\bibinfo  {publisher}
  {Perseus/Taylor and Francis},\ \bibinfo {address} {Reading, MA},\ \bibinfo
  {year} {1998})\BibitemShut {NoStop}%
\end{thebibliography}%
\end{document}